\documentclass[11pt]{article}
\pdfoutput = 1
\DeclareUnicodeCharacter{202F}{\,} 

\PassOptionsToPackage{svgnames}{xcolor}
\usepackage{xcolor}

\definecolor{cottoncandy}{rgb}{1.0, 0.74, 0.85}
\definecolor{cornellred}{rgb}{0.7, 0.11, 0.11}
\definecolor{darktangerine}{rgb}{1.0, 0.66, 0.07}

\usepackage[utf8]{inputenc}
\usepackage{color,graphicx}
\usepackage{verbatim}
\usepackage{enumerate}
\usepackage{amssymb}
\usepackage{stmaryrd}
\usepackage{physics}
\usepackage{amsfonts}
\usepackage{cite}
\usepackage{array}
\usepackage{setspace}
\usepackage{float}
\usepackage{url}
\usepackage{mathtools}
\usepackage{bbold}
\usepackage{tikz}
\usetikzlibrary{decorations.pathmorphing}
\tikzset{node distance=2cm, auto}
\tikzset{snake it/.style={decorate, decoration=snake}}
\usepackage{graphicx}
\usepackage{wrapfig}
\usepackage{subfigure}
\usepackage[labelfont=bf,font={rm},font={small}]{caption}
\usepackage{tikz}
\usepackage{cite}
\usepackage{cancel}
\usepackage{amsmath}
\usepackage{nicematrix}

\allowdisplaybreaks
\usepackage[most]{tcolorbox}

\usepackage[T1]{fontenc} 

\usepackage{braket}
\usepackage{todonotes}
\usepackage{braket}
\usepackage{appendix}
\usepackage{float}
\usepackage{array}
\usepackage{comment}
\usepackage{xcolor}
\usepackage{amsmath}

\usepackage{amsmath, amssymb, graphics, setspace}
\usepackage{tcolorbox} 
\usepackage[margin = 2.2cm]{geometry}
\newcommand{\mathsym}[1]{{}}
\newcommand{\unicode}[1]{{}}

\usepackage[margin = 2.2cm]{geometry}
\usepackage[ragged]{footmisc}
\usepackage{hyperref}
\hypersetup{
    colorlinks,
    citecolor=blue,
    filecolor=black,
    linkcolor=blue,
    urlcolor=blue,
    linktocpage=true
}

\newcommand{\rmi}{{\mathrm{i}}}
\newcommand{\rmd}{{\mathrm{d}}}
\newcommand{\CR}{{\mathcal R}}

\def\e{{\rm e}}

\newcommand{\be}{\begin{equation}}
\newcommand{\ee}{\end{equation}}
\newcommand{\bea}{\begin{align}}
\newcommand{\eea}{\end{align}}
\newcommand{\bi}{\begin{itemize}}
\newcommand{\ei}{\end{itemize}}

\def\XXint#1#2#3{{\setbox0=\hbox{$#1{#2#3}{\int}$}
     \vcenter{\hbox{$#2#3$}}\kern-.5\wd0}}

\numberwithin{equation}{section}

\title{Template}

\begin{document}

\thispagestyle{empty}
\begin{center}
\vspace*{.4cm}
 
    {\LARGE \bf 
Non-perturbative data for  \\ Weil-Petersson volumes and intersection numbers\\\bigskip  using ordinary differential equations
  }
    
    \vspace{0.4in}
    {\bf Clifford V. Johnson$^{\dagger}$, 
      João Rodrigues$^{\ddagger,\dagger}$
    }

\bigskip\bigskip

    {
$^{\dagger}$
Physics Department, Broida Hall, University of California,\\ Santa Barbara, CA 93106, USA \\ 
$^{\ddagger}$CAMGSD, Departamento de Matem\'atica, Instituto Superior T\'ecnico,\\ Universidade de Lisboa, 1049-001 Lisboa, Portugal}
    \vspace{0.1in}

    {\tt cliffordjohnson@ucsb.edu},  {\tt joao.carlos.rodrigues@tecnico.ulisboa.pt}
\end{center}

\vspace{0.4in}
\begin{abstract}
\noindent 
Recently, a new method was  introduced for computing  $V_{g,1}(b)$,  the Weil-Petersson volumes of the moduli space of Riemann surfaces of genus $g$ with one geodesic boundary of length~$b$,  various supersymmetric generalizations of them, as well as  analogous quantities in intersection theory. The physical setting is the computation of a certain one-point function in a variety of models of 2D gravity for which there is a double-scaled random matrix model (RMM) description.  The  method   combines perturbative solutions of two ordinary differential equations (ODEs), the  Gel'fand-Dikii resolvent equation, and   the RMM's string equation.  In this paper, we extend the method  to extract   non-perturbative information about the $V_{g,1}(b)$  (and their analogues) that is  naturally contained in the full ODEs, providing an  efficient prescription for  computing the transseries coefficients of  the one-point correlation  function, fully incorporating ZZ-brane and  FZZT-brane effects, and for the first time, mixed ZZ-FZZT-effects. We use as a case study the $(2,3)$ minimal string,  computing  perturbative  and non-perturbative quantities,  comparing them to perturbative results from  topological recursion,  and to  results from the recent non-perturbative topological recursion framework. As a particularly powerful further application we provide general predictions for the large order in $g$ growth of  $V_{g,1}(b)$, and apply them to JT gravity, finding agreement with known results,  and for analogous quantities in $\mathcal{N} {=} 1$ JT supergravity, proving a conjecture of Stanford and Witten. Our predictions yield  new growth formulae for the cases of  $\mathcal{N} {=} 2$ and ${\cal N}{=}4$ JT supergravity.
\end{abstract}
\pagebreak

\setcounter{page}{1}

\tableofcontents

\newpage


\section{Introduction}


\subsection{Background and Motivation}
Random matrix models, recognized for  half a century as powerful tools for enumerating properties of certain  classes of 2D surfaces~\cite{Brezin:1977sv}, have long played a highly instructive role in quantum gravity. These models are typically built from $N \times N$ matrices $H$, drawn from an ensemble with probability  $P(H)=\exp\{-N\mathrm{Tr}[V(H)]\}$, for some suitable potential $V(H)$ (discussed further below). The partition function of such a model has a  Feynman diagram expansion, and when organized using ’t Hooft’s double-line notation~\cite{'tHooft:1973jz},  maps naturally  onto discretized two-dimensional surfaces. For large $N$, this expansion arranges itself as a topological series where each diagram corresponds to a surface with a given number of handles ($g$) and boundaries ($n$), and contributes a factor proportional to $N^{\chi}$, where $\chi {=} 2 {-} 2g {-} n$ is the Euler characteristic of the discretized surface. 
The results for the Euclidean path integral over smooth two-dimensional geometries are  recovered by taking the limit $N {\to} \infty$ while simultaneously tuning the parameters of the potential $V(H)$ to focus on the  contributions from surfaces that are infinitely large in comparison to the typical scale size of the discretization~\cite{d90b,g90,b90}. This ``double scaling limit'' (DSL) is a continuum limit that transforms random matrix models into theories that fully capture the physics of a large class of Euclidean two-dimensional quantum gravity theories, depending  on the choice of potential~$V(H)$, and the properties of the ensemble from which $H$ is drawn.

Crucially, the random matrix model (RMM) does more than just capture topological perturbation theory, as useful as that alone can be.\footnote{Even at a given  order in the genus expansion, random matrix models are powerful and swift methods for computing quantities that are frequently much more difficult to calculate using continuum gravity or conformal field theory techniques.} It supplies {\it non-perturbative} information as well. This non-perturbative aspect will be a primary focus of this paper. It has implications for several important  geometrical quantities, such as the symplectic invariants that capture the intersection numbers of various Chern classes on the moduli space of marked Riemann surfaces, the Weil-Petersson volumes $V_{g,n}(b)$ of bordered Riemann surfaces of genus $g$ and $n$ boundaries, and various supersymmetric generalizations. 

Our work  extends the recently introduced alternative method~\cite{Johnson:2024bue} for efficiently computing  $V_{g,1}(b)$ (and various extensions and intersection theory analogues) by extracting them perturbatively from an   ordinary differential equation (ODE), the  Gel'fand-Dikii resolvent equation~\cite{gd75} combined with the ``string equation'' ODE of the relevant matrix model. We use transseries methods to extract  the {\it  non-perturbative information} about the quantities of interest  that is naturally contained in the full solutions of the ODEs, and thereby obtain a wealth of valuable information. The method is quite general, with applications to a wide class of models, and it gives access to many features that have so far been hard to obtain by other non-perturbative methods. 

In what follows next, we  unpack the necessary background and context for our methods and results in  Subsections~\ref{sec:RMM-and-2D-gravity} to \ref{sec:resurgence}. Readers already familiar with the setting can skip to Section~\ref{sec:results-tour} for a detailed guide to the results obtained in this paper.

\subsubsection{Random matrix models and 2D gravity}
\label{sec:RMM-and-2D-gravity}
The simplest examples of the RMM/2D-gravity pairing are provided by the infinite subfamily (denoted $(2,2k-1)$ for $k\in\mathbb{N}$) of ``minimal string theories'', which have a continuum description as the $(2,2k-1)$ minimal conformal field theories coupled to Liouville gravity\cite{d88,d89}. They are dual to double scaled Hermitian one-matrix models with special polynomial potentials~\cite{d90b,g90,b90}. (See {\it e.g., } ref.~\cite{Ginsparg:1993is} for a classic review and {\it e.g.,} refs.~\cite{Seiberg:2003nm,Seiberg:2004at,mmsss04} for several further aspects.)
More recently, the perturbative expansion of the Euclidean gravitational path integral (GPI) approach to Jackiw–Teitelboim (JT) gravity \cite{t83,j84} was shown to be equivalent to a double-scaled Hermitian one-matrix model \cite{Saad:2019lba}. The correspondence was uncovered by  using an equivalence between RMM loop equations and topological recursion \cite{eo07}. (Some of this will be reviewed later). The potential for the model can be described in terms of a particular combination of the potentials for the $(2,2k-1)$ models. Similar matrix model realizations have since been identified for various supersymmetric extensions, including  fully non-perturbative definitions. In particular, various $\mathcal{N} {=} 1$ JT supergravity models were shown in \cite{Stanford:2019vob} to be perturbatively equivalent to  double-scaled positive Hermitian matrix models, with non-perturbative formulations derived in refs.~\cite{Johnson:2019eik, Johnson:2020mwi, Johnson:2020exp, Johnson:2021owr}.  Meanwhile $\mathcal{N} {=} 2$ JT supergravity was later argued to have a matrix model description\cite{Turiaci:2023jfa},  with further study and a non-perturbative completion presented in ref.~\cite{j23b}. Most recently, fully non-perturbative random matrix model descriptions of various ${\cal N}{=}4$ and ${\cal N}{=}3$ JT supergravity theories~\cite{Heydeman:2020hhw,Heydeman:2025vcc} have been presented in refs.~\cite{Johnson:2024tgg,Johnson:2025oty}. In the context of critical string theories, double-scaled Hermitian matrix models have also seen recent developments. They were shown to  reproduce the perturbative genus expansion of the Virasoro minimal string \cite{cemr23,Castro:2024kpj,Johnson:2024bue}, as well as various $\mathcal{N} = 1$ supersymmetric extensions \cite{Johnson:2024fkm,Johnson:2025vyz}. More recently, the complex Liouville string, yet another critical string theory, was discovered and found to be dual to a suitable double-scaling limit of a Hermitian two-matrix model \cite{cemr24a,cemr24b}. 

A rather different perspective on the duality between double-scaled random matrix models and two-dimensional quantum gravity emerges from a holographic viewpoint. Inspired by higher-dimensional holography~\cite{tHooft:1993dmi,Susskind:1994vu,Maldacena:1997re}, it is to be expected that two-dimensional quantum gravity should be dual to a one-dimensional, non-gravitational theory governed by a fixed Hamiltonian $H$ with a discrete spectrum. However, the Euclidean path integral's definition is formulated in terms of smooth  geometry, and so might be suspected as providing only a coarse-grained description of the underlying microscopic physics. This is fully borne out by the matrix model correspondence: It shows that the coarse-graining in terms of geometry is precisely achieved by a statistical sampling from an ensemble of  Hamiltonians. Rather than pointing to a particular spectrum, it yields average properties of the dual spectrum, while nevertheless still incorporating microscopic features attributable to discreteness such as aspects of the late time spectral form factor~\cite{Saad:2018bqo,Saad:2019lba}, probability peaks for individual energy levels~\cite{Johnson:2021zuo}, {\it etc.}.

In particular, given  a definite Hamiltonian $H$, with eigenvalues $\left\{E_i \hspace{1pt} \vert \hspace{1pt} i {\in} \text{I}\right\}\subset \mathbb{R}$, we might consider the associated canonical partition function at some inverse temperature $\beta$:
\begin{equation}
    Z(\beta) = \sum_{i \in \text{I}} e^{-\beta E_i} = \text{Tr}[e^{-\beta H}]\ .
\end{equation}
Working instead with  the  ensemble of Hamiltonians, $H$, we no longer have access to the canonical partition function of a single Hamiltonian, but only to its ensemble average. The beauty of the class of double scaled random matrix models under consideration is that, once the dust settles, the  computation of averages can be performed in an elegant auxiliary framework~\cite{Banks:1990df}. For example,   the average partition function is the expectation value of a ``macroscopic loop operator'', computed as follows:
\begin{equation}
    \left\langle Z (\beta)\right\rangle = \int_{-\infty}^{\mu} \!\!\rmd x\, \langle x \vert e^{-\beta \mathcal{H}} \vert x \rangle
     \label{eq:int1}
\end{equation}
where $\cal H$ is the  Hamiltonian of an auxiliary quantum mechanics problem on the real line $x\in\mathbb{R}$:
\begin{equation}
    \mathcal{H} = -\hbar^2 \frac{\partial^2}{\partial x^2} + u(x) \ ,
    \label{eq:auxiliary-hamiltonian}
\end{equation}
where $\hbar$ is a (renormalized) $1/N$ expansion parameter, $\mu$ is a parameter to be discussed later, and  $u(x)$ is a solution of the ``string equation'' of the matrix model \cite{d90b,b90,g90}.  In the bosonic case (using a convenient unified set of conventions to be unpacked later),
the string equation can generically be written in the following form:
\begin{equation}
    \mathcal{R} = \sum_{k=1}^{\infty} t_k R_k[u] + x = 0
    \label{eq:int2}
\end{equation}
where $R_k[u]$ is the $k$th Gel’fand-Dikii polynomial in $u(x)$ and its 
$x$-derivatives\cite{gd75}, normalized here so that the coefficient of the purely polynomial part is unity. The generic double-scaled model defined by the string equation above can be interpreted as a combination of ``multicritical'' models generalizing~\cite{k89} the classic Gaussian case, weighted by the coefficients 
$\{ t_k \mid k \in \mathbb{N} \} \subset \mathbb{R}$.  Different classes of  matrix ensemble yield different kinds of string equation (another kind, useful for supersymmetric models, will be discussed  below), while making different choices for  the set $\{t_k\}$ is equivalent to specifying (fully non-perturbatively through the string equation) the  matrix model potential $V(H)$. Both of these aspects  result in different two-dimensional quantum gravity models. 

As a result of averaging over the ensemble, $\langle Z(\beta)\rangle$ can be written as  the Laplace transform of a continuous density of states $\rho(E)$:
\begin{equation}
    \label{eq:define-spectral-denisty}
    \langle Z(\beta)\rangle=\int dE \rho(E) \e^{-\beta E}\ .
\end{equation}
These quantities have a perturbative  part which is a topological expansion, as well as non-perturbative pieces:
\begin{equation}
    \langle Z(\beta) \rangle = \sum_{g=0}^{\infty} \hbar^{2g-1} \langle Z(\beta) \rangle_g  +\cdots\ ,\qquad \rho(E) =
    \sum_{g=0}^{\infty} \hbar^{2g-1} \rho_g(E)\ , 
    \label{eq:int3}
\end{equation}
where $\hbar$ is the (renormalized) $1/N$ parameter that counts topology, and $g$ counts genus.
An important quantity of interest is the leading (at large $N$) spectral density of the model, $\rho_0(E)$, which comes from the disc topology. It has a representation in terms of the leading solution $u_0(x)$ to the string equation: 
\begin{equation}
   \rho_0(E) = \frac{1}{2\pi\hbar}\int_{-\infty}^\mu \frac{dx}{\sqrt{E-u_0(x)}}
   =
   \frac{1}{2\pi\hbar}\int_{E_0}^E \frac{f(u_0)du_0}{\sqrt{E-u_0}}\ ,\quad
   \text{where}\quad f\equiv -\frac{dx}{du_0}
   \quad\text{and}\quad E_0=u_0(\mu)\ .\label{eq:density-leading}
\end{equation}
In the last expression, it can be seen that the parameter $\mu$ sets the edge or threshold energy of the model $E_0$, {\it i.e.,} $E_0 \equiv u_0(\mu)$. As an example, consider the old ``pure gravity'' model where the leading string equation can be taken to be (in the conventions of \cite{gs21}):
\begin{equation}
\frac{3}{4\sqrt{2}}u_0^2+x=0\ ,
\quad\text{and so}\quad 
    f= -\frac{dx}{du_0} = \frac{3}{2\sqrt{2}}u_0\ ,
    \label{eq:2-3-model-string-eqn}
\end{equation}
and in  the convention choice where $E_0 = 1$, the leading string equation gives 
\begin{equation}
    \mu = -\frac{3}{4\sqrt{2}}
    \label{eq:toprec4}
\end{equation}
and so:
\begin{equation}
    \rho_0(E)=\frac{1}{2\sqrt{2}\pi\hbar}(1+2E)\sqrt{E-1}\ .
    \label{eq:density-result}
\end{equation}
Very central later on will be the spectral curve of the matrix model, which follows from $\rho_0(E)$ by going to the cover of the cut $E$ plane {\it via} $1-E=2z^2$, and writing $\rho_0(z)=\frac{y(z)}{2\pi{\rm i}\hbar}$, we have the pair:
\begin{eqnarray}
         (-)E(z) &=&  2z^2-1=T_2(z) \ , \label{eq:chebychev-pure-gravity-x}\\
         y(z) &=&  4 z^3-3z=  T_3(z) \ .
    \label{eq:chebychev-pure-gravity}
\end{eqnarray}
where $T_m(z)$ is the $m$th Chebychev polynomial of the first kind: $T_m(\cos\theta){\equiv}\cos(m\theta)$. This model, which we will return to later for some illustration of our methods, is the $l=2$ entry in the $(2,2l-1)$ minimal string family, which  more generally has~\cite{Seiberg:2004at}:
\begin{eqnarray}
 \label{eq:chebychev-minimal-series}
         (-)E(z) &=& T_2(z) = 2z^2-1\ , \nonumber\\
         y(z) &=&  T_{2l-1}(z) \ .
\end{eqnarray}
Note that there is a different set of conventions that is also useful, which can be obtained by shifting and rescaling according to~\cite{Mertens:2020hbs}:
\begin{equation}
\label{eq:change-of-conventions}
    E-1\longrightarrow \frac{8\pi^2}{(2l-1)^2}E\ ;\qquad  u(x)-1\longrightarrow \frac{8\pi^2}{(2l-1)^2}u(x)\ .
\end{equation}
This set of conventions yields a different decomposition of the $u(x)$ in string equation~(\ref{eq:int2}) in terms of multicritical models. The set $(\{t_k\},\mu)$  is now given by:
\begin{equation}
\label{eq:minimal-model-teekay}
    t_k=\frac{\pi^{2k-2}}{2k!(k-1)!}\frac{4^{k-1}(l+k-2)!}{(l-k)!(2l-1)^{2k-2}}\ ,\quad\mu=0\ .
\end{equation}
This tees up  nicely the  case of JT gravity. It is a model of 2D gravity coupled in a special way to a scalar, and with a parameter $S_0$ setting the strength of the (topological in 2D) Einstein-Hilbert term. Such models are interesting in their own right as toy quantum gravity models, but also naturally arise as the leading physics of the near-horizon, low-energy limit of higher dimensional charged black holes, and are therefore of considerable broader interest for quantum gravity. At $T{=}0$, in a semi-classical treatment,  the Bekenstein-Hawking ``extremal'' entropy of such black holes is $S_0$.

\begin{figure}
    \centering
\begin{tikzpicture}

\def\bangle{90};
\draw[fill = violet,fill opacity=0.2,line width = 0pt] (0,1.5) to[out = -90+\bangle, in = 90-\bangle] (0,0) to[out = 90+\bangle, in = -90-\bangle]cycle;

\draw[fill = violet,fill opacity=0.2,line width = 0pt] (0,1.5) to[out = 0, in = 180] (1,1)to[out = 0,in = 180] (3,2.5)to[out = 0,in = 180] (5,2)to[out = 0,in = 180] (7,2.5)to[out = 0,in = 90] (9,1)to[out = -90,in = 0] (7,-0.5)to[out = 180,in = 0] (5,0)to[out = 180,in = 0] (3,-0.5)to[out = 180,in = 0] (1,0.5)to[out = 180,in = 0] (0,0) to[out = 90+\bangle, in = -90-\bangle]cycle;

\draw[line width = 2pt] (0,1.5) to[out = 0, in = 180] (1,1)to[out = 0,in = 180] (3,2.5)to[out = 0,in = 180] (5,2)to[out = 0,in = 180] (7,2.5)to[out = 0,in = 90] (9,1)to[out = -90,in = 0] (7,-0.5)to[out = 180,in = 0] (5,0)to[out = 180,in = 0] (3,-0.5)to[out = 180,in = 0] (1,0.5)to[out = 180,in = 0] (0,0);
\draw[line width = 2pt] (0,0) to[out = 90-\bangle, in = -90+\bangle](0,1.5)to[out = 90+\bangle, in = -90-\bangle]cycle;

\node at (-0.8,0.7) {$\beta$};

\draw[line width = 0pt,fill = white] (2.5+0.2,1-0.1) to[out = 30, in = 180-30] (4-0.2,1-0.1) to[out = 180+25, in = -25]cycle;
\draw[line width = 2pt] (2.5,1) to[out = -30, in = 180+30] (4,1);
\draw[line width = 2pt] (2.5+0.2,1-0.1) to[out = 30, in = 180-30] (4-0.2,1-0.1);

\def\hspace{2}

\draw[line width = 0pt,fill = white] (2.5+0.2+\hspace,1-0.1) to[out = 30, in = 180-30] (4-0.2+\hspace,1-0.1) to[out = 180+25, in = -25]cycle;
\draw[line width = 2pt] (2.5+\hspace,1) to[out = -30, in = 180+30] (4+\hspace,1);
\draw[line width = 2pt] (2.5+0.2+\hspace,1-0.1) to[out = 30, in = 180-30] (4-0.2+\hspace,1-0.1);

\def\hspace{4}

\draw[line width = 0pt,fill = white] (2.5+0.2+\hspace,1-0.1) to[out = 30, in = 180-30] (4-0.2+\hspace,1-0.1) to[out = 180+25, in = -25]cycle;
\draw[line width = 2pt] (2.5+\hspace,1) to[out = -30, in = 180+30] (4+\hspace,1);
\draw[line width = 2pt] (2.5+0.2+\hspace,1-0.1) to[out = 30, in = 180-30] (4-0.2+\hspace,1-0.1);
\end{tikzpicture}
    \caption{Illustration of a genus three hyperbolic Riemann surface with a single  boundary of length $\beta$.}
    \label{fig:RiemannSurface}
\end{figure}
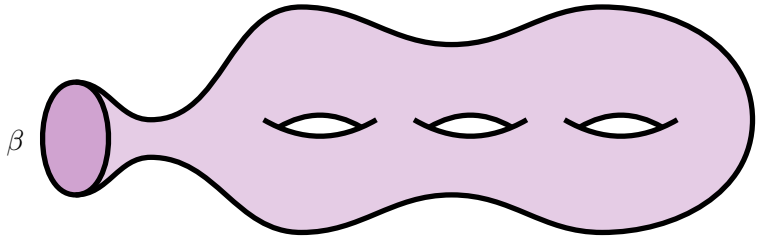

The two-dimensional Euclidean gravitational path integral (GPI) treatment of JT gravity becomes an  integral over all hyperbolic Riemann surfaces with an asymptotic boundary of length $\beta$, with a Schwarzian action for the boundary dynamics. (An illustration  is shown in figure \ref{fig:RiemannSurface}.) Again there is a natural  topological sum:
\begin{equation}
    Z^{\rm JT}(\beta) = \sum_{g=0}^{\infty} \hbar^{2g-1}  Z_g^{\rm JT}(\beta)  +\cdots\ ,
    \label{eq:GPI-sum}
\end{equation}
where $ Z_g^{\rm JT}(\beta) $ is the contribution to the GPI from hyperbolic Riemann surfaces of genus $g$. This time, $\hbar{\equiv}\e^{-S_0}$, so that large $S_0$ (appropriate when this is the extremal entropy in some semi-classical treatment of a near-horizon geometry of a black hole) allows a reliable perturbative treatment.
The striking observation of ref.~\cite{Saad:2019lba} is that there is a double-scaled random matrix model description of the GPI results. The precise statement in terms of the  language reviewed above,  the random matrix model naturally computes the same (topological) perturbative quantities through the perturbative (in~$\hbar$) expansion of the loop observable in equation~(\ref{eq:int1}), where $u(x)$ solves  the string equation~(\ref{eq:int2}), with the $(\{t_k\},\mu)$ now given by the infinite set~\cite{Dijkgraaf:2018vnm,os19,Johnson:2019eik}:
\begin{equation}
\label{eq:JT-teekay}
    t_k=\frac{\pi^{2k-2}}{2k!(k-1)!}\ ,\quad\mu=0\ . 
\end{equation}
Comparing to the expression~(\ref{eq:minimal-model-teekay}), JT gravity can also be interpreted as a large $l$ limit of the $(2,2l-1)$ minimal string series~\cite{Saad:2019lba}. Equivalently, it can simply be thought of as an infinite sum of multicritical matrix models. The leading JT gravity spectral density $\rho_0(E)=\sinh(2\pi\sqrt{E})/4\pi^2\hbar$ becomes, under the uniformization map $E=-z^2$, the spectral curve:
\begin{equation}
    y(z)=\frac{\sin(2\pi z)}{2\pi}\ ,
    \label{eq:JT-y}
\end{equation}
long known in the literature to arise from a special choice of couplings/deformations in the topological gravity spectral curve approach~\cite{e07,Eynard:2016yaa}.

Following up on remarks made a few pages above, it is in this precise sense that the random matrix model, an averaging over spectral properties of an ensemble of matrices in a certain class, can be  (at least perturbatively in topology\footnote{Notice that the statement of equivalence is to all orders in perturbation theory. Sometimes the matrix model itself is such that the double scaling limit tunes its potential to a place where it is non-perturbatively ill-defined. Even then, useful non-perturbative data can be extracted, for example from the nature of the asymptotic perturbative series itself.  Sometimes embedding into wider classes of  random matrix models can provide non-perturbative completions to JT gravity~\cite{Johnson:2019eik}. The situation is much more natural in many  cases pertaining to JT supergravity, where the matrix models often are well-defined at the outset. See early work and discussion in refs.~\cite{Johnson:2020exp,Johnson:2020heh,Johnson:2022wsr,Johnson:2021tnl}.}) equivalent to a  quantum gravity problem, defined through a Euclidean gravitational path integral. It is natural to interpret this equivalence as a precise realization of the idea that the GPI is a coarse-grained definition of  quantum gravity.

\subsubsection{Weil-Petersson volumes and topological recursion}
At any genus, the contribution to the partition function can be expressed as:
\begin{align}
     \langle Z(\beta) \rangle_g  & = \int_0^{+\infty}\rmd b \hspace{1pt} b Z_{\text{tr}}(\beta,b) V_{g,1}(b) 
     \label{eq:int5}
\end{align}
where:
\begin{equation}
    Z_{\text{tr}}(\beta,b) = \frac{1}{2\sqrt{\pi \beta}}\exp\left(-\frac{b^2}{4\beta}\right)
    \label{eq:trumpet-form}
\end{equation}
is the ``trumpet partition function'' \cite{Saad:2019lba}, and $V_{g,1}(b)$ denotes the Weil–Petersson volume associated with the Deligne–Mumford compactification of the moduli space of compact hyperbolic Riemann surfaces of genus $g$ with a single geodesic boundary of length $b$. A couple of examples are~\cite{Wolpert1983,penner1992weil,m06a}:
\begin{equation}
    \label{eq:sample-WP-volumes}
      V_{1,1}(b)=\frac{(4\pi^2+b^2)}{48}\ , \qquad  V_{2,1}(b) = \frac{\left(4 \pi^{2}+b^{2}\right) \left(12 \pi^{2}+b^{2}\right) \left(6960 \pi^{4}+384 \pi^{2} b^{2}+5 b^{4}\right)}{2211840} \ .
\end{equation}
The integral over $b$ can be regarded as the procedure of  gluing   the trumpet to the bulk geometry represented by $V_{g,1}(b)$.  The trumpet is universal in form, and more generally, correlation functions of $n$ copies of the partition function, $\langle Z(\beta_1)\cdots Z(\beta_n)\rangle$ are computed by gluing $n$ trumpets in the analogous way to $V_{g,n}(b_1,\cdots,b_n)$. These volumes are famously governed by Mirzakhani’s recursion relations \cite{m06a,m06b}, which with an integral kernel of a particular functional form,  determine all $V_{g,n}(b_1,\cdots,b_n)$ from just knowledge of $V_{0,3}(b_1,b_2,b_3) {=} 1$, and $V_{1,1}(b)$ in equation~(\ref{eq:sample-WP-volumes}) (or alternatively, the formal ``$V_{0,2}(b_1,b_2){=}(b_1^{-1}\delta(b_1-b_2))$''). We will not list the relations here, but refer the reader to ref.~\cite{m06a}.  After an appropriate Laplace transform, the recursion relations were shown by Eynard and Orantin to be equivalent to a form of  topological recursion \cite{eo07,e07} following from loop equations of  the  double-scaled matrix model. Concretely, we have the natural objects:
\begin{equation}
    \widehat{W}_{g,n}(z_1,\cdots,z_n) = \left[ \prod_{j = 1}^n\int_{0}^{+\infty} \rmd b_j \, b_j e^{-b_jz_j} \right] V_{g,n}(b_1,\cdots,b_n)
    \label{eq:int4}
\end{equation}
where the functions:
\begin{equation}
    \left\{\widehat{W}_{g,n}(z_1,\cdots,z_n) \hspace{1pt} \vert \hspace{1pt} g \in \mathbb{N}_0, n \in \mathbb{N} \right\}
\end{equation}
are the resolvent correlation functions of the double scaled matrix model, usually obtained by means of the topological recursion.
A couple of examples are:
\begin{equation}
    \label{eq:sample-W-resolvents}
      \widehat{W}_{1,1}(z)=\frac{\pi^2}{12z^2}+\frac{1}{8z^4}\ , \quad  \widehat{W}_{2,1}(z) =  \frac{105}{128 z^{10}}+\frac{203 \pi^{2}}{192 z^{8}}+\frac{139 \pi^{4}}{192 z^{6}}+\frac{169 \pi^{6}}{480 z^{4}}+\frac{29 \pi^{8}}{192 z^{2}}\ .
\end{equation}
The hat superscript indicates that they are written with respect to the uniformization variable, including the appropriate Jacobians.\footnote{The variables $\{z_i\}$ are related to the regular eigenvalue variables $\{E_i\}$ {\it via} a uniformization map, whose definition will be given later in the paper. For a detailed exposition, we refer the reader to \cite{ekr2018}.} Indeed, we can write:
\begin{equation}
    \widehat{W}_{n,g}(z_1,\cdots,z_n) = \left[\prod_{j=1}^n\frac{\rmd E}{\rmd z}(E_j)\right] W_{n,g}(E_1,\cdots,E_n)\ ,
\end{equation}
where the functions: 
\begin{equation}
    \left\{W_{n,g}(E_1,\cdots,E_n) \hspace{1pt} \vert \hspace{1pt} g \in \mathbb{N}_0, n \in \mathbb{N} \right\}
\end{equation}
are the regular resolvent correlation functions of the double scaled matrix model \cite{ekr2018}.

All the $\widehat{W}_{g,n}(z_!,\ldots,z_n)$ are determined by the topological recursion procedure, starting with the data in just  two basic objects:
\begin{equation}
  \widehat{W}_{0,1}(z)=2zy(z)\ , \quad \text{and}\quad  \widehat{W}_{0,2}(z_1,z_2)=\frac{1}{(z_1-z_2)^2}\ , 
\end{equation}
where the second  is the counterpart of the formal object $V_{0,2}(b_1,b_2)$ given above (\ref{eq:int4}). The first enters the integral kernel of the recursion, which will be reviewed in Section \ref{sec:gde}. The key point here is that while $\widehat{W}_{0,2}(z_1,z_2)$ is universal, $\widehat{W}_{0,1}(z)$ is determined by $y(z)$, the spectral curve of the model, which in turn is obtained from the leading spectral density of the model $\rho_0(E)$, as described above. 
So knowledge of the matrix model's leading spectral density (which is just the inverse Laplace transform of $\langle Z(\beta)\rangle_0$) determines all matrix model correlators, built from trumpets and the $V_{g,n}(b_1,\cdots,b_n)$. But this is the same structure as seen in the continuum approach to the GPI JT gravity model, and hence the two approaches are computing the same thing (for some implicitly defined matrix model potential)~\cite{Saad:2019lba}.

Analogous connections arise in the aforementioned supersymmetric extensions of JT gravity~\cite{Mertens:2017mtv,Stanford:2017thb,Forste:2017kwy,Heydeman:2020hhw,Heydeman:2025vcc}, where now the relevant volumes correspond to the compactification of the moduli space of compact super Riemann surfaces with a single geodesic boundary of length $b$ \cite{Stanford:2019vob,tw23}.
Moreover, ref.~\cite{Ahmed:2025lxe} has recently used random matrix model techniques to define the $V_{g,1}(b)$ for various random matrix realizations~\cite{Johnson:2024tgg,Johnson:2025oty}  of ${\cal N}{=}2$ and   ${\cal N}{=}4$ JT supergravity.

\subsubsection{Weil-Petersson volumes and an ODE method}
In  ref.~\cite{Johnson:2024bue}, it was  shown how to compute $V_{g,1}(b)$ efficiently by an alternative method to topological recursion. One instead  recursively solves  the Gel-fand-Dikii equation~\cite{gd75}, an ordinary differential equation (ODE)  for the diagonal resolvent  ${\widehat R}(x,E)\equiv\hbar\langle x|({\cal H}-E)^{-1}|x\rangle$ of the auxiliary Hamiltonian~(\ref{eq:auxiliary-hamiltonian}): 
\begin{equation}
\label{eq:gelfand-dikii}
    4(u-E)\widehat{R}^2-2\hbar^2\widehat{R}\widehat{R}''+\hbar^2(\widehat{R}')^2=1\ ,
\end{equation}
which takes $u(x)$, which itself solves the random matrix model's string equation,  as input. A prime denotes an $x$-derivative. Define the series expansions:
 \begin{equation}
 \label{eq:u-and-R-expansions}
     \widehat{R}(x,E)=\sum_{g=0}^{+\infty} \widehat{R}_g(x,E) \hbar^{2g}+\cdots\ ,\qquad  u(x){=}\sum_{g = 0}^{+\infty}u_{2g}(x)\hbar^{2g}+\cdots\ , 
 \end{equation}
(where again the ellipses denote possible non-perturbative parts).  The leading contribution is $\widehat{R}_0(x,E)=\pm\frac12(u_0(x)-E)^{-\frac12}$, and higher orders are made of terms involving $(u_0(x)-E)^{-\frac12}$ to some integer power, with factors built from the $u_g(x)$ and their derivatives. (The reader can look ahead to {\it e.g.,} equation~(\ref{eq:expansion-of-R}).) For general $u(x)$, the $\widehat{R}_g(x,E)$ become increasingly complicated as $g$ grows. 

Remarkably, however~\cite{Johnson:2024bue,Lowenstein:2024gvz,Ahmed:2025lxe}, for $g>0$, if $u(x)$ satisfies the string equation to some order in the $\hbar$ expansion, then by expanding the Gel'fand-Dikii equation to that order, the resulting ${\widehat R}_g(x,E)$ is a total $x$-derivative! This is very handy, since the physics of interest follows from integrating it with respect to~$x$. In fact the following identification with the quantities of the previous Section can be made:
\begin{equation}
    \label{eq:W-R-relation}
 W_{g,1}(E) = \int_{-\infty}^\mu\!\! {\widehat R}_g(x,E)\,\rmd x\quad  {\it i.e.,}\quad  \widehat{W}_{g,1}(z) = -2z\int_{-\infty}^\mu\!\! {\widehat R}_g(x,E)\,\rmd x\ ,
\end{equation} after translating to the double cover of the (cut) $E$-plane using  the uniformization variable $z$ defined here as $z^2=u_0(\mu)-E$.
Mirzakhani's crucial observation that the $V_{g,n}(\{ b_i\})$ are polynomial in the~$b_i$, {\it i.e.,} that in this case the $\widehat{W}_{g,1}(z)$ are polynomial in  $1/z$,  is guaranteed in this method by the fact that the ${\widehat R}_g(x,E)$ are total derivatives: It follows since the ${\widehat R}_g(x,E)$ are  {\it by construction} power series in $(u_0(x)-E)^{-\frac12}$, with coefficients  that can all be written in terms of the derivatives of $u_0(x)$.  Furthermore $ u_0(\mu)\equiv E_0$, the threshold energy of the leading distribution $\rho_0(E)$, and the derivatives all vanish at $x=-\infty$. So evaluating at the boundaries of the integral simply yields a power series in $(E_0-E)^{-\frac12}$, which can be replaced by the inverse of the uniformization variable $z^{-1}$. In the conventions chosen here, $E_0=0$, but this works more generally~\cite{Lowenstein:2024gvz,Ahmed:2025lxe}, as will be seen later.

A key motivation for the present paper is the fact that the  Gel'fand-Dikii ODE~(\ref{eq:gelfand-dikii}) also contains non-perturbative information, and so by extending the method just described to extract it, valuable information about the $V_{g,1}(b)$ beyond order by order perturbation theory should be accessible.

In fact, this presents an interesting opportunity, as there have been steady developments of matrix integral saddle-point expansion techniques that go beyond the standard topological recursion. This line of work began with the foundational analysis of free energy instantons in \cite{msw07}, was later refined in \cite{mss22}, and more recently generalised to correlation functions in \cite{eggls23}, where the framework was extended by leveraging the loop insertion operator \cite{Eynard:2007kz} in what was dubbed the ``non-perturbative topological recursion''.

A natural question therefore is how the non-perturbative information that can be found by digging deeper into  the ODE method compares with that defined by non-perturbative topological recursion. In principle they {\it could} yield different results, since (at least on the face of it) they are different recursive methods, non-perturbatively extended using distinct techniques!

Our paper will address this directly, showing agreement whenever the techniques overlap, but our methods in fact give  {\it more} non-perturbative data than currently is possible using the techniques of ref.~\cite{eggls23}.

It is important to note that the ODE method we will describe is not the first approach to leverage recursive relations among Weil-Petersson volumes in order to extract non-perturbative information about the theory. Indeed, the volumes $V_{0,n}(b_1,\cdots,b_n)$ satisfy a recursion relation, first derived in \cite{z93}, which was subsequently employed in \cite{m94} to obtain a continuum, non-perturbative formulation of the specific heat $u(x)$ associated with $(2,3)$ minimal string theory. Unlike our approach, which relies crucially on the matrix model duality, this method is instead rooted in the connection between Liouville theory and the uniformization theory of Riemann surfaces \cite{zt88a,zt88b,t89,z90}. Closely related techniques were later applied to Seiberg-Witten theory in \cite{b04}, where they reproduced the recursion relations governing the instanton contributions to the prepotential of $\mathcal{N}=2$ super Yang-Mills theory with gauge group $\text{SU}(2)$ \cite{m95}.

Before describing our results it is worth reviewing the techniques of recursion and transseries.

\subsubsection{Asymptotic series, non-perturbative effects, and resurgence}
\label{sec:resurgence}

Notably, the topological series expansion for the one-point correlator of the  partition function \eqref{eq:int3} is typically asymptotic, signalling the necessity to incorporate additional non-perturbative information beyond the $\langle Z(\beta)\rangle_g$. (The same is true, through~(\ref{eq:int5}), for the asymptotic series built from Weil-Petersson volumes $V_{g,1}$.) Resurgence techniques (see ref.~\cite{abs19})  offer a systematic approach to incorporating non-perturbative effects by organizing them into a hierarchy of asymptotic series, each weighted by exponential terms taking the generic form
%
 $   \exp\left(-{A}/{\hbar}\right)$ 
%
for some instanton action $A \in \mathbb{C}$, known as ``transmonomials''. These weighted series are known as transseries sectors and together, they form what is known as a transseries, which captures both perturbative and non-perturbative contributions within a unified analytic framework. For the topological problems of interest here, the methods of topological recursion provide an efficient way to compute the usual perturbative sector of the transseries while the non-perturbative sectors, until recently, could only be derived indirectly by studying the large order growth of the perturbative coefficients~\cite{abs19}.

The transseries of the one--point function \eqref{eq:int1} is naturally inherited from those of the connected correlation functions \eqref{eq:int4}, whose specific forms were examined in an initial study in \cite{eggls23}. These transseries feature sectors that are associated with distinct instanton actions, which can be classified according to the non-perturbative D-brane configuration they describe in the context of the Hermitian random matrix model descriptions of minimal strings. Concretely, there are transseries sectors associated with ZZ branes \cite{zz01}, FZZT branes \cite{fzz00,t00} and  combinations of both (ZZ-FZZT).

The non-perturbative topological recursion procedure yields analytical predictions for the transseries coefficients associated with ZZ-sectors \cite{eggls23}. Moreover, an analytical prescription to compute the one-point function transseries coefficients associated with the FZZT-sector has been developed for JT gravity in \cite{Saad:2019lba} and further employed in \cite{os19}. However, generically computing transseries coefficients associated with ZZ-FZZT-sectors remains an open problem. {\it The methods we develop in this paper based on the ODE method will capture all of these sectors with equal ease.}

\subsection{The results of this paper}
\label{sec:results-tour}
In this work, we present a systematic and efficient method to compute all transseries sectors appearing in the one-point correlation function, thereby obtaining a full non-perturbative transseries completion of the canonical partition function \eqref{eq:int1}. Our approach relies on the Gel’fand–Dikii equation. As mentioned (and briefly reviewed) earlier, in ref.\cite{Johnson:2024bue}, a method was established to compute the perturbative expansion of the one-point correlation function by means of a perturbative series ansatz for the solution of the Gel’fand–Dikii equation~(\ref{eq:gelfand-dikii}), supplemented by information from the string equation. It was explored further and refined in refs.~\cite{Johnson:2024fkm,Lowenstein:2024gvz,Ahmed:2025lxe}. Here, we extend the method to compute all transseries sectors of the one-point correlation function, including ZZ, FZZT, and ZZ–FZZT-effects alike. This is achieved by generalising the perturbative series ansatz~(\ref{eq:u-and-R-expansions}) for the Gel’fand–Dikii resolvent to a transseries form, whose coefficients can be solved for recursively, using the Gel’fand–Dikii equation.

This result is interesting and extremely useful in its own right, as no other systematic method for computing the full non-perturbative transseries content of the one-point correlation function was, to the author’s best knowledge, known in the literature. As an application, we provide fully generic predictions for the large order growth of moduli space volumes in two-dimensional gravity described by Hermitian matrix models, and apply them to JT gravity, finding agreement with known results \cite{eggls23}. The same approach is then applied to $\mathcal{N} = 1$ JT supergravity, proving a conjecture put forward in \cite{Stanford:2019vob}, to $\mathcal{N} = 2$ JT supergravity and finally to small and large $\mathcal{N} = 4$ JT supergravity.

In {\bf Section \ref{sec:gde}}, we address the perturbative genus expansion of the one-point correlation function and introduce two distinct recursive methods to compute its coefficients: the topological recursion reviewed in Subsection \ref{subsec:toprec}, and the Gel’fand–Dikii resolvent equation, whose connection to the genus expansion of the one-point correlation function is reviewed in Subsection \ref{subsec:ode}. We illustrate this in the context of the $(2,3)$ minimal model, a simple yet remarkably instructive example that we adopt as the working toy model throughout this work.

In {\bf Section \ref{sec:nonpert}}, we address the non-perturbative completion of the one-point correlation function perturbative genus expansion. In Subsection \ref{subsec:transseries}, we give concrete analytical meaning to non-perturbative completion of an asymptotic perturbative series, by reviewing the concept of a transseries. We further identify all distinct non-perturbative effects contributing to the one-point correlation function and show how they naturally lead to a transseries structure. In Subsection \ref{sec:nonperttopred}, we review the non-perturbative topological recursion framework and show how it can be used to predict part of the transseries structure presented in \ref{subsec:transseries}. In Subsection \ref{subsec:WKBexpansion}, we review the WKB expansion introduced in \cite{sss19} and subsequently applied in \cite{os19}, and show how it can be used to derive another part of the transseries structure presented in \ref{subsec:transseries}. Finally, in Subsection \ref{sec:transseriesfromgel}, we extend the method outlined in Subsection \ref{subsec:ode} to allow for the systematic computation of all transseries coefficients, highlighting the non-perturbative predictive power of the Gel’fand-Dikii equation. We further validate our results by matching them with the ones obtained via the non-perturbative topological recursion and the WKB expansion.




In {\bf Section \ref{sec:largeorder}}, we apply our results to obtain fully generic and analytic expressions for the large-order growth of the perturbative genus expansion of the one-point correlation function, as well as for moduli space volumes associated with a single asymptotic boundary. In Subsection \ref{subsec:largeorderformula}, we briefly review the interplay between transseries sectors and the large-order behaviour of asymptotic perturbative series. In Subsection \ref{subsec:largecorrelation}, we derive a fully generic and analytic expression for the large-order growth of the perturbative genus expansion of the one-point correlation function. Finally, in Subsection \ref{subsec:volumes}, we extend these results to moduli space volumes associated with a single asymptotic boundary. We further specialise our results to JT gravity and to $\mathcal{N}=1,2,4$ JT supergravity.

In {\bf Section \ref{sec:conclusion}}, we make several concluding remarks.

In {\bf Appendix \ref{appendix:Transseriesdata}}, we collect several transseries coefficients associated with the Gel’fand–Dikii resolvent and the one-point correlation function for $(2,3)$ minimal string theory, computed following the procedure outlined in Subsection \ref{sec:transseriesfromgel}. We also collect several transseries coefficients associated with the string equation solution $u(x)$ (sometimes called the ``specific heat''), the free energy, and the partition function of $(2,3)$ minimal string theory. Several of the free energy and partition function transseries coefficients presented here are used in the comparison with non-perturbative topological recursion carried out in Subsection \ref{sec:transseriesfromgel}.

In {\bf Appendix \ref{appendix:Matrixintegrals}}, we perform several truncated saddle-point expansions associated with matrix integrals arising in the non-perturbative topological recursion procedure reviewed in Subsection \ref{sec:nonperttopred}. The resulting formulae are then employed in the match with the non-perturbative topological recursion carried out in Subsection \ref{sec:transseriesfromgel}.

In {\bf Appendix \ref{appendix:BorelSummation}}, we provide a brief review of the Borel summation procedure. This procedure is required in order to extract exact predictions from the transseries constructed in Subsection \ref{sec:transseriesfromgel} and displayed in equation \eqref{eq:nonperef1} (see equation \eqref{eq:BorelSum}).


\section{The perturbative one-point function {\it via} two approaches}

\label{sec:gde}

In this Section, we study the random matrix model one-point function $\langle Z(\beta)\rangle$ in perturbation theory and show explicitly in a simple example the two different methods: topological recursion, and the  method of ref.~\cite{Johnson:2024bue} that uses the Gel'fand-Dikii ODE combined with the string equation. The two approaches will give the same answers  in quite different ways. This will be a good preparation for then going beyond perturbation theory, which is quite natural in the ODE approach. Since by now in a series of papers the method has been illustrated for the Airy (1,2) model, for JT gravity and various JT supergravity  and Virasoro minimal string variants~\cite{Johnson:2024bue,Johnson:2024fkm,Lowenstein:2024gvz,Ahmed:2025lxe}),  we will use  a new example for our case study,  the $(2,3)$ minimal string model.


\subsection{Topological recursion approach}

\label{subsec:toprec}
Let's compute the one-point correlation function for the $(2,3)$ minimal string model, starting with the topological recursion approach.
The core piece of data is the random matrix model's spectral curve.
%
The spectral curve is the Riemann surface $\Sigma$ defined by the algebraic equation 
\begin{equation}
    y^2(E) = \left(V'(E)-2W_{0,1}(E)\right)^2
\end{equation}
where $V(E)$ is the polynomial potential and $W_{0,1}(E)$ is the planar coefficient featuring in the one-point correlation function perturbative expansion. More generally, we have
\begin{equation}
    W_{n}(E_1,\cdots,E_n) =\left\langle \prod_{i=1}^n\text{Tr}\left[\frac{1}{E_i-M}\right]\right\rangle_c = \sum_{g=0}^{+\infty} W_{g,n}(E_1,\cdots,E_n)\hbar^{2g+n-2}.
    \label{eq:toprec1}
\end{equation}
The topological recursion \cite{eo07} offers a systematic and recursive method to compute the coefficients appearing in the perturbative expansion \eqref{eq:toprec1}. To formulate this procedure, it is convenient to introduce the meromorphic immersion
$i: \mathbb{P}^1 \to \Sigma \subset \mathbb{C} \times \mathbb{C}$
which maps the uniformization of the spectral curve onto itself. This immersion enables a redefinition of the (one-cut) spectral curve coordinates on $\mathbb{P}^1$. More concretely, we have
\begin{equation}
   \Sigma = \left\{ i(z) = (E(z),y(z)) \in \mathbb{C} \times \mathbb{C} \hspace{1pt} \vert \hspace{1pt} z \in \mathbb{P}^1 \right\}
\end{equation}
where the coordinate $z$ is commonly referred to as the uniformization variable. Using this variable, we can define the meromorphic differential forms
\begin{equation}
    \omega_{g,n}(z_1,\cdots,z_n) = W_{g,n}(z_1,\cdots,z_n) \rmd E_1(z_1)\cdots \rmd E_n(z_n) 
    \label{eq:omega-meets-W}
\end{equation}
for $2g+n-2 >0$ and
\begin{equation}
    \omega_{0,1}(z_1) = \frac{1}{2}y(z_1) \rmd E(z_1)
    \label{eq:omega-y-relation}
\end{equation}
as well as
\begin{equation}
     \omega_{0,2}(z_1,z_2) = B(z_1,z_2)
\end{equation}
where $B(z_1,z_2)$ is the Bergmann kernel of $\mathbb{P}^1$ \cite{ekr2018}. Taking as initial data the planar coefficients above, the recursion reads
\begin{equation}
\begin{split}
    \omega_{g,h} \left(z_1,J\right) = \sum_{r \in \CR}\underset{z \to r}{\text{Res}}\, \Bigg\{  K_r(z_1,z)\Big( \omega_{g-1,h+1} \left(z,\sigma(z),J\right) 
+ \sum_{\substack{m+m'=g \\ I\sqcup I'=J}}' \omega_{m,|I|+1} \left(z,I\right)\, \omega_{m',|I'|+1} \left(\sigma(z),I'\right) \Big) \Bigg\}
\end{split}
\label{eq:toprec2}
\end{equation}
where $J = \{z_2,\cdots,z_h\}$, $\CR$ indicates the set of all branch points of $E(z)$, $\sigma:\mathbb{P}^1 \to \mathbb{P}^1$ stands for the Galois involution \cite{ekr2018} associated with $i$ and the prime in the summation indicates that we should not include the index combinations $(I,m) = (J,g)$ and $(I',m') = (J,g)$. Moreover, we defined the recursion kernel
\begin{equation}
    K_r(z_1,z) =  \frac{1}{\omega_{0,1} \left(z\right) - \omega_{0,1} \left(\sigma(z)\right)}\, \left[\frac{1}{2} \int_{\sigma(z)}^{z}\hspace{1pt}\omega_{0,2} \left(z_{1},\bullet\right)\right]
\end{equation}
where the bullet/dot in the integrand marks the argument with respect to which the integration is performed.  Using the recursion \eqref{eq:toprec2}, we are able to recover all of the coefficients appearing in \eqref{eq:toprec1}. If the claim of the previous Subsection is to hold, then this means that topological recursion should produce the same coefficients computed recursively by plugging a perturbative power series ansatz in the Gel’fand-Dikii equation \eqref{eq:gde6}. In what follows, we will show this.


At this point, we are ready to  specialize to the $(2,3)$ minimal string theory, discussed in the Introduction. Given its spectral curve specified in~(\ref{eq:chebychev-pure-gravity}), we can  write:
\begin{equation}
\label{eq:omega_defintion}
    \omega_{0,1}(z_1) = 2 z_1^2(4z_1^2-3)\rmd z_1    
%
\ ,\quad \text{and}
\quad
%
    \omega_{0,2}(z_1,z_2)= \frac{\rmd z_1 \rmd z_2}{(z_1-z_2)^2}\ .
\end{equation}
Running the topological recursion \eqref{eq:toprec2} for the first couple of coefficients yields
\begin{align}
    &\omega_{1,1}(z_1) = \frac{4 z_1^2+3}{144 z_1^4}\rmd z_1 \label{eq:check-recur-1}\\ 
    & \omega_{2,1}(z_1) =\frac{7  \left(1024 z_1^8+768 z_1^6+576 z_1^4+348 z_1^2+135\right)}{248832 z_1^{10}} \rmd z_1\label{eq:check-recur-2}\\ &
     \omega_{3,1}(z_1) = \frac{7 }{644972544 z_1^{16}}\Big( 22937600 z_1^{14}+17203200 z_1^{12}+12902400 z_1^{10}+8893440 z_1^8+5495040 z_1^6+ \nonumber \\ &2908224 z_1^4+1189188 z_1^2+289575\Big) \rmd z_1.\label{eq:check-recur-3}
\end{align}
In each case, the coefficients of the polynomial in $1/z_1^2$ have a natural interpretation in intersection numbers arising from integrating two types of Chern classes (the $\psi$-classes and $\kappa$-classes\footnote{See {\it e.g.} ref.~\cite{Eynard:2016yaa} for a thorough review and ref.~\cite{Do2008TouristGuide} for a swift illuminating tour of key concepts.} over the moduli space of Riemann surfaces of genus $g$ with $n$ punctures. In this example, there is the explicit decomposition
\begin{eqnarray}
 \omega_{1,1}(z_1) &=& \left(\frac{1}{6}\right)\left(3\langle \tau_{1}\rangle_1\frac{1}{z_1^4}
 +
 4\langle \kappa_{1}\rangle_1\frac{1}{z_1^2}
 \right)dz_1\ ,\quad\text{with}\quad
 \langle \tau_{1}\rangle_1=\langle \kappa_{1}\rangle_1=\frac{1}{24}\ , \quad\text{and}
 \nonumber\\
    \omega_{g,1}(z_1) &=& \left(\frac{1}{6^{2g-1}}\right)\sum_{m=0}^{3g-2}\frac{4^m}{m!}(2(3g-2-m)+1)!!\langle \kappa_1^m\tau_{3g-2-m}\rangle_g\frac{dz_1}{z_1^{2(3g-2-m)+2}}\ ,
    \end{eqnarray}
    which yields, {\it e.g.,} for $g=2$, the numbers:
    \begin{eqnarray}
 \langle \tau_{4}\rangle_2=\frac{1}{(24)^2\cdot 2}\ ,\,
 \langle \kappa_1\tau_{3}\rangle_2=\frac{29}{5760}\ ,\,
  \langle \kappa_1^2\tau_{2}\rangle_2=\frac{7}{240}\ ,\,
   \langle \kappa_1^3\tau_{1}\rangle_2=\frac{7}{48}\ ,\quad\text{and}\quad
    \langle \kappa_1^4\tau_{0}\rangle_2=\frac{7}{12}\ .
\end{eqnarray}
In general, the highest  order in the expansion in $1/z_1^2$ contains the  pure $\psi$-class intersection: $\langle \tau_{3g-2}\rangle_g=\frac{1}{(24)^gg!}$ (as computed for example in the Airy model {\it aka} the (2,1) model or pure topological gravity\cite{Witten:1989ig,Witten:1990hr,Kontsevich:1992ti}), while the other orders are mixed terms involving contributions from both classes. (In physical terms, the extra class appears because the $(2,3)$ model has the gravitationally dressed identity CFT operator present, and  so is in fact a $\kappa$-deformation of the pure topological gravity case.\footnote{More generally, each gravitationally dressed primary field in the $(p,q)$ minimal string model introduces elements of certain higher Chern classes~\cite{Witten:1991mk}, for which this framework naturally computes intersection numbers.})

The ODE approach to be reviewed next will give a swift means of obtaining the $\omega_{g,n}(z_1)$ (for $n=1$)  for a large class of models. The striking thing, and the main subject of this paper, is that the ODE gives a very natural notion of how to go beyond genus perturbation theory, yielding valuable non-perturbative information (such as large $g$ growth) that is much harder to obtain by other methods.

\subsection{ODE  approach}

\label{subsec:ode}
Let's now turn to the ODE method of ref.~\cite{Johnson:2024bue} (explored further in refs.~\cite{Johnson:2024fkm,Lowenstein:2024gvz,Ahmed:2025lxe}) as a means of swiftly computing the same perturbative quantities. The point of this paper is that this method will allow for a very natural approach for obtaining information beyond just the perturbative expansion, through the use of the Gel'fand-Dikii resolvent and the ODE it satisfies.

First,   the reason that the Gel'fand-Dikii resolvent is directly relevant here is instructive to recall, as done in ref.~\cite{Ahmed:2025lxe}. Key is the fact that the basic one-point correlator $\langle Z(\beta)\rangle$ in equation~(\ref{eq:int1}) is built from the auxiliary quantum mechanics' Hamiltonian~(\ref{eq:auxiliary-hamiltonian}). It can be written as:
\begin{align}
  \langle Z(\beta)\rangle=    \int_{-\infty}^{\mu} \rmd x \left\langle x \vert e^{-\beta \mathcal{H}} \vert x \right\rangle &= \frac{1}{2\pi \rmi } \int_{\gamma-\infty \rmi}^{\gamma +\infty\rmi} \rmd E \exp\left(-\beta E\right) \int_{-\infty}^{\mu} \rmd x \left\langle x \left\vert \frac{1}{\mathcal{H}-E} \right \vert x \right \rangle \nonumber \\& = \frac{1}{ \hbar}\frac{1}{2\pi \rmi } \int_{\gamma-\infty \rmi}^{\gamma +\infty\rmi} \rmd E \exp\left(-\beta E\right) \int_{-\infty}^{\mu} \rmd x \widehat{R}(E,x)
    \label{eq:gde1}
\end{align}
where  the Gel’fand-Dikii
diagonal resolvent is:
\begin{equation}
    \widehat{R}(E,x) = \hbar\left\langle x \left\vert \frac{1}{\mathcal{H}-E} \right \vert x \right \rangle
    \label{eq:gde2}
\end{equation}
and $\gamma \in \mathbb{R}$ is chosen to be less than the real part of all roots of $ \widehat{R}(E,x) $. It is important to note what is being called~$E$ here. Since  $\langle Z(\beta)\rangle$ is the Laplace transform~(\ref{eq:define-spectral-denisty}) of the spectral density $\rho(E)$ where~$E$ is the energy, there is a relation between $\rho(E)$ and the $x$-integral of $ \widehat{R}(E,x) $, after a continuation on $E$, as will become clear later. 
Central to what is to come is that the resolvent~(\ref{eq:gde2}) satisfies the ordinary differential equation (ODE) that was recognized long ago  by Gel’fand and Dikii  as important for the study of integrable systems  \cite{gd75}, repeated here for convenience:
\begin{equation}
    4(u(x)-E)\widehat{R}^2(E,x) -2\hbar^2 \widehat{R}(E,x) \widehat{R}^{\prime\prime}(E,x) + \hbar^2 \left(\widehat{R}^\prime(E,x)\right)^2 = 1\ ,
    \label{eq:gde6}
\end{equation}
along with the  perturbative form of the solution, and the input potential $u(x)$:
 \begin{equation}
     \widehat{R}(x,E)=\sum_{g=0}^{+\infty} \widehat{R}_g(x,E) \hbar^{2g}+\cdots\ ,\qquad  u(x){=}\sum_{g = 0}^{+\infty}u_{2g}(x)\hbar^{2g}+\cdots\ , 
 \end{equation}
(where  the ellipses denote possible non-perturbative parts). 
%
%
Finally, because of the miracle that each ${\widehat{R}}_g(x,E)$ becomes a total derivative when $u(x)$ satisfies the string equation, there is the relation~(\ref{eq:W-R-relation}):
\begin{equation}
    \int_{-\infty}^{\mu} \rmd x \widehat{R}_g(E,x) = W_{g,1}(E) = \frac{d z}{dE}(z)\widehat{W}_{g,1}(z)
\ ,     \label{eq:gde5}
\end{equation}
after a change of variable from $E$ to what will turn out to be the natural uniformizing variable $z=z(E)$.
Generally speaking, for the appropriate $u(x)$, this allows a very efficient computation of the symplectic invariants $\widehat{W}_{g,1}(z)$ (built from intersection numbers) for a wide class of theories.

%
%
Continuing the development of the correlator, we see that at each genus:
\begin{align}
    \frac{1}{ \hbar}\frac{1}{2\pi \rmi } \int_{\gamma-\infty \rmi}^{\gamma +\infty\rmi} \rmd E \exp\left(-\beta E\right) \int_{-\infty}^{\mu} \rmd x \widehat{R}_g(E,x)  & =   \frac{1}{ \hbar}\frac{1}{2\pi \rmi } \int_{\gamma-\infty \rmi}^{\gamma +\infty\rmi} \rmd E \exp\left(-\beta E\right) \frac{d z}{dE}(E)\widehat{W}_{g,1}(E) \nonumber \\ & = \frac{1}{ \hbar}\frac{1}{2\pi \rmi } \int_{z(\gamma-\infty \rmi)}^{{z(\gamma+\infty \rmi)}} \rmd z \exp\left(\beta z^2\right)\widehat{W}_{g,1}(z) \nonumber  \\ &= \frac{1}{ \hbar}\frac{1}{2\pi \rmi } \int_{z(\gamma-\infty \rmi)}^{{z(\gamma+\infty \rmi)}} \rmd z \exp\left(\beta z^2\right)\int_{0}^{+\infty} \rmd b \hspace{1pt} b \exp\left(-b z\right) V_{g,1}(b) \nonumber \\ & = \frac{1}{ \hbar}\int_{0}^{+\infty} \rmd b \hspace{1pt} b Z_{\text{tr}}(\beta,b) V_{g,1}(b)\ ,
    \label{eq:gde3}
\end{align}
where we used that ({\it e.g.} for the case of JT gravity) the uniformization variable satisfies the simple relation $E(z) = -z^2$ and
we recover the trumpet~(\ref{eq:trumpet-form})  that features heavily in the GPI approach of ref.~\cite{Saad:2019lba} because:
\begin{equation}
    \frac{1}{2\pi \rmi } \int_{z(\gamma-\infty \rmi)}^{{z(\gamma+\infty \rmi)}} \rmd z \exp\left(\beta z^2-bz\right) = Z_{\text{tr}}(\beta,b)\ .
\end{equation}
Hence, using equation \eqref{eq:gde3}, we can finally write
\begin{equation}
\langle Z(\beta)\rangle = \sum_{g=0}^{+\infty} \hbar^{2g-1}\int_{0}^{+\infty} \rmd b \hspace{1pt} b Z_{\text{tr}}(\beta,b) V_{g,1}(b)
\end{equation}
which is in agreement with equation \eqref{eq:int5}. So given a $u(x)$ that satisfies a string equation, all we need is that the ${\widehat R}(x,E)$ satisfies the  ODE~(\ref{eq:gde6}),
and one can obtain the $\widehat{W}_{g,1}(z)$ and hence the $V_{g,1}(b)$ at any $g$, by just expanding the ODE in $\hbar$.


\label{subsec:odeapproach}

Let's now focus on the $(2,3)$ example to see how things work in detail. 
At leading order we have (picking a conventional sign) the following leading solution to~(\ref{eq:gde6}):
\begin{equation}
    \widehat{R}_0(E,x) = -\frac12\frac{1}{[u_0(x)-E]^{1/2}}
    \ ,
\end{equation}
and from the leading string equation~(\ref{eq:2-3-model-string-eqn}) we have $u_0(x) = \frac{2\sqrt[4]{2}}{\sqrt{3}}\sqrt{-x}$.
Recall that $\mu=-\frac{3}{4\sqrt{2}}$, and that $u_0(\mu)=1$, and so we have:
\begin{equation}
    \int_{-\infty}^\mu \widehat{R}_0(E,x) \rmd x = -\frac{3}{4\sqrt{2}}\int_{1}^E  \frac{u_0}{[u_0-E]^{1/2}}\rmd u_0=\frac{1}{2\sqrt{2}}(1+2E)\sqrt{1-E}=-\frac12(4z^3-3z)\ ,
\end{equation}
where we substituted\footnote{Compare this to equation~(\ref{eq:density-result}) and the transformation $E=1+2z^2$ done there. So $\rho(E)=\pi^{-1}{\rm Im}\int \widehat{R}(E,x)\rmd x$.}  $E=1-2z^2$. From this, we see that $\rmd E=-4z\rmd z$, and hence  we verify the $g = 0$ case of:
%
\begin{equation}
   \left(\int_{-\infty}^\mu\rmd x\widehat{R}_g(E(z),x)\right) \rmd E(z) = \omega_{g,1}(z)\ ,
   \label{eq:toprec3}
\end{equation}
obtained by rewriting equation~\eqref{eq:gde5}.
To see how this works  to higher order, note that the ${\widehat R}_g(E,x)$ generically depend on the all  $u_{2i}(x)$ from $i=0,\cdots,g$, (and their derivatives). However, the string equation yields relations between the $u_{2i}(x)$ and $u_0(x)$ and its derivatives, such as:
\begin{eqnarray}
 u_2(x)&=&-\frac{1}{12}\frac{\rmd^2}{\rmd x^2}\ln(u_0'(x))=\frac{u_0''^2-u_0'u_0'''}{12u_0'^2}\ ,\nonumber\\
 \quad\text{and}\quad
     u_4(x)&=&\frac{\rmd^2}{\rmd x^2}\left[\frac{u_0''(x)^3}{90 u_0'(x)^4}-\frac{7 u_0^{(3)}(x)
   u_0''(x)}{480 u_0'(x)^3}+\frac{u_0^{(4)}(x)}{288
   u_0'(x)^2}\right]\ ,
\end{eqnarray}
and when these are used, the ${\widehat R}_g(E,x)$ arrange themselves into total derivatives. For example:
%
\begin{eqnarray}
\label{eq:expansion-of-R}
    &&\hskip-0.8cm \widehat{R}_1(E,x) 
=\frac{7 u_0(x) (4 E -5 u_0(x))-8 E ^2}{216 u_0(x)^4 (u_0(x)-E )^{7/2}}
=\frac{\rmd}{\rmd x}\left(-\frac{u_0^{\prime\prime}(x)}{48 u_0^\prime(x)[u_0(x)-E]^{3/2}}+\frac{u_0^\prime(x)}{32[u_0(x)-E]^{5/2}}\right)
\ ,
\\ 
   &&\hskip-0.8cm \widehat{R}_2(E,x) =
 \nonumber\\&& \frac{7 \left(-5824 E ^4 u_0(x)+16256 E ^3 u_0(x)^2-25184 E ^2 u_0(x)^3+22792 E  u_0(x)^4-10421 u_0(x)^5+896 E ^5\right)}{46656 u_0(x)^9 (u_0(x)-E)^{13/2}}\nonumber\\
 &&
=\frac{\rmd}{\rmd x}
\Biggl(\Biggr.\frac{105u_0'^3}{2048[u_0-E]^{11/2}}-\frac{203u_0'u_0''}{3072[u_0-E]^{9/2}}+\frac{29u_0'u_0'''-3u_0''^2}{1536u_0'[u_0-E]^{7/2}} \nonumber\\
        &&\hskip0.5cm-\frac{17u_0''^3-34u_0'u_0''u_0'''+15u_0'^2u_0^{(4)}}{3840u_0'^3[u_0-E]^{5/2}}-\frac{64u_0''^4-111u_0'u_0''^2u_0'''+21u_0'^2u_0'''^2+31u_0'^2u_0''u_0^{(4)}-5u_0'^3u_0^{(5)}}{5760u_0'^5[u_0-E]^{3/2}}
\Biggl.\Biggr)\ .  \nonumber
 \end{eqnarray}  
Given the total derivative result, however, all that is needed to evaluate their integrals is to evaluate the quantities in brackets at the boundaries. The derivatives of $u_0(x)$ at $x\to-\infty$ are zero and so the contributions there all vanish. So only  $x=\mu$ is of interest, where, {\it e.g.}:
\begin{equation}
\label{eq:useful-u-derivatives}
    u_0^{\prime}(\mu)=-\frac{2\sqrt{2}}{3}\ ,\quad
    u_0^{\prime\prime}(\mu)=-\frac{8}{9}\ ,\quad
    u_0^{\prime\prime\prime}(\mu)=-\frac{16\sqrt{2}}{9}\ ,\quad
    u_0^{(4)}(\mu)=-\frac{320}{27}\ ,
\end{equation}
yielding:
\begin{equation}
    \int_{-\infty}^\mu \widehat{R}_1(E,x) dx = 
-\frac{\sqrt{2}}{72 [1-E]^{3/2}}-\frac{\sqrt{2}}{48[1-E]^{5/2}}    
    =-\frac{1}{144z^3}-\frac{1}{192z^5}\ ,
\end{equation}
 using again $1-E=2z^2$. Using~(\ref{eq:toprec3}), gives  exact agreement with the topological recursion result~(\ref{eq:check-recur-1})!
 Similar substitutions of~(\ref{eq:useful-u-derivatives}) into the above expression for $\widehat{R}_2(E,x)$ verifies agreement with the topological recursion result~(\ref{eq:check-recur-2}).
 Finally, we have, directly:
\begin{align}
    \widehat{R}_3(E,x) =&
    \frac{\rmd}{\rmd x}
\left({\widehat Q}_3(E,x)\right)\\
=&-\frac{7}{5038848 u_0(x)^{14} (u_0(x)-E )^{19/2}}\Big(-13619200 E ^7 u_0(x)+58329600 E ^6 u_0(x)^2\nonumber \\ & -148531840 E ^5 u_0(x)^3+249366080 E ^4 u_0(x)^4-288259008 E ^3 u_0(x)^5+230346296 E ^2 u_0(x)^6\nonumber \\ &-120770908 E  u_0(x)^7+33346305 u_0(x)^8+1433600 E ^8\Big)\ ,
\end{align}
where the expression for ${\widehat Q}_3(E,x)={\widehat Q}_3(E,u_0(x),u^{\prime}_0(x),u^{\prime\prime}_0(x),\ldots)$, while readily computed, is cumbersome and not needed here. It is displayed in ref.~\cite{Ahmed:2025lxe}. Putting in the $x=\mu$ values~(\ref{eq:useful-u-derivatives}) (and a few more),  and using equation~(\ref{eq:toprec3}), gives agreement with~(\ref{eq:check-recur-3}).

So we see that the ODE method directly (without needing to compute intermediate quantities such as $W_{0,3}, W_{1,2}$) yields the $W_{g,1}$ (and all the intersection theory data contained within them) by straightforward recursive expansion of the Gel'fand-Dikii equation. 
Upon reflection, the appearance of the Gel'fand-Dikii equation is quite natural here. On the one hand, it is at the heart of the structure of the KdV integrable hierarchy~\cite{Gelfand:1976B}, but on the other hand, intersection theory on the moduli space of marked Riemann surfaces is also organized, through the aforementioned work of Witten and Kontsevich~\cite{Witten:1990hr,Kontsevich:1992ti}, by the KdV hierarchy. 

%
%
%


\section{Non-perturbative corrections to the one-point correlation function}

\label{sec:nonpert}

In the previous Section, we explored two distinct approaches to deriving all perturbative coefficients appearing in the genus expansion of the one-point correlation function: the Gel’fand–Dikii resolvent equation \eqref{eq:gde6} and topological recursion \eqref{eq:toprec2}. In this Section, we aim to extend the former approach to allow for the computation of non-perturbative corrections. 

In Subsection \ref{subsec:transseries}, we give concrete analytical meaning to non-perturbative corrections of the one-point correlation function, by introducing the concept of a transseries. We further identify all distinct non-perturbative effects contributing to the one-point correlation function and show how they naturally lead to a transseries structure.

In Subsection \ref{sec:nonperttopred}, we review the non-perturbative topological recursion and show how it can be used to predict part of the transseries structure presented in \ref{subsec:transseries}. We introduce this method at this stage with the aim of testing ours results obtained in Subsection \ref{sec:transseriesfromgel}.

In Subsection \ref{subsec:WKBexpansion}, we review the WKB expansion introduced in \cite{sss19} and subsequently applied in \cite{os19}, and show how it can be used to derive another part of the transseries structure presented in \ref{subsec:transseries}. As in Subsection \ref{sec:nonperttopred}, we introduce this method at this stage with the aim of testing the results obtained in Subsection \ref{sec:transseriesfromgel}.

Finally, in Subsection \ref{sec:transseriesfromgel}, we extend the method outlined in Subsection \ref{subsec:ode} to allow for the systematic computation of all transseries coefficients, highlighting the non-perturbative predictive power of the Gel’fand-Dikii equation. We further validate our results by matching them with the ones obtained via the non-perturbative topological recursion and the WKB expansion.

\subsection{Non-perturbative effects and transseries}

\label{subsec:transseries}

The genus expansion of connected correlation functions is known to be asymptotic in nature, as their coefficients exhibit factorial growth \cite{beg24}. This factorial divergence signals the necessity of supplementing the perturbative genus expansion with additional asymptotic series associated with instanton configurations, encoding all possible non-perturbative effects. The resulting object, incorporating both the perturbative genus expansion and the non-perturbative instanton corrections, is referred to as a ``transseries'' and the respective asymptotic series are called ``transseries sectors'' \cite{abs19}.

Following \cite{eggls23}, two distinct types of non-perturbative effects appear as transseries sectors: ``ZZ'' contributions and ``FZZT'' contributions~\cite{fzz00,t00,zz01}. The former arise from instanton configurations associated with the tunnelling of eigenvalues and anti-eigenvalues from the spectral distribution of the matrix model to its various saddle points \cite{d90a,d93,msw07,mss22}. As first observed in \cite{m08} and further developed in \cite{gikm12,asv12,bssv22}, in the case of the free energy these instantons may be alternatively derived directly from the string equation by plugging a suitable transseries and recursively computing it's coefficients (in Subsection \ref{sec:transseriesfromgel} we will extend this method to the Gel’fand-Dikii resolvent equation \eqref{eq:gde6}). The name is appropriate  since the corresponding transseries sectors may be interpreted, particularly within Hermitian matrix models dual to minimal string theories, as the asymptotic genus expansion obtained by expanding the correlation functions \eqref{eq:toprec1} around backgrounds including ZZ D-brane configurations \cite{zz01} (see \cite{sst23} for a detailed review of this correspondence in the context of the partition function transseries). 

In the particularly simple case of a double scaled Hermitian matrix model whose spectral curve features a cut extending from $\infty$ to $E_0 \in \mathbb{C}$ and a single saddle $E^\star \in \mathbb{C}$, the associated transseries sectors are proportional to the transmonomial
\begin{equation}
    \exp\left(-\frac{(n_+-n_-) A_{\text{ZZ}}}{\hbar}\right)
    \label{eq:nonperef6}
\end{equation}
where $n_+,n_- \in \mathbb{N}_0$ count the number of eigenvalues and anti-eigenvalues tunnelled to the saddle~$E^\star$ \cite{mss22}, or, equivalently, the number of positive and negative tension ZZ D-branes included in the background~\cite{sst23}. Moreover, the instanton action reads: 
\begin{equation}
    A_{\text{ZZ}} = \oint_B \omega_{0,1}(\bullet)
    \label{eq:nonperef10}
\end{equation}
\begin{figure}
    \centering

    \begin{tikzpicture}
        \draw[fill = LightBlue,fill opacity=0.2,line width = 0pt] (0,0)to[out = 90, in = 180] (3,2)to[out = 0, in = 90] (6-0.5,0)to[out = 90, in = 90]  (2.3,0)to[out = 180+30, in = -30]cycle;
        \draw[fill = darktangerine,fill opacity=0.2,line width = 0pt] (0,0)to[out = -90, in = 180] (3,-2)to[out = 0, in = -90] (6-0.5,0)to[out = -90, in = -90] (2.3,0)to[out = 180+30, in = -30]cycle;
          \draw[line width = 2pt] (0,0)to[out = 90, in = 180] (3,2)to[out = 0, in = 90] (6-0.5,0)to[out = -90, in = 0] (3,-2)to[out = 180, in = -90] cycle;
          \draw[line width = 2pt] (2.3,0)to[out = 90, in = 90] (6-0.5,0)to[out = -90, in = -90]  cycle;
          

\draw[line width = 2pt,color = ForestGreen](2.3,0)to[out = 180+30, in = -30](0,0);
\draw[line width = 2pt,color = ForestGreen,dashed](2.3,0)to[out = 180-30, in = 30](0,0);




\draw[line width = 2pt] (1,0) to[out = 90, in = 180] (3,1.5) to[out = 0, in  = 90] (5.5,0) to[out = -90, in = 0] (3,-1.5) to [out = 180, in = -90] cycle;

  \draw[line width = 2pt,->] (1,-0.01) -- (1,0.01);

\node at (2.4,1) {$B$};


        \draw[color = ForestGreen, fill= ForestGreen, line width=1pt] (2.3,0) circle (0.7ex);
\draw[color = ForestGreen, fill= ForestGreen, line width=1pt] (0,0) circle (0.7ex);
\draw[color = cornellred, fill=  cornellred, line width=1pt] (6-0.5,0) circle (0.7ex);

\node at (3,2.5) {\scalebox{1.3}{$\Sigma$}
};

\node[color = cornellred] at (6,0) {$E^\star$
};

\node[color = ForestGreen] at (0-0.5,0) {$\infty$
};
\node[color = ForestGreen] at (0+0.5+2.3,0) {$E_0$
};


\def\hspace{8};

        \draw[fill = LightBlue,fill opacity=0.2,line width = 0pt] (0+\hspace,0)to[out = 90, in = 180] (3+\hspace,2)to[out = 0, in = 90] (6-0.5+\hspace,0)to[out = 90, in = 90]  (2.3+\hspace,0)to[out = 180+30, in = -30]cycle;
        \draw[fill = darktangerine,fill opacity=0.2,line width = 0pt] (0+\hspace,0)to[out = -90, in = 180] (3+\hspace,-2)to[out = 0, in = -90] (6-0.5+\hspace,0)to[out = -90, in = -90] (2.3+\hspace,0)to[out = 180+30, in = -30]cycle;
          \draw[line width = 2pt] (0+\hspace,0)to[out = 90, in = 180] (3+\hspace,2)to[out = 0, in = 90] (6-0.5+\hspace,0)to[out = -90, in = 0] (3+\hspace,-2)to[out = 180, in = -90] cycle;
          \draw[line width = 2pt] (2.3+\hspace,0)to[out = 90, in = 90] (6-0.5+\hspace,0)to[out = -90, in = -90]  cycle;
          

\draw[line width = 2pt,color = ForestGreen](2.3+\hspace,0)to[out = 180+30, in = -30](0+\hspace,0);
\draw[line width = 2pt,color = ForestGreen,dashed](2.3+\hspace,0)to[out = 180-30, in = +30](0+\hspace,0);




\draw[line width = 2pt] (0.5+\hspace,0) to[out = 90, in = 180] (2+\hspace,1.5)to[out =0, in =90] (2+\hspace+0.5,1.5-0.3)to[out =-90, in =180+30] (2+\hspace+0.5+0.7,1.5-0.2);

\draw[line width = 2pt] (0.5+\hspace,0) to[out = -90, in = 180] (2+\hspace,-1.5)to[out =0, in =-90] (2+\hspace+0.5,-1.5+0.3)to[out =90, in =180-30] (2+\hspace+0.5+0.7,-1.5+0.2);

\filldraw[line width=1pt] (2+\hspace+0.5+0.7,-1.5+0.2) circle (0.7ex);
\filldraw[line width=1pt] (2+\hspace+0.5+0.7,1.5-0.2) circle (0.7ex);

\draw[line width = 2pt,->] (0.5+\hspace,-0.001) -- (0.5+\hspace,0.001);


        \draw[color = ForestGreen, fill= ForestGreen, line width=1pt] (2.3+\hspace,0) circle (0.7ex);
\draw[color = ForestGreen, fill= ForestGreen, line width=1pt] (0+\hspace,0) circle (0.7ex);
\draw[color = cornellred, fill=  cornellred, line width=1pt] (6-0.5+\hspace,0) circle (0.7ex);

\node at (3+\hspace,2.5) {\scalebox{1.3}{$\Sigma$}
};

\node[color = cornellred] at (6+\hspace,0) {$E^\star$
};

\node[color = ForestGreen] at (0+\hspace-0.5,0) {$\infty$
};
\node[color = ForestGreen] at (0+\hspace+0.5+2.3,0) {$E_0$
};

\node at (2+\hspace+0.5+0.7+0.4,1.5-0.2) {$z$
};

\node at (2+\hspace+0.5+0.7+0.7,-1.5+0.2) {$\sigma(z)$
};

\node at (\hspace+1.5,0.8) {$\gamma(z)$
};

    \end{tikzpicture}
    \caption{On the left, a pictorial representation of the cycle $B \subset \Sigma$ and on the right, of the contour $\gamma(z)\subset \Sigma$. The cut connecting $\infty$ to $E_0$ is denoted by a green line while the saddle $E^\star$ (red dot) is represented by a pinch.}
    \label{fig:Cycles}
\end{figure}
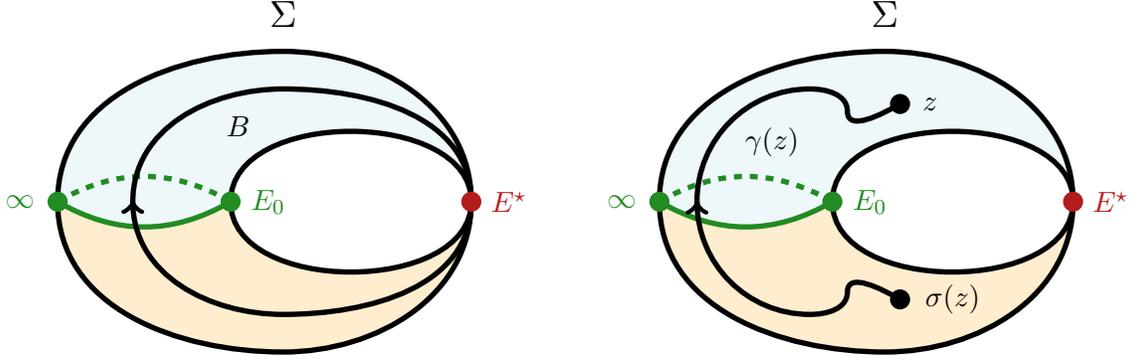
where $B \subset \Sigma$ is schematically depicted in figure \ref{fig:Cycles}. (We remind the reader that the bullet dot in the integrand marks the argument with respect to which the integration is performed.) 

The second kind of non-perturbative effects, FZZT contributions,  correspond to the insertion of $\det(E-M)$ in the Hermitian matrix model and can roughly by interpreted as a brane probing the insertion point $E$ \cite{mmsss04}. In this case, the corresponding transseries sectors may be interpreted, particularly within Hermitian matrix models dual to minimal string theories, as the asymptotic genus expansion obtained by expanding the correlation functions \eqref{eq:toprec1} around a background including FZZT D-branes \cite{fzz00,t00,Saad:2019lba}. In the particular case of the one-point correlation function, the associated transseries sector is proportional to the transmonomial
\begin{equation}
    \exp\left( \frac{A_{\text{FZZT}}(E)}{\hbar}\right)
\end{equation}
where the instanton action reads
\begin{equation}
    A_{\text{FZZT}}(E) = \oint_{\gamma(z(E))} \omega_{0,1}(\bullet)
    \label{eq:vol4}
\end{equation}
and $\gamma(z) \subset \Sigma$ is schematically depicted in figure \ref{fig:Cycles}. 

Before proceeding, it is important to note that FZZT D-branes possess a ghost partner, obtained by inserting $\det(E-M)^{-1}$ in the Hermitian one-matrix model (see \cite{ekr2018} for further details). This object, known as the ghost FZZT D-brane, mirrors many features of its regular counterpart. In particular, it also generates a non-perturbative sector in the transseries for the one-point correlation function, whose associated transmonomial is
\begin{equation}
    \exp\left(- \frac{A_{\text{FZZT}}(E)}{\hbar}\right).
\end{equation}

While ZZ contributions are associated with an infinite family of distinct transmonomials, labelled by the two natural numbers $(n_+,n_-)$ (see the transmonomial \eqref{eq:nonperef6}), the FZZT contributions involve only two such transmonomials. This difference stems from the fact that ZZ contributions originate in the non-perturbative completion of the string equation solution, whose non-linear structure forces the transseries to incorporate an infinite tower of transmonomials (see \cite{abs19} for a concise explanation, or \cite{gikm12,asv12,bssv22} for a more comprehensive account). On the other hand, the FZZT D-brane and its ghost partner are simpler and arise naturally as two linearly independent WKB solutions of the linear wave equation
\begin{equation}
    \frac{d^2 \psi}{dx^2}(x,E)  -\frac{1}{\hbar^2}(u(x)+E)\psi(x,E) = 0
\end{equation}
obtained by double scaling the three-term recursion of orthogonal polynomials \cite{Douglas:1989ve,Brezin:1990rb,Gross:1990aw,Banks:1990df,ekr2018}.

Assembling all of the non-perturbative effects above, the one-point correlation function transseries takes the shape
\begin{align}
    &W_1(E;\sigma_{\text{ZZ}_\pm},\sigma_{\text{FZZT}_\pm}) = W^{\text{pert}}(E) + \sigma_{\text{FZZT}_+}\exp\left(+\frac{A_{\text{FZZT}}(E)}{\hbar}\right) W^{\text{FZZT}_+}(E) \nonumber\\ & +\sigma_{\text{FZZT}_-}\exp\left(-\frac{A_{\text{FZZT}}(E)}{\hbar}\right)W^{\text{FZZT}_-}(E)  +\sum_{\substack{n_\pm \in \mathbb{N}_0 }}'\sigma_{\text{ZZ}_+}^{n_+}\sigma_{\text{ZZ}_-}^{n_-}\exp\left(-(n_+-n_-) \frac{A_{\text{ZZ}}}{\hbar}\right) W_{(n_+,n_-)}^{\text{ZZ}}(E)   \nonumber\\ & +\sigma_{\text{FZZT}_+}\exp\left(+\frac{A_{\text{FZZT}}(E)}{\hbar}\right)\sum_{\substack{n_\pm \in \mathbb{N}_0 }}'\sigma_{\text{ZZ}_+}^{n_+}\sigma_{\text{ZZ}_-}^{n_-}\exp\left(-(n_+-n_-) \frac{A_{\text{ZZ}}}{\hbar}\right) W_{(n_+,n_-)}^{\text{ZZ-FZZT}_+}(E)
   \nonumber \\ & + \sigma_{\text{FZZT}_-}\exp\left(-\frac{A_{\text{FZZT}}(E)}{\hbar}\right)\sum_{\substack{n_\pm \in \mathbb{N}_0}}'\sigma_{\text{ZZ}_+}^{n_+}\sigma_{\text{ZZ}_-}^{n_-}\exp\left(-(n_+-n_-) \frac{A_{\text{ZZ}}}{\hbar}\right) W_{(n_+,n_-)}^{\text{ZZ-FZZT}_-}(E)
    \label{eq:nonperef1}
\end{align}
where $\sigma_{\text{ZZ}_\pm},\sigma_{\text{FZZT}_\pm} \in \mathbb{C}$ are transseries parameters \cite{abs19}, the prime in the summations above indicates we should not include the index combination $(n_+,n_-) = (0,0)$ and the perturbative sector reads
\begin{equation}
    W^{\text{pert}}(E) = \sum_{g \in \mathbb{N}_0}W_{g,1}(E)\hbar^{2g-1}.
    \label{eq:nonperef11}
\end{equation}

In the transseries \eqref{eq:nonperef1}, we have also included non-perturbative sectors associated with the combination of both FZZT and ZZ transmonomials (see the third and fourth lines). This is motivated by the fact that, in resurgence theory, distinct non-perturbative effects often mix. As we will show later (see Subsection \ref{subsubsec:zz-fzzt}), these mixed sectors are indeed non-vanishing. We will refer to them as non-perturbative ZZ-FZZT contributions, as they should be understood, particularly within Hermitian matrix models dual to minimal string theories, as arising from backgrounds featuring non-trivial combinations of both ZZ and FZZT D-branes.

In order to obtain the full non-perturbative expression for the correlation function $W_1(E)$, resurgence instructs us to Borel-resum the transseries. More concretely, we have
\begin{align}
    &W_1(E) = \mathcal{S}\left[W_1\right](E;\sigma_{\text{ZZ}_\pm},\sigma_{\text{FZZT}_\pm}) = \mathcal{S}\left[W^{\text{pert}}\right](E) + \sigma_{\text{FZZT}_+}\exp\left(+\frac{A_{\text{FZZT}}(E)}{\hbar}\right) \mathcal{S}\left[W^{\text{FZZT}_+}\right](E) \nonumber\\ & +\sigma_{\text{FZZT}_-}\exp\left(-\frac{A_{\text{FZZT}}(E)}{\hbar}\right)\mathcal{S}\left[W^{\text{FZZT}_-}\right](E)  +\sum_{\substack{n_\pm \in \mathbb{N}_0 }}'\sigma_{\text{ZZ}_+}^{n_+}\sigma_{\text{ZZ}_-}^{n_-}\exp\left(-(n_+-n_-) \frac{A_{\text{ZZ}}}{\hbar}\right) \mathcal{S}\left[W_{(n_+,n_-)}^{\text{ZZ}}\right](E)   \nonumber\\ & +\sigma_{\text{FZZT}_+}\exp\left(+\frac{A_{\text{FZZT}}(E)}{\hbar}\right)\sum_{\substack{n_\pm \in \mathbb{N}_0 }}'\sigma_{\text{ZZ}_+}^{n_+}\sigma_{\text{ZZ}_-}^{n_-}\exp\left(-(n_+-n_-) \frac{A_{\text{ZZ}}}{\hbar}\right) \mathcal{S}\left[W_{(n_+,n_-)}^{\text{ZZ-FZZT}_+}\right](E)
   \nonumber \\ & + \sigma_{\text{FZZT}_-}\exp\left(-\frac{A_{\text{FZZT}}(E)}{\hbar}\right)\sum_{\substack{n_\pm \in \mathbb{N}_0}}'\sigma_{\text{ZZ}_+}^{n_+}\sigma_{\text{ZZ}_-}^{n_-}\exp\left(-(n_+-n_-) \frac{A_{\text{ZZ}}}{\hbar}\right) \mathcal{S}\left[W_{(n_+,n_-)}^{\text{ZZ-FZZT}_-}\right](E)
   \label{eq:BorelSum}
\end{align}
where $\mathcal{S}$ denotes the Borel summation operator along a direction in the Borel plane free of singularities. See Appendix \ref{appendix:BorelSummation} for a compact introduction to the Borel summation procedure and \cite{abs19} for a more pedagogical and comprehensive review. Although exact, the expression above is only valid locally, as it fails to be well defined for arbitrary values of $E$ or $\hbar$. This is due to obstructions in the analytic continuation of the Borel resummation that arise when crossing Stokes lines in the complex $E$ or $\hbar$ planes \cite{abs19,krsst25a}. Extending the resummation beyond these lines requires shifting the transseries parameters in accordance with the appropriate connection formulae \cite{bssv22}, an effect known as the Stokes phenomenon. In this paper, with the exception of Section \ref{sec:largeorder}, we restrict ourselves to the local transseries expansion \eqref{eq:nonperef1} and will therefore disregard Stokes phenomena entirely.

\paragraph{Transseries of the resolvent:} The transseries structure \eqref{eq:nonperef1} propagates naturally to the Gel'fand-Dikii resolvent \eqref{eq:gde2}. Indeed, {\it we propose the following transseries ansatz for it}:
\begin{align}
    &\widehat{R}(E,x;\sigma_{\text{ZZ}_\pm},\sigma_{\text{FZZT}_\pm}) = \widehat{\mathcal{R}}^{\text{pert}}(E,x) + \sigma_{\text{FZZT}_+}\exp\left(+\frac{\mathcal{A}_{\text{FZZT}}(E,x)}{\hbar}\right) \widehat{\mathcal{R}}^{\text{FZZT}_+}(E,x) \nonumber\\ & +\sigma_{\text{FZZT}_-}\exp\left(-\frac{\mathcal{A}_{\text{FZZT}}(E,x)}{\hbar}\right)\widehat{\mathcal{R}}^{\text{FZZT}_-}(E,x)  +\sum_{\substack{n_\pm \in \mathbb{N}_0 }}'\sigma_{\text{ZZ}_+}^{n_+}\sigma_{\text{ZZ}_-}^{n_-}\exp\left(-(n_+-n_-) \frac{\mathcal{A}_{\text{ZZ}}(x)}{\hbar}\right) \widehat{\mathcal{R}}_{(n_+,n_-)}^{\text{ZZ}}(E,x)   \nonumber\\ & +\sigma_{\text{FZZT}_+}\exp\left(+\frac{\mathcal{A}_{\text{FZZT}}(E,x)}{\hbar}\right)\sum_{\substack{n_\pm \in \mathbb{N}_0 }}'\sigma_{\text{ZZ}_+}^{n_+}\sigma_{\text{ZZ}_-}^{n_-}\exp\left(-(n_+-n_-) \frac{\mathcal{A}_{\text{ZZ}}(x)}{\hbar}\right) \widehat{\mathcal{R}}_{(n_+,n_-)}^{\text{ZZ-FZZT}_+}(E,x)
   \nonumber \\ & + \sigma_{\text{FZZT}_-}\exp\left(-\frac{\mathcal{A}_{\text{FZZT}}(E,x)}{\hbar}\right)\sum_{\substack{n_\pm \in \mathbb{N}_0}}'\sigma_{\text{ZZ}_+}^{n_+}\sigma_{\text{ZZ}_-}^{n_-}\exp\left(-(n_+-n_-) \frac{\mathcal{A}_{\text{ZZ}}(x)}{\hbar}\right) \widehat{\mathcal{R}}_{(n_+,n_-)}^{\text{ZZ-FZZT}_-}(E,x)
    \label{eq:nonperef2}
\end{align}
for some functions $\mathcal{A}_{\text{FZZT}}(E,x)$ and $\mathcal{A}_{\text{ZZ}}(x)$ such that
\begin{align}
   &  \mathcal{A}_{\text{FZZT}}(E,\mu) = A_{\text{FZZT}}(E) \\ 
    &  \mathcal{A}_{\text{ZZ}}(\mu) = A_{\text{ZZ}}
\end{align}
and where the perturbative sector reads
\begin{equation}
   \widehat{\mathcal{R}}^{\text{pert}}(E,x) = \sum_{g\in \mathbb{N}_0}\widehat{R}_g(E,x)\hbar^{2g}\ .
   \label{eq:nonperef12}
\end{equation}
Using equation \eqref{eq:gde5}, we can write
\begin{equation}
  W_1(E;\sigma_{\text{ZZ}_\pm},\sigma_{\text{FZZT}_\pm}) =\frac{1}{\hbar}\int_{-\infty}^{\mu} \rmd x \widehat{R}(E,x;\sigma_{\text{ZZ}_\pm},\sigma_{\text{FZZT}_\pm}).
  \label{eq:nonperef3}
\end{equation}
The equation above allows us to set up a recursive procedure for computing the transseries coefficients of the one-point correlation function, assuming the transseries expansion of the resolvent is known. We will carry out this computation later in the paper, highlighting the non-perturbative predictive power of the Gel’fand-Dikii equation.


\subsection{The non-perturbative topological recursion}

\label{sec:nonperttopred}

There have been many great developments in our understanding of how to go beyond topological recursion, starting with the foundational work of \cite{msw07} in which it was understood how to systematically compute instanton corrections for the free energy, by means of certain matrix integral saddle-point expansions (some of which we will review later). This framework was extended in \cite{mss22} to account for the entire free energy transseries. These ideas have recently culminated in \cite{eggls23}, in which the loop insertion operator (which we will define shortly) was used to extend the previous results to compute correlation function ZZ non-perturbative transseries sectors in what was called the non-perturbative topological recursion.

Below, we provide a brief review of this procedure and show how it can be used to derive analytical expressions for the ZZ non-perturbative transseries sectors of \eqref{eq:nonperef1}. The purpose of this review is to set the stage for an independent consistency check of the novel results obtained in Subsection \ref{sec:transseriesfromgel} by resorting to the Gel’fand–Dikii resolvent equation \eqref{eq:gde6}. For further details on the non-perturbative topological recursion, see \cite{eggls23}.

We begin by introducing the wave function, an object that plays a central role in the non-perturbative topological recursion. It is constructed from perturbative expansion coefficients of the connected correlation functions \eqref{eq:omega-meets-W} and reads
\begin{equation}
    \psi(\gamma) = \exp\left(\mathbb{S}(\gamma)\right)
    \label{eq:nonperef13}
\end{equation}
for some continuous (possibly disconnected) path $\gamma \subset \Sigma$ where
\begin{equation}
    \mathbb{S}(\gamma) = \sum_{\chi \in \mathbb{N}_0} \mathbb{S}_\chi(\gamma)\hbar^{\chi-1}
\end{equation}
and
\begin{equation}
    \mathbb{S}_\chi(\gamma) = \sum_{\substack{2g-2+n = \chi -1 \\ g \ge0 \hspace{1pt},\hspace{1pt} n \ge 1}} \frac{F_{g,n}(\gamma)}{n!}
\end{equation}
as well as
\begin{equation}
    F_{g,n}(\gamma) = \left[\prod_{i=1}^n \int_{\gamma}  \right] \omega_{g,n}(\bullet,\cdots,\bullet).
    \label{eq:nonperef14}
\end{equation}
The computation of $F_{0,2}(\gamma)$ requires a regularization that depends on the chosen path $\gamma$.

Much like the one-point correlation function, the partition function features a transseries representation incorporating ZZ non-perturbative contributions (see \cite{krsst25a} for a comprehensive account of the partition function transseries structure). However, unlike transseries associated with correlation functions, this transseries does not receive non-perturbative contributions of the FZZT type.\footnote{A simple yet heuristic reason is that, while the FZZT D-brane depends on the insertion variable $E$, the partition function does not. Consequently, there is no natural variable with which to associate $A_{\text{FZZT}}(E)$ in this case.} More concretely, in the example we have been following, we have
\begin{equation}
Z(\hbar;\sigma_{\text{ZZ}_\pm}) = Z^{\text{pert}}(\hbar)+\sum_{n_\pm \in \mathbb{N}_0}'\sigma_{\text{ZZ}_+}^{n_+}\sigma_{\text{ZZ}_-}^{n_-}\exp\left(-\frac{n_+-n_-}{\hbar} A_{\text{ZZ}}\right)Z_{(n_+,n_-)}^{\text{ZZ}}(\hbar).
\label{eq:nonperef9}
\end{equation}
Using the equation above, we can write the free energy transseries as
\begin{equation}
    F(\hbar;\sigma_{\text{ZZ}_\pm}) = \log\left(Z(\hbar;\sigma_{\text{ZZ}_\pm})\right) = F^{\text{pert}}(\hbar)+\sum_{n_\pm \in \mathbb{N}_0}'\sigma_{\text{ZZ}_+}^{n_+}\sigma_{\text{ZZ}_-}^{n_-}\exp\left(-\frac{n_+-n_-}{\hbar} A_{\text{ZZ}}\right)F_{(n_+,n_-)}^{\text{ZZ}}(\hbar)
    \label{eq:nonperef7}
\end{equation}
where the perturbative sector reads
\begin{equation}
  F^{\text{pert}}(\hbar) = \sum_{g\in \mathbb{N}_0} F_{g}\hbar^{2g-2}.
  \label{eq:nonperef8}
\end{equation}

The coefficients featured in the series above are canonically related to the topological recursion coefficients \eqref{eq:omega-meets-W} by the formula \cite{ekr2018}
\begin{equation}
    F_g = \frac{1}{2-2g}\sum_{r \in \mathcal{R}}\underset{z \to r}{\text{Res}} \left(\int^z_c \omega_{0,1}(\bullet)\right)\omega_{g,1}(z)
    \label{eq:nonperef5}
\end{equation}
for all $g \ge 2$ and for any $c \in \mathbb{C}$.\footnote{Due to the presence of the residue, the expression \eqref{eq:nonperef5} is independent of $c$.} The coefficients $F_0$ and $F_1$ are defined somewhat independently, and explicit expressions can be found in \cite{ekr2018}. 

The last ingredient we will need is the loop insertion operator. For any $z \in \Sigma$, this operator is denoted by $\Delta_z$ and acts as a derivation (i.e., it obeys the Leibniz rule) in the topological recursion coefficients. We refer the reader to \cite{eo07} for a concrete definition. Importantly, one can show that
\begin{equation}
    \Delta_{z(E)} F_g = W_{g,1}(E)\frac{\rmd E}{\rmd z}(E).
\end{equation}

One can now conjecture that the equation above should also hold non-perturbatively, in which case one would have
\begin{equation}
    \hbar \Delta_{z(E)} F^{\text{ZZ}}_{(n_+,n_-)}(\hbar) = W_{(n_+,n_-)}^{\text{ZZ}}(E) \frac{\rmd E}{\rmd z}(E).
\end{equation}
The equation above admits a straightforward generalisation to higher-point correlation functions via iterative applications of loop insertion operators at distinct points (see \cite{eggls23} for an application in the context of JT gravity). However, these cases will not be our focus, as we are primarily interested in the one-point correlation function. 

The equation above enables a direct extraction of the non-perturbative ZZ-effects in the one-point correlation function from those in the free energy. Consequently, the final step reduces to computing explicit analytical expressions for free energy non-perturbative contributions. This has been fully understood in \cite{mss22}, and in what follows we review some of the associated results.

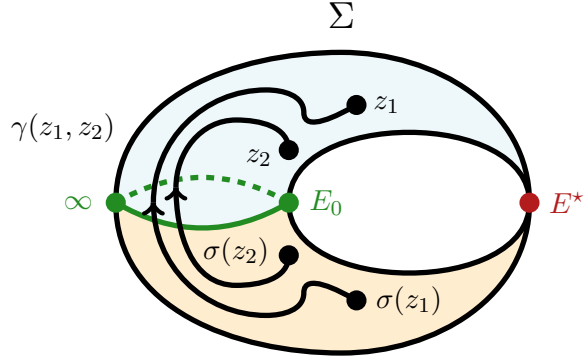
\begin{figure}
    \centering
    \begin{tikzpicture}
    
\def\hspace{0};

\draw[fill = LightBlue,fill opacity=0.2,line width = 0pt] (0+\hspace,0)to[out = 90, in = 180] (3+\hspace,2)to[out = 0, in = 90] (6-0.5+\hspace,0)to[out = 90, in = 90]  (2.3+\hspace,0)to[out = 180+30, in = -30]cycle;
\draw[fill = darktangerine,fill opacity=0.2,line width = 0pt] (0+\hspace,0)to[out = -90, in = 180] (3+\hspace,-2)to[out = 0, in = -90] (6-0.5+\hspace,0)to[out = -90, in = -90] (2.3+\hspace,0)to[out = 180+30, in = -30]cycle;
\draw[line width = 2pt] (0+\hspace,0)to[out = 90, in = 180] (3+\hspace,2)to[out = 0, in = 90] (6-0.5+\hspace,0)to[out = -90, in = 0] (3+\hspace,-2)to[out = 180, in = -90] cycle;
\draw[line width = 2pt] (2.3+\hspace,0)to[out = 90, in = 90] (6-0.5+\hspace,0)to[out = -90, in = -90]  cycle;

\draw[line width = 2pt,color = ForestGreen](2.3+\hspace,0)to[out = 180+30, in = -30](0+\hspace,0);
\draw[line width = 2pt,color = ForestGreen,dashed](2.3+\hspace,0)to[out = 180-30, in = 30](0+\hspace,0);


\draw[line width = 2pt] (0.5+\hspace,0) to[out = 90, in = 180] (2+\hspace,1.5)to[out =0, in =90] (2+\hspace+0.5,1.5-0.3)to[out =-90, in =180+30] (2+\hspace+0.5+0.7,1.5-0.2);

\draw[line width = 2pt] (0.5+\hspace,0) to[out = -90, in = 180] (2+\hspace,-1.5)to[out =0, in =-90] (2+\hspace+0.5,-1.5+0.3)to[out =90, in =180-30] (2+\hspace+0.5+0.7,-1.5+0.2);

\filldraw[line width=1pt] (2+\hspace+0.5+0.7,-1.5+0.2) circle (0.7ex);
\filldraw[line width=1pt] (2+\hspace+0.5+0.7,1.5-0.2) circle (0.7ex);

\draw[line width = 2pt,->] (0.5+\hspace,-0.001) -- (0.5+\hspace,0.001);


\draw[line width = 2pt] (1+\hspace-0.2,0+0.2) to[out = 90, in = 180] (2+\hspace-0.4,1.1)to[out =0, in =90] (2+\hspace+0.1+0.2,1.5-0.3-0.7+0.2);

\draw[line width = 2pt] (1+\hspace-0.2,0+0.2) to[out = -90, in = 180] (2+\hspace-0.4,-1.1)to[out =0, in =-90] (2+\hspace+0.1+0.2,-1.5+0.3+0.7-0.2);

\filldraw[line width=1pt] (2+\hspace+0.1+0.2,1.5-0.3-0.7+0.2) circle (0.7ex);
\filldraw[line width=1pt] (2+\hspace+0.1+0.2,-1.5+0.3+0.7-0.2) circle (0.7ex);

\draw[line width = 2pt,->] (0.8+\hspace,-0.001+0.2) -- (0.8+\hspace,0.001+0.2);


\draw[color = ForestGreen, fill= ForestGreen, line width=1pt] (2.3+\hspace,0) circle (0.7ex);
\draw[color = ForestGreen, fill= ForestGreen, line width=1pt] (0+\hspace,0) circle (0.7ex);
\draw[color = cornellred, fill=  cornellred, line width=1pt] (6-0.5+\hspace,0) circle (0.7ex);

\node at (3+\hspace,2.5) {\scalebox{1.3}{$\Sigma$}
};

\node at (2+\hspace+0.5+0.7+0.4,1.5-0.2) {$z_1$
};
\node at (2+\hspace+0.1-0.2,1.5-0.3-0.7+0.1) {$z_2$
};

\node at (2+\hspace+0.1-0.5,-1.55+0.3+0.7-0.1) {$\sigma(z_2)$
};

\node at (2+\hspace+0.5+0.7+0.7,-1.5+0.2) {$\sigma(z_1)$
};

\node at (-0.7,1) {$\gamma(z_1,z_2)$};

\node[color = ForestGreen] at (0+\hspace-0.5,0) {$\infty$
};

\node[color = ForestGreen] at (2.3+\hspace+0.5,0) {$E_0$};

\node[color = cornellred] at (6-0.5+\hspace+0.5,0) {$E^\star$};
    \end{tikzpicture}
    \caption{Pictorial representation of the contour $\gamma(z_1,z_2) \subset \Sigma$.}
    \label{fig:nCycle}
\end{figure}

\paragraph{Positive instantons:} We start by addressing the sector $(n_+,n_-) = (1,0)$. Following \cite{mss22}, one can write
\begin{equation}
   \frac{Z^{\text{ZZ}}_{(1,0)}(\hbar)}{Z^{\text{pert}}(\hbar)}  =F^{\text{ZZ}}_{(1,0)}(\hbar) =  \int_{\mathcal{C}^\star} \frac{\rmd E}{2\pi}\psi(\gamma(z(E)))
\end{equation}
where $\gamma(z)$ is depicted in figure \ref{fig:Cycles} and $\mathcal{C}^\star\subset \mathbb{C}$ is the steepest descent contour associated with the saddle~$E^\star$. Using the matrix integral formula displayed in equation (3.10) of \cite{mss22}, we can straightforwardly generalize the result above to the instanton sector $(n_+,n_-) = (n,0)$. Indeed, we can write
\begin{equation}
    \frac{Z_{(n,0)}^{\text{ZZ}}(\hbar)}{Z^{\text{pert}}(\hbar)} = \frac{1}{n!}\left[\prod_{j=1}^n \int_{\mathcal{C}^\star} \frac{\rmd E_j}{2\pi}\right]\Delta_{n}^2(E_1,\cdots,E_n)\psi(\gamma(z(E_1),\cdots,z( E_n)))
\end{equation}
where 
\begin{equation}
    \gamma(z_1,\cdots,z_n) = \bigsqcup_{i=1}^n \gamma(z_i).
    \label{eq:nonpertopef2}
\end{equation}
An example of this contour for $n = 2$ is depicted in figure \ref{fig:nCycle}.
To relate the transseries sectors above to those of the free energy, we expand the logarithm in \eqref{eq:nonperef7} connecting the two quantities in powers of the transseries parameters. Once this expansion is carried out, we obtain
\begin{equation}
   \frac{Z_{(n,0)}^{\text{ZZ}}(\hbar)}{Z^{\text{pert}}(\hbar)} = \sum_{\sigma \in \Gamma(n)} \prod_{\substack{1 \le k \le n }}\frac{1}{\sigma_k!} \left(F_{(k,0)}^{\text{ZZ}}(\hbar)\right)^{\sigma_k}
    \label{eq:nonpertopef1}
\end{equation}
where $\Gamma(n)$ denotes the set of Young diagrams whose total number of boxes is $n$ and $\sigma_k$ denotes the number of arrays in $\sigma\in \Gamma(n)$ with $k$ boxes. The equation above can be recursively solved for all free energy transseries sectors.

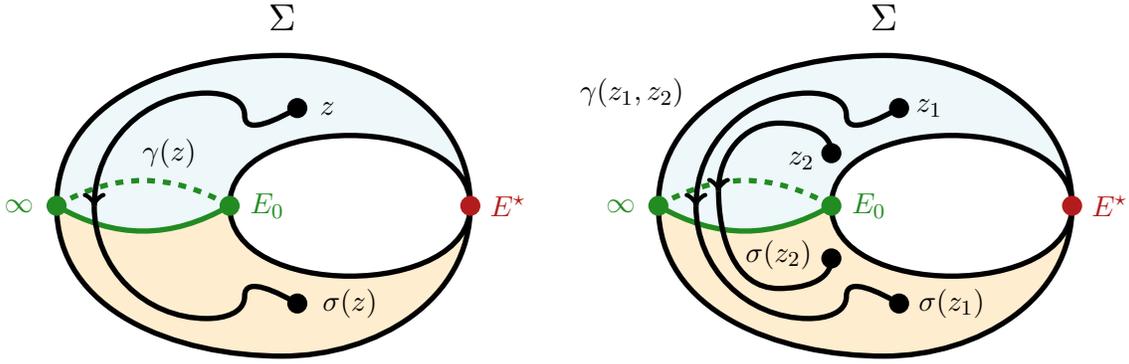
\begin{figure}
    \centering
    \begin{tikzpicture}
    
\def\hspace{0};

\draw[fill = LightBlue,fill opacity=0.2,line width = 0pt] (0+\hspace,0)to[out = 90, in = 180] (3+\hspace,2)to[out = 0, in = 90] (6-0.5+\hspace,0)to[out = 90, in = 90]  (2.3+\hspace,0)to[out = 180+30, in = -30]cycle;
\draw[fill = darktangerine,fill opacity=0.2,line width = 0pt] (0+\hspace,0)to[out = -90, in = 180] (3+\hspace,-2)to[out = 0, in = -90] (6-0.5+\hspace,0)to[out = -90, in = -90] (2.3+\hspace,0)to[out = 180+30, in = -30]cycle;
\draw[line width = 2pt] (0+\hspace,0)to[out = 90, in = 180] (3+\hspace,2)to[out = 0, in = 90] (6-0.5+\hspace,0)to[out = -90, in = 0] (3+\hspace,-2)to[out = 180, in = -90] cycle;
\draw[line width = 2pt] (2.3+\hspace,0)to[out = 90, in = 90] (6-0.5+\hspace,0)to[out = -90, in = -90]  cycle;

\draw[line width = 2pt,color = ForestGreen](2.3+\hspace,0)to[out = 180+30, in = -30](0+\hspace,0);

\draw[line width = 2pt,color = ForestGreen,dashed](2.3+\hspace,0)to[out = 180-30, in = 30](0+\hspace,0);


\draw[line width = 2pt] (0.5+\hspace,0) to[out = 90, in = 180] (2+\hspace,1.5)to[out =0, in =90] (2+\hspace+0.5,1.5-0.3)to[out =-90, in =180+30] (2+\hspace+0.5+0.7,1.5-0.2);

\draw[line width = 2pt] (0.5+\hspace,0) to[out = -90, in = 180] (2+\hspace,-1.5)to[out =0, in =-90] (2+\hspace+0.5,-1.5+0.3)to[out =90, in =180-30] (2+\hspace+0.5+0.7,-1.5+0.2);

\filldraw[line width=1pt] (2+\hspace+0.5+0.7,-1.5+0.2) circle (0.7ex);
\filldraw[line width=1pt] (2+\hspace+0.5+0.7,1.5-0.2) circle (0.7ex);

\draw[line width = 2pt,<-] (0.5+\hspace,-0.001) -- (0.5+\hspace,0.001);







\draw[color = ForestGreen, fill= ForestGreen, line width=1pt] (2.3+\hspace,0) circle (0.7ex);
\draw[color = ForestGreen, fill= ForestGreen, line width=1pt] (0+\hspace,0) circle (0.7ex);
\draw[color = cornellred, fill=  cornellred, line width=1pt] (6-0.5+\hspace,0) circle (0.7ex);

\node at (3+\hspace,2.5) {\scalebox{1.3}{$\Sigma$}
};
\node at (2+\hspace+0.5+0.7+0.4,1.5-0.2) {$z$
};


\node at (2+\hspace+0.5+0.7+0.7,-1.5+0.2) {$\sigma(z)$};

\node at (2+\hspace-0.5,-1.5+0.2+2) {$\gamma(z)$};


\def\hspace{8};

\draw[fill = LightBlue,fill opacity=0.2,line width = 0pt] (0+\hspace,0)to[out = 90, in = 180] (3+\hspace,2)to[out = 0, in = 90] (6-0.5+\hspace,0)to[out = 90, in = 90]  (2.3+\hspace,0)to[out = 180+30, in = -30]cycle;
\draw[fill = darktangerine,fill opacity=0.2,line width = 0pt] (0+\hspace,0)to[out = -90, in = 180] (3+\hspace,-2)to[out = 0, in = -90] (6-0.5+\hspace,0)to[out = -90, in = -90] (2.3+\hspace,0)to[out = 180+30, in = -30]cycle;
\draw[line width = 2pt] (0+\hspace,0)to[out = 90, in = 180] (3+\hspace,2)to[out = 0, in = 90] (6-0.5+\hspace,0)to[out = -90, in = 0] (3+\hspace,-2)to[out = 180, in = -90] cycle;
\draw[line width = 2pt] (2.3+\hspace,0)to[out = 90, in = 90] (6-0.5+\hspace,0)to[out = -90, in = -90]  cycle;

\draw[line width = 2pt,color = ForestGreen](2.3+\hspace,0)to[out = 180+30, in = -30](0+\hspace,0);
\draw[line width = 2pt,color = ForestGreen,dashed](2.3+\hspace,0)to[out = 180-30, in = 30](0+\hspace,0);


\draw[line width = 2pt] (0.5+\hspace,0) to[out = 90, in = 180] (2+\hspace,1.5)to[out =0, in =90] (2+\hspace+0.5,1.5-0.3)to[out =-90, in =180+30] (2+\hspace+0.5+0.7,1.5-0.2);

\draw[line width = 2pt] (0.5+\hspace,0) to[out = -90, in = 180] (2+\hspace,-1.5)to[out =0, in =-90] (2+\hspace+0.5,-1.5+0.3)to[out =90, in =180-30] (2+\hspace+0.5+0.7,-1.5+0.2);

\filldraw[line width=1pt] (2+\hspace+0.5+0.7,-1.5+0.2) circle (0.7ex);
\filldraw[line width=1pt] (2+\hspace+0.5+0.7,1.5-0.2) circle (0.7ex);

\draw[line width = 2pt,<-] (0.5+\hspace,-0.001) -- (0.5+\hspace,0.001);


\draw[line width = 2pt] (1+\hspace-0.2,0+0.2) to[out = 90, in = 180] (2+\hspace-0.4,1.1)to[out =0, in =90] (2+\hspace+0.1+0.2,1.5-0.3-0.7+0.2);

\draw[line width = 2pt] (1+\hspace-0.2,0+0.2) to[out = -90, in = 180] (2+\hspace-0.4,-1.1)to[out =0, in =-90] (2+\hspace+0.1+0.2,-1.5+0.3+0.7-0.2);

\filldraw[line width=1pt] (2+\hspace+0.1+0.2,1.5-0.3-0.7+0.2) circle (0.7ex);
\filldraw[line width=1pt] (2+\hspace+0.1+0.2,-1.5+0.3+0.7-0.2) circle (0.7ex);

\draw[line width = 2pt,<-] (0.8+\hspace,-0.001+0.2) -- (0.8+\hspace,0.001+0.2);


\draw[color = ForestGreen, fill= ForestGreen, line width=1pt] (2.3+\hspace,0) circle (0.7ex);
\draw[color = ForestGreen, fill= ForestGreen, line width=1pt] (0+\hspace,0) circle (0.7ex);
\draw[color = cornellred, fill=  cornellred, line width=1pt] (6-0.5+\hspace,0) circle (0.7ex);

\node at (3+\hspace,2.5) {\scalebox{1.3}{$\Sigma$}
};
\node at (2+\hspace+0.5+0.7+0.4,1.5-0.2) {$z_1$
};
\node at (2+\hspace+0.1-0.2,1.5-0.3-0.7+0.1) {$z_2$};

\node at (2+\hspace+0.1-0.5,-1.55+0.3+0.7-0.1) {$\sigma(z_2)$};

\node at (2+\hspace+0.5+0.7+0.7,-1.5+0.2) {$\sigma(z_1)$};

\node at (2+\hspace-2.35,1.5-0.3-0.7+1) {$\gamma(z_1,z_2)$};


\node[color = ForestGreen] at (0+\hspace-0.5,0) {$\infty$
};

\node[color = ForestGreen] at (2.3+\hspace+0.5,0) {$E_0$};

\node[color = cornellred] at (6-0.5+\hspace+0.5,0) {$E^\star$};

\node[color = ForestGreen] at (0-0.5,0) {$\infty$
};

\node[color = ForestGreen] at (2.3+0.5,0) {$E_0$};

\node[color = cornellred] at (6-0.5+0.5,0) {$E^\star$};

    \end{tikzpicture}
   \caption{On the left, a pictorial representation of the contour $\gamma(z) \subset \Sigma$ and on the right, of the contour $\gamma(z_1,z_2)\subset \Sigma$.}
    \label{fig:AntiCycle}
\end{figure}

\paragraph{Negative instantons:} Now, we consider the sector $(n_+,n_-) = (0,1)$. Following \cite{mss22}, one can write
\begin{equation}
   \frac{Z^{\text{ZZ}}_{(0,1)}(\hbar)}{Z^{\text{pert}}(\hbar)}  =F^{\text{ZZ}}_{(0,1)}(\hbar) =  \int_{\bar{\mathcal{C}}^\star} \frac{\rmd \bar{E}}{2\pi}\psi(\gamma(z(\bar{E})))
\end{equation}
where $\gamma(z)$ is depicted in figure \ref{fig:AntiCycle} (compare with figure \ref{fig:Cycles} and notice that only the orientation changes) and $\bar{\mathcal{C}}^\star \subset \mathbb{C}$ is the steepest-ascent contour associated with the saddle $E^\star$.
Similarly to what we did before, we can generalise the result above for multi-instanton configurations by resorting to the matrix integral formula displayed in equation (3.10) of \cite{mss22}. Indeed, we can write
\begin{equation}
    \frac{Z_{(0,n)}^{\text{ZZ}}(\hbar)}{Z^{\text{pert}}(\hbar)} = \frac{1}{n!}\left[\prod_{j=1}^n \int_{\bar{\mathcal{C}}^\star} \frac{\rmd \bar{E}_j}{2\pi}\right]\Delta_{n}^2(\bar{E}_1,\cdots,\bar{E}_n)\psi(\gamma(z(\bar{E}_1),\cdots,z(\bar{E}_n)))
\end{equation}
where
\begin{equation}
    \gamma(z_1,\cdots,z_n) = \bigsqcup_{i=1}^n \gamma(z_i).
    \label{eq:nonpertopef3}
\end{equation}
An example of this contour for $n = 2$ is depicted in figure \ref{fig:AntiCycle} (compare with figure \ref{fig:nCycle}). Performing an expansion identical to the one leading up to equation \eqref{eq:nonpertopef1} yields
\begin{equation}
    \frac{Z_{(0,n)}^{\text{ZZ}}(\hbar)}{Z^{\text{pert}}(\hbar)} = \sum_{\sigma \in \Gamma(n)} \prod_{\substack{1 \le k \le n }}\frac{1}{\sigma_k!} \left(F_{(0,k)}^{\text{ZZ}}(\hbar)\right)^{\sigma_k}.
    \label{eq:nonpertopef5}
\end{equation}
The equation above can be recursively solved for all free energy transseries sectors.

\paragraph{Bulk instantons:} Finally, we consider the very first bulk transseries sector \cite{bssv22} $(n_+,n_-) = (1,1)$. Resorting to the matrix integral formula displayed in equation (3.10) of \cite{mss22}, it is straightforward to write
\begin{equation}
    \frac{Z^{\text{ZZ}}_{(1,1)}(\hbar)}{Z^{\text{pert}}(\hbar)} = \int_{\mathcal{C}^\star+\mathcal{C}_{\text{res}}} \frac{\rmd E}{2\pi}\int_{\bar{\mathcal{C}}^\star} \frac{\rmd \bar{E}}{2\pi} \frac{1}{(E-\bar{E})^2}\psi(\gamma(z(E),z(\bar{E})))
\end{equation}
where the contour $\gamma(z_1,z_2)$ is depicted in figure \ref{fig:BulkCycle} (compare with figures \ref{fig:nCycle} and \ref{fig:AntiCycle}, and note that each contour now has a different orientation) and the contour $\mathcal{C}_{\text{res}} \subset\mathbb{C}$ is depicted in figure \ref{fig:ResContour}.

The need to include the integration contour $\mathcal{C}_{\text{res}}$ arises from an obstruction to the analytical extension of the steepest-descent contour associated with an eigenvalue tunnelling to the saddle $E^\star$ in the presence of an anti-eigenvalue. We refer the reader to \cite{mss22} for a detailed derivation of the equation above.

\begin{figure}
    \centering
    \begin{tikzpicture}



   \draw[ decorate, decoration={snake, segment length=9, amplitude=4},line width=2pt, color = ForestGreen] (-4.24-2,0)--(-2-0.4242*0,0);


\draw[ForestGreen, fill=ForestGreen] (-2-0.4242*0,0) circle (0.7ex);
\draw[ForestGreen, fill=ForestGreen] (-4.24-2,0) circle (0.7ex);
\draw[cornellred, fill=cornellred] (2+0.4242*0,0) circle (0.7ex);


\node[color = ForestGreen] at (-4.24-2-0.5,0){$\infty$};
\node[color = ForestGreen] at (-2-0.4242*0+0.5,0){$E_0$};
\node[color = blue!100!white] at (0,0.9){$\mathcal{C}_{\text{res}}$};

\node[color = cornellred] at (2+0.4242*0-0.5,0){$E^\star$};

\draw[color = blue, line width = 2pt](-2-0.4242*0,-0.5) -- (2+0.4242*0,-0.5) to[out = 0, in  = -90] (2+0.4242*0+0.5,0)to[out = 90, in  = 0] (2+0.4242*0,0.5) -- (-2-0.4242*0,0.5)to[out = 180, in  = 90] (-2-0.4242*0-0.5,0)to[out = -90, in  = 180]cycle;


	\end{tikzpicture}
    \caption{Pictorial representation of the contour $\mathcal{C}_{\text{res}}$.}
    \label{fig:ResContour}
\end{figure}
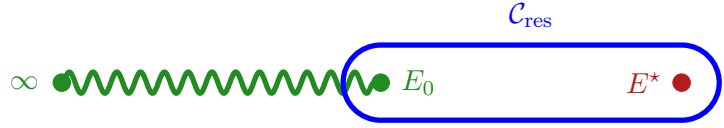
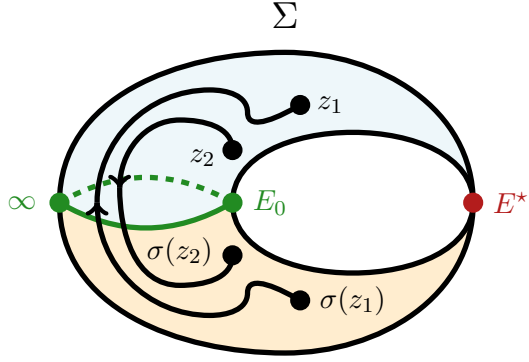
\begin{figure}
    \centering
    \begin{tikzpicture}

\def\hspace{0};

\draw[fill = LightBlue,fill opacity=0.2,line width = 0pt] (0+\hspace,0)to[out = 90, in = 180] (3+\hspace,2)to[out = 0, in = 90] (6-0.5+\hspace,0)to[out = 90, in = 90]  (2.3+\hspace,0)to[out = 180+30, in = -30]cycle;
\draw[fill = darktangerine,fill opacity=0.2,line width = 0pt] (0+\hspace,0)to[out = -90, in = 180] (3+\hspace,-2)to[out = 0, in = -90] (6-0.5+\hspace,0)to[out = -90, in = -90] (2.3+\hspace,0)to[out = 180+30, in = -30]cycle;
\draw[line width = 2pt] (0+\hspace,0)to[out = 90, in = 180] (3+\hspace,2)to[out = 0, in = 90] (6-0.5+\hspace,0)to[out = -90, in = 0] (3+\hspace,-2)to[out = 180, in = -90] cycle;
\draw[line width = 2pt] (2.3+\hspace,0)to[out = 90, in = 90] (6-0.5+\hspace,0)to[out = -90, in = -90]  cycle;

\draw[line width = 2pt,color = ForestGreen](2.3+\hspace,0)to[out = 180+30, in = -30](0+\hspace,0);
\draw[line width = 2pt,color = ForestGreen,dashed](2.3+\hspace,0)to[out = 180-30, in = 30](0+\hspace,0);


\draw[line width = 2pt] (0.5+\hspace,0) to[out = 90, in = 180] (2+\hspace,1.5)to[out =0, in =90] (2+\hspace+0.5,1.5-0.3)to[out =-90, in =180+30] (2+\hspace+0.5+0.7,1.5-0.2);

\draw[line width = 2pt] (0.5+\hspace,0) to[out = -90, in = 180] (2+\hspace,-1.5)to[out =0, in =-90] (2+\hspace+0.5,-1.5+0.3)to[out =90, in =180-30] (2+\hspace+0.5+0.7,-1.5+0.2);

\filldraw[line width=1pt] (2+\hspace+0.5+0.7,-1.5+0.2) circle (0.7ex);
\filldraw[line width=1pt] (2+\hspace+0.5+0.7,1.5-0.2) circle (0.7ex);

\draw[line width = 2pt,->] (0.5+\hspace,-0.001) -- (0.5+\hspace,0.001);


\draw[line width = 2pt] (1+\hspace-0.2,0+0.2) to[out = 90, in = 180] (2+\hspace-0.4,1.1)to[out =0, in =90] (2+\hspace+0.1+0.2,1.5-0.3-0.7+0.2);

\draw[line width = 2pt] (1+\hspace-0.2,0+0.2) to[out = -90, in = 180] (2+\hspace-0.4,-1.1)to[out =0, in =-90] (2+\hspace+0.1+0.2,-1.5+0.3+0.7-0.2);

\filldraw[line width=1pt] (2+\hspace+0.1+0.2,1.5-0.3-0.7+0.2) circle (0.7ex);
\filldraw[line width=1pt] (2+\hspace+0.1+0.2,-1.5+0.3+0.7-0.2) circle (0.7ex);

\draw[line width = 2pt,<-] (0.8+\hspace,-0.001+0.2) -- (0.8+\hspace,0.001+0.2);


\draw[color = ForestGreen, fill= ForestGreen, line width=1pt] (2.3+\hspace,0) circle (0.7ex);
\draw[color = ForestGreen, fill= ForestGreen, line width=1pt] (0+\hspace,0) circle (0.7ex);
\draw[color = cornellred, fill=  cornellred, line width=1pt] (6-0.5+\hspace,0) circle (0.7ex);

\node at (3+\hspace,2.5) {\scalebox{1.3}{$\Sigma$}
};
\node at (2+\hspace+0.5+0.7+0.4,1.5-0.2) {$z_1$
};
\node at (2+\hspace+0.1-0.2,1.5-0.3-0.7+0.1) {$z_2$};

\node at (2+\hspace+0.1-0.5,-1.55+0.3+0.7-0.1) {$\sigma(z_2)$};

\node at (2+\hspace+0.5+0.7+0.7,-1.5+0.2) {$\sigma(z_1)$};


\node[color = ForestGreen] at (0+\hspace-0.5,0) {$\infty$
};

\node[color = ForestGreen] at (2.3+\hspace+0.5,0) {$E_0$};

\node[color = cornellred] at (6-0.5+\hspace+0.5,0) {$E^\star$};

    \end{tikzpicture}
   \caption{Pictorial representation of the contour $\gamma(z_1,z_2) \subset \Sigma$.}
    \label{fig:BulkCycle}
\end{figure}
Expanding the logarithm in \eqref{eq:nonperef7} in powers of the transmonomials yields 
\begin{equation}
  \frac{Z_{(1,1)}^{\text{ZZ}}(\hbar)}{Z^{\text{pert}}(\hbar)} = F_{(1,0)}^{\text{ZZ}}(\hbar)F_{(0,1)}^{\text{ZZ}}(\hbar) + F_{(1,1)}^{\text{ZZ}}(\hbar).
  \label{eq:nonpertopef4}
\end{equation}
Similar expressions apply to the remaining bulk transseries sectors. However, they quickly become unwieldy as the instanton number increases \cite{krsst25a} and will therefore be omitted here.


\subsection{The WKB expansion}

\label{subsec:WKBexpansion}

An analytical approach to computing the FZZT non-perturbative transseries sectors of \eqref{eq:nonperef1} was developed in \cite{Saad:2019lba} and subsequently applied in \cite{os19}. In this Subsection, we present a brief overview of this method, referred to as the WKB expansion, and show how it can be used to derive analytical formulae for FZZT transseries sector coefficients. The purpose of this review is to set the stage for an independent consistency check of the novel results obtained in Subsection \ref{sec:transseriesfromgel} by resorting to the Gel’fand–Dikii resolvent equation \eqref{eq:gde6}. 

Simply put, the WKB expansion claims that 
\begin{equation}
    W^{\text{FZZT}_+}(z) = \left\langle\frac{\det\left(E(z)-M\right)}{\det\left(E(\sigma(z))-M\right)}\right\rangle.
    \label{eq:fzzt4}
\end{equation}
Using the wave function \eqref{eq:nonperef13}, we can write  
\begin{align}
    \left\langle\frac{\det\left(E(z)-M\right)}{\det\left(E(\sigma(z))-M\right)}\right\rangle = \psi(\gamma(z))
\end{align}
where the contour $\gamma(z) \subset \Sigma$ is depicted in figure \ref{fig:Cycles}. Performing an expansion in powers of $\hbar$ further yields
\begin{equation}
    \psi(\gamma(z)) = \exp\left(\frac{A_{\text{FZZT}}(z)}{\hbar}\right)\sum_{g \in \mathbb{N}_0}\psi_g^+(z)\hbar^g
\end{equation}
where the first few coefficients read
\begin{align}
    &\psi_0^+(z) = \exp\left(\frac{F_{0,2}(\gamma(z))}{2}\right) \\ 
   & \psi_1^+(z) =  \exp\left(\frac{F_{0,2}(\gamma(z))}{2}\right)\left(\frac{F_{0,3}(\gamma(z))}{6}+F_{1,1}(\gamma(z))\right) \\
    &\psi_2^+(z) =  \exp\left(\frac{F_{0,2}(\gamma(z))}{2}\right)\left(\frac{1}{2}\left(\frac{F_{0,3}(\gamma(z))}{6}+F_{1,1}(\gamma(z))\right)^2+\frac{F_{0,4}(\gamma(z))}{24}+\frac{F_{1,2}(\gamma(z))}{2}\right).
\end{align}

Similarly, the ghost FZZT D-brane non-perturbative transseries sector can be reproduced as
\begin{equation}
    W^{\text{FZZT}_-}(z) = \left\langle  \frac{\det\left(E(\sigma(z))-M\right)}{\det\left(E(z)-M\right)}\right\rangle.
\end{equation}
Indeed, one can show that the ghost FZZT D-brane contribution is an analytic continuation of the FZZT D-brane contribution onto the second sheet of $x(z)$ (compare with the expectation value \eqref{eq:fzzt4}). We can further write
\begin{equation}
    \left\langle  \frac{\det\left(E(\sigma(z))-M\right)}{\det\left(E(z)-M\right)}\right\rangle = \psi(\gamma(z))
\end{equation}
where the contour $\gamma(z) \subset\Sigma$ is depicted in figure \ref{fig:AntiCycle}. Performing an expansion in powers of $\hbar$ yields
\begin{equation}
    \psi(\gamma(z)) = \exp\left(-\frac{A_{\text{FZZT}}(z)}{\hbar}\right)\sum_{g \in \mathbb{N}_0}\psi_g^-(z)\hbar^g
    \label{eq:wkb5}
\end{equation}
where 
\begin{equation}
    \psi^-_g(z) = (-1)^{g}\psi^+_g(z).
    \label{eq:wkb4}
\end{equation}
The equation above provides a straightforward relation between the coefficients of the FZZT transseries sector and those of its ghost counterpart.

The expressions above are fully generic. In order to proceed, we must specify a double-scaled Hermitian matrix model, which is encoded in the spectral curve on which we run the topological recursion \eqref{eq:toprec2} to generate the coefficients featuring in the definition \eqref{eq:nonperef14}. Choosing our model to be $(2,3)$ minimal string theory and running the topological recursion for the associated spectral curve (see equations \eqref{eq:chebychev-pure-gravity-x} and \eqref{eq:chebychev-pure-gravity}) yields
\begin{align}
    & F_{0,2}(\gamma(z)) = -2\log\left(4z^2\right) \\ 
    & F_{0,3}(\gamma(z)) = \int_{-z}^{z} \rmd z_1 \int_{-z}^{z} \rmd z_2 \int_{-z}^{z} \rmd z_3 \hspace{1pt} \omega_{0,3}(z_1,z_2,z_3)= -\frac{4}{3 z^3} \\ 
    & F_{1,1}(\gamma(z)) = \int_{-z}^{z} \rmd z_1 \hspace{1pt} \omega_{1,1}(z_1)= -\frac{4 z^2+1}{72 z^3} \\ 
     & F_{0,4}(\gamma(z)) = \int_{-z}^{z} \rmd z_1 \int_{-z}^{z} \rmd z_2 \int_{-z}^{z} \rmd z_3 \int_{-z}^{z} \rmd z_4 \hspace{1pt} \omega_{0,4}(z_1,z_2,z_3,z_4)= \frac{16 \left(z^2+1\right)}{9 z^6}  \\  & F_{1,2}(\gamma(z)) = \int_{-z}^{z} \rmd z_1 \int_{-z}^{z} \rmd z_2 \hspace{1pt} \omega_{1,2}(z_1,z_2)= \frac{32 z^4+16 z^2+7}{216 z^6}.
\end{align}
Using the expressions above, we can finally write
\begin{align}
   & \psi_0^+(z) = \frac{1}{4z^2} \label{eq:wkb1} \\ 
    & \psi_1^+(z) = -\frac{4 z^2+17}{288 z^5} \label{eq:wkb2}\\ 
    &  \psi_2^+(z) = \frac{7 \left(112 z^4+184 z^2+175\right)}{41472 z^8}.\label{eq:wkb3}
\end{align}
The equations above yield analytical formulae for the first three coefficients of the FZZT non-perturbative transseries sector in \eqref{eq:nonperef1} for $(2,3)$ minimal string theory. These expressions will be matched against our novel results in Subsection \ref{sec:transseriesfromgel}, providing a highly non-trivial consistency check of our method. 

Yet another alternative approach for the analytical computation of the FZZT D-brane non-perturbative transseries sectors involves making use of the determinantal formulae \cite{ekr2018}, which relate the FZZT D-brane wave function and its ghost partner to multi-point correlation functions. Although valid, this strategy entails computing the perturbative expansion of the wave function and its ghost partner, and then propagating them through the relevant determinantal formula. The alternative method we present in \ref{sec:transseriesfromgel} is much more direct.

\subsection{Non-perturbative ODE approach}

\label{sec:transseriesfromgel}

In this Subsection, we demonstrate how to derive {\it all} non-perturbative transseries sectors in \eqref{eq:nonperef1} by resorting to the Gel’fand-Dikii equation \eqref{eq:gde6}. 
We start by substituting our transseries ansatz \eqref{eq:nonperef2}, into equation \eqref{eq:gde6} and solving for its coefficients recursively. We then employ equation \eqref{eq:nonperef3} to extract the transseries coefficients of the one-point correlation function directly from those of the resolvent. While our derivation remains largely general, we will illustrate it through a simple, concrete matrix model to aid the reader’s understanding. We pick $(2,3)$ minimal string theory to fill this role.

In Appendix \ref{appendix:Transseriesdata}, we collect several transseries coefficients associated with the Gel’fand–Dikii resolvent and the one-point correlation function, computed following the procedure outlined in this Subsection.


\subsubsection{ZZ-effects}
\label{subsec:zzeffects}

We start by showing how to derive the coefficients associated with ZZ transseries sectors featured in \eqref{eq:nonperef2}. For convenience of the reader, we rewrite here the Gel’fand-Dikii equation 
\begin{equation}
    4(u(x)-E)\widehat{R}^2(E,x)-2\hbar^2 \widehat{R}(E,x)\widehat{R}''(E,x)+\hbar^2\left(\widehat{R}'(E,x)\right)^2 = 1.
    \label{eq:zz12}
\end{equation}

Before proceeding, it is important to unpack the transseries representation of the specific heat~$u(x)$. Much like the free energy, this transseries only incorporates ZZ non-perturbative contributions and can be written as 
\begin{equation}
    u(x;\sigma_{\text{ZZ}_\pm})  = u^{\text{pert}}(x)+ \sum_{\substack{n_\pm \in \mathbb{N}_0}}' \sigma_{\text{ZZ}_+}^{n_+}\sigma_{\text{ZZ}_-}^{n_-}\exp\left(-\frac{n_+-n_-}{\hbar} \mathcal{A}_{\text{ZZ}}(x)\right)u_{(n_+,n_-)}^{\text{ZZ}}(x)
    \label{eq:zz15}
\end{equation}
where the instanton action reads
\begin{equation}
    \mathcal{A}_{\text{ZZ}}(x) =\frac{3\sqrt{3}}{5} u_0(x)^{\frac{5}{2}}.
    \label{eq:zz18}
\end{equation}
The perturbative sector can be written as
\begin{equation}
   u^{\text{pert}}(x) = \sum_{g \in \mathbb{N}_0}u_g(x)\hbar^{2g}.
   \label{eq:zz17}
\end{equation}

The instanton action above, together with the transseries sectors in \eqref{eq:zz15}, can be obtained by inserting the transseries \eqref{eq:zz15} into the string equation 
\begin{equation}
    -\frac{3}{4\sqrt{2}}\left(u(x)^2-\frac{\hbar^2}{3}  u''(x)\right) = x
    \label{eq:stringeq}
\end{equation}
and recursively solving for the coefficients of the transseries sectors (see Subsection 3.4 of \cite{gs21} for further details on this procedure). Analytical expressions for these coefficients may also be derived by relating the specific heat to the partition function and using the closed-form expression recently obtained in \cite{krsst25a}. The first few transseries coefficients are displayed in Appendix \ref{appendix:Transseriesdata}.

Substituting the transseries ansatz \eqref{eq:nonperef2} and the transseries \eqref{eq:zz15} into the equation \eqref{eq:zz12} and collecting terms by powers of the transseries parameters yields an infinite set of equations. To derive a recursion relation for the coefficients appearing in the ZZ transseries sectors, it suffices to consider only those equations associated with the transseries parameters $\sigma_{\text{ZZ}_\pm}$. Doing so quickly reveals the structure
\begin{align}
   & \widehat{\mathcal{R}}_{(n,0)}^{\text{ZZ}}(E,x) = \hbar^{\frac{n}{2}}\sum_{g \in \mathbb{N}_0} \widehat{\mathcal{R}}_{g,(n,0)}^{\text{ZZ}}(E,x)\hbar^{g} \label{eq:zz19}\\
   & \widehat{\mathcal{R}}_{(0,n)}^{\text{ZZ}}(E,x) = \hbar^{\frac{n}{2}}\sum_{g \in \mathbb{N}_0} \widehat{\mathcal{R}}_{g,(0,n)}^{\text{ZZ}}(E,x)\hbar^{g} \label{eq:zz20}\\& \widehat{\mathcal{R}}_{(1,1)}^{\text{ZZ}}(E,x) = \hbar\sum_{g \in \mathbb{N}_0}\widehat{\mathcal{R}}_{g,(1,1)}^{\text{ZZ}}(E,x)\hbar^{2g}
   \label{eq:zz14}
\end{align}
for all $n \in \mathbb{N}$. The leading coefficients of the first few transseries sectors associated with positive multiples of the instanton action $-\mathcal{A}_{\text{ZZ}}(x)$ read
\begin{align}
     & \widehat{\mathcal{R}}_{0,(1,0)}^{\text{ZZ}}(E,x) =-\frac{1}{2 u_0(x)^{1/4} \sqrt{u_0(x)-E } (u_0(x)+2 E )} \\ 
     & \widehat{\mathcal{R}}_{0,(2,0)}^{\text{ZZ}}(E,x) =-\frac{1}{12 u_0(x)^{3/2} \sqrt{u_0(x)-E } (u_0(x)+2 E )}\\ 
    & \widehat{\mathcal{R}}_{0,(3,0)}^{\text{ZZ}}(E,x) = -\frac{1}{96 u_0(x)^{11/4} \sqrt{u_0(x)-E } (u_0(x)+2 E )}
\end{align}
and several more coefficients can be found in Appendix \ref{appendix:Transseriesdata}. The coefficients of the first few transseries sectors associated with positive multiples of the instanton action $\mathcal{A}_{\text{ZZ}}(x)$ follow the rather simple structure
\begin{equation}
    \widehat{\mathcal{R}}_{g,(0,n)}(E,x) = (-1)^g\widehat{\mathcal{R}}_{g,(n,0)}(E,x).
    \label{eq:negativepositive}
\end{equation}

The leading coefficient associated with the first bulk instanton sector reads
\begin{equation}
     \widehat{\mathcal{R}}_{0,(1,1)}(E,x) = \frac{E  u_0(x)-2 u_0(x)^2-2 E ^2}{2 u_0(x)^{3/2} (u_0(x)-E )^{3/2} (u_0(x)+2 E )^2}.
     \label{eq:zz13}
\end{equation}

Resorting to equation \eqref{eq:nonperef3} one can then derive a simple recursion relation for the one-point correlation function coefficients. In order to do so, it is useful to define an auxiliary transseries
\begin{align}
    &\mathcal{W}_1(E,x;\sigma_{\text{ZZ}_\pm},\sigma_{\text{FZZT}_\pm}) = \mathcal{W}^{\text{pert}}(E,x) + \sigma_{\text{FZZT}_+}\exp\left(+\frac{A_{\text{FZZT}}(E,x)}{\hbar}\right) \mathcal{W}^{\text{FZZT}_+}(E,x) \nonumber\\ & +\sigma_{\text{FZZT}_-}\exp\left(-\frac{A_{\text{FZZT}}(E,x)}{\hbar}\right)\mathcal{W}^{\text{FZZT}_-}(E,x)  +\sum_{\substack{n_\pm \in \mathbb{N}_0 }}'\sigma_{\text{ZZ}_+}^{n_+}\sigma_{\text{ZZ}_-}^{n_-}\exp\left(-(n_+-n_-) \frac{A_{\text{ZZ}}(x)}{\hbar}\right) \mathcal{W}_{(n_+,n_-)}^{\text{ZZ}}(E,x)   \nonumber\\ & +\sigma_{\text{FZZT}_+}\exp\left(+\frac{A_{\text{FZZT}}(E,x)}{\hbar}\right)\sum_{\substack{n_\pm \in \mathbb{N}_0 }}'\sigma_{\text{ZZ}_+}^{n_+}\sigma_{\text{ZZ}_-}^{n_-}\exp\left(-(n_+-n_-) \frac{A_{\text{ZZ}}(x)}{\hbar}\right) \mathcal{W}_{(n_+,n_-)}^{\text{ZZ-FZZT}_+}(E,x)
   \nonumber \\ & + \sigma_{\text{FZZT}_-}\exp\left(-\frac{A_{\text{FZZT}}(E,x)}{\hbar}\right)\sum_{\substack{n_\pm \in \mathbb{N}_0}}'\sigma_{\text{ZZ}_+}^{n_+}\sigma_{\text{ZZ}_-}^{n_-}\exp\left(-(n_+-n_-) \frac{A_{\text{ZZ}}(x)}{\hbar}\right) \mathcal{W}_{(n_+,n_-)}^{\text{ZZ-FZZT}_-}(E,x)
    \label{eq:zz1}
\end{align}
obtained by encoding the dependence on the Fermi surface value on $x$. In particular, we have
\begin{equation}
   \mathcal{W}_1(E,\mu;\sigma_{\text{ZZ}_\pm},\sigma_{\text{FZZT}_{\pm}})  = W_1(E;\sigma_{\text{ZZ}_\pm},\sigma_{\text{FZZT}_{\pm}}). 
\end{equation}
Differentiating both sides of \eqref{eq:nonperef3}, yields the structure
\begin{align}
   & \mathcal{W}_{(n,0)}^{\text{ZZ}}(E,x) = \hbar^{\frac{n}{2}}\sum_{g \in \mathbb{N}_0} \mathcal{W}_{g,(n,0)}^{\text{ZZ}}(E,x) \hbar^g 
   \label{eq:zz21}\\ 
    & \mathcal{W}_{(0,n)}^{\text{ZZ}}(E,x) = \hbar^{\frac{n}{2}}\sum_{g \in \mathbb{N}_0}\mathcal{W}_{g,(0,n)}^{\text{ZZ}}(E,x) \hbar^g \label{eq:zz22}\\ 
     & \mathcal{W}_{(1,1)}^{\text{ZZ}}(E,x) = \sum_{g \in \mathbb{N}_0}\mathcal{W}_{g,(1,1)}^{\text{ZZ}}(E,x) \hbar^{2g} \label{eq:zz23}
\end{align}
where
\begin{align}
  &  \widehat{\mathcal{R}}_{g,(n,0)}^{\text{ZZ}}(E,x) = -\mathcal{A}_{\text{ZZ}}'(x)\mathcal{W}_{g,(n,0)}^{\text{ZZ}}(E,x) + \left(\mathcal{W}_{g-1,(n,0)}^{\text{ZZ}}\right)'(E,x) \label{eq:zz24} \\ &
  \widehat{\mathcal{R}}_{g,(0,n)}^{\text{ZZ}}(E,x) = \mathcal{A}_{\text{ZZ}}'(x)\mathcal{W}_{g,(0,n)}^{\text{ZZ}}(E,x) + \left(\mathcal{W}_{g-1,(0,n)}^{\text{ZZ}}\right)'(E,x)\label{eq:zz25} \\ 
  &
  \widehat{\mathcal{R}}_{g,(1,1)}^{\text{ZZ}}(E,x) = \left(\mathcal{W}_{g,(1,1)}^{\text{ZZ}}\right)'(E,x)
  \label{eq:zz16}
\end{align}
and where we adopt the convention that the correlation function coefficient is taken to be zero whenever $g < 0$. 

Using the equations above, we can easily compute several coefficients of the auxiliary transseries \eqref{eq:zz1}. The leading coefficients of the first few transseries sectors associated with positive multiples of the instanton action $-\mathcal{A}_{\text{ZZ}}(x)$ read
\begin{align}
    & \mathcal{W}_{0,(1,0)}^{\text{ZZ}}(E,x) =-\frac{1}{2 \sqrt{6} u_0(x)^{3/4} \sqrt{u_0(x)-E } (u_0(x)+2 E )}\\ 
    & \mathcal{W}_{0,(2,0)}^{\text{ZZ}}(E,x) = -\frac{1}{24 \sqrt{6} u_0(x)^2 \sqrt{u_0(x)-E } (u_0(x)+2 E )}\\ 
    & \mathcal{W}_{0,(3,0)}^{\text{ZZ}}(E,x) = -\frac{1}{288 \sqrt{6} u_0(x)^{13/4} \sqrt{u_0(x)-E } (u_0(x)+2 E )}.
\end{align}
 The coefficients of the first few transseries sectors associated with positive multiples of the instanton action $\mathcal{A}_{\text{ZZ}}(x)$ follow the rather simple structure
\begin{equation}
    \mathcal{W}_{g,(0,n)}^{\text{ZZ}}(E,x) = (-1)^{g+1}\mathcal{W}_{g,(n,0)}^{\text{ZZ}}(E,x).
    \label{eq:zz6}
\end{equation} 

To obtain the coefficients of the first bulk transseries sector, we must integrate the coefficients in \eqref{eq:zz14} (see equation \eqref{eq:zz16}). In particular, the right-hand side of equation \eqref{eq:zz13} must be integrated to obtain the leading coefficient. Therefore, as in the case of the perturbative sector of the Gel’fand-Dikii resolvent discussed in Subsection \ref{subsec:odeapproach}, we expect the coefficients in \eqref{eq:zz14} to be total derivatives. This can be readily confirmed for the leading coefficient by noting that
\begin{equation}
  \widehat{\mathcal{R}}_{0,(1,1)}^{\text{ZZ}}(E,x) =-\frac{3}{\sqrt{2}}    \frac{\rmd}{\rmd x}\left(\frac{\sqrt{u_0(x)}}{\sqrt{u_0(x)-E } (2 u_0(x)+4 E )}\right)
\end{equation}
from which it follows immediately that
\begin{equation}
     \mathcal{W}_{0,(1,1)}^{\text{ZZ}}(E,x) =  -\frac{3}{\sqrt{2}} \frac{\sqrt{u_0(x)}}{\sqrt{u_0(x)-E } (2 u_0(x)+4 E )}.
\end{equation}

Finally, we have to tune $x$ to the Fermi surface value \eqref{eq:toprec4} in order to obtain the transseries coefficients of \eqref{eq:nonperef1}. Doing so yields
\begin{align}
    & W_{0,(1,0)}^{\text{ZZ}}(E) =-\frac{1}{2 \sqrt{6} \sqrt{1-E } (2 E +1)} \label{eq:zz2}\\ 
    & W_{0,(2,0)}^{\text{ZZ}}(E) =-\frac{1}{24 \sqrt{6} \sqrt{1-E } (2 E +1)}\\ 
    & W_{0,(3,0)}^{\text{ZZ}}(E) = -\frac{1}{288 \sqrt{6} \sqrt{1-E } (2 E +1)}\label{eq:zz3}\\ 
      & W_{0,(1,1)}^{\text{ZZ}}(E) =-\frac{3}{2\sqrt{2} \sqrt{1-E } (2 E +1)}.
      \label{eq:zz10}
\end{align}
Furthermore, equation \eqref{eq:zz6} translates naturally to
\begin{equation}
    W_{g,(0,n)}^{\text{ZZ}}(E) = (-1)^{g+1}W_{g,(n,0)}^{\text{ZZ}}(E).
    \label{eq:zz7}
\end{equation}

\paragraph{Matching with the non-perturbative topological recursion:} We can now attempt to match the coefficients of the one-point correlation function computed above with those obtained via the non-perturbative topological recursion. We start with the coefficients associated with positive multiples of the instanton action $-\mathcal{A}_{\text{ZZ}}(x)$. Invoking the derivation property of the loop insertion operator and using equation \eqref{eq:nonpertopef1} yields
\begin{align}
     \hbar\Delta_{z(E)} \frac{Z^{\text{ZZ}}_{(1,0)}(\hbar)}{Z^{\text{pert}}(\hbar)}  &=\hbar \Delta_{z(E)} F^{\text{ZZ}}_{(1,0)}(\hbar) = W_{(1,0)}^{\text{ZZ}}(E)\frac{\rmd E}{\rmd z}(E)  \label{eq:zz4} \\ 
    \hbar \Delta_{z(E)} \frac{Z^{\text{ZZ}}_{(2,0)}(\hbar)}{Z^{\text{pert}}(\hbar)}  &= \left(\hbar\Delta_{z(E)} F^{\text{ZZ}}_{(1,0)}(\hbar)\right) F^{\text{ZZ}}_{(1,0)}(\hbar) +\hbar\Delta_{z(E)} F^{\text{ZZ}}_{(2,0)}(\hbar) \nonumber \\&= \left(W_{(1,0)}^{\text{ZZ}}(E)F_{(1,0)}^{\text{ZZ}}(\hbar)+W_{(2,0)}^{\text{ZZ}}(E)\right)\frac{\rmd E}{\rmd z}(E) \\ 
    \hbar \Delta_{z(E)} \frac{Z^{\text{ZZ}}_{(3,0)}(\hbar)}{Z^{\text{pert}}(\hbar)}  &= \left(\frac{1}{2}W_{(1,0)}^{\text{ZZ}}(E)\left(F_{(1,0)}^{\text{ZZ}}(\hbar)\right)^2 + W_{(2,0)}^{\text{ZZ}}(E)F_{(1,0)}^{\text{ZZ}}(\hbar)+W_{(3,0)}^{\text{ZZ}}(E)\right)\frac{\rmd E}{\rmd z}(E) \label{eq:zz5}.
\end{align}
In Appendix \ref{appendix:Matrixintegrals}, we compute truncated saddle-point expansions of the matrix integrals featuring in the left-hand side of the equations above. In particular, using the expansions \eqref{eq:matrixintegral1}, \eqref{eq:matrixintegral2} and \eqref{eq:matrixintegral3}, we can write
\begin{align}
     & \frac{Z^{\text{pert}}(\hbar)}{Z_{(1,0)}^{\text{ZZ}}(\hbar)} \hbar\Delta_{z(E)} \frac{Z^{\text{ZZ}}_{(1,0)}(\hbar)}{Z^{\text{pert}}(\hbar)}  = \frac{4 \sqrt{3}}{2E -1}\hbar^{\frac{1}{2}}  -\frac{ 68 E ^3-51 E +161}{6 (E +1) (2 E -1)^3}\hbar^{\frac{3}{2}} +\mathcal{O}\left(\hbar^{\frac{5}{2}}\right) \\ &
 \frac{Z^{\text{pert}}(\hbar)}{Z_{(2,0)}^{\text{ZZ}}(\hbar)} \hbar\Delta_{z(E)} \frac{Z^{\text{ZZ}}_{(2,0)}(\hbar)}{Z^{\text{pert}}(\hbar)}  =\frac{8 \sqrt{3}}{2 E -1}\hbar  -\frac{2  \left(37 E  \left(4 E ^2-3\right)+85\right)}{(E +1) (2 E -1)^3}\hbar^{2} +\mathcal{O}\left(\hbar^{3}\right) \\ 
  &
 \frac{Z^{\text{pert}}(\hbar)}{Z_{(3,0)}^{\text{ZZ}}(\hbar)} \hbar\Delta_{z(E)} \frac{Z^{\text{ZZ}}_{(3,0)}(\hbar)}{Z^{\text{pert}}(\hbar)}  =\frac{12 \sqrt{3}}{2 E -1}\hbar^{\frac{3}{2}} +\mathcal{O}\left(\hbar^{\frac{5}{2}}\right).
\end{align}
On the other hand, using the transseries coefficients \eqref{eq:zz2} to \eqref{eq:zz3} along with several subleading coefficients as well as free energy and partition function transseries coefficients found in Appendix \ref{appendix:Transseriesdata}, we can write
\begin{align}
    & \frac{Z^{\text{pert}}(\hbar)}{Z_{(1,0)}^{\text{ZZ}}(\hbar)}W_{(1,0)}^{\text{ZZ}}(E)\frac{\rmd E}{\rmd z}(E) = \frac{4 \sqrt{3}}{2E -1}\hbar^{\frac{1}{2}}  -\frac{68 E ^3-51 E +161}{6 (E +1) (2 E -1)^3}\hbar^{\frac{3}{2}} +\mathcal{O}\left(\hbar^{\frac{5}{2}}\right) \\ & \frac{Z^{\text{pert}}(\hbar)}{Z_{(2,0)}^{\text{ZZ}}(\hbar)}\left(W_{(1,0)}^{\text{ZZ}}(E)F_{(1,0)}^{\text{ZZ}}(\hbar)+W_{(2,0)}^{\text{ZZ}}(E)\right)\frac{\rmd E}{\rmd z}(E) = \frac{8 \sqrt{3}}{2 E -1}\hbar  -\frac{2  \left(37 E  \left(4 E ^2-3\right)+85\right)}{(E +1) (2 E -1)^3}\hbar^{2} +\mathcal{O}\left(\hbar^{3}\right) \\
    &  \frac{Z^{\text{pert}}(\hbar)}{Z_{(3,0)}^{\text{ZZ}}(\hbar)}\left(\frac{1}{2}W_{(1,0)}^{\text{ZZ}}(E)\left(F_{(1,0)}^{\text{ZZ}}(\hbar)\right)^2 + W_{(2,0)}^{\text{ZZ}}(E)F_{(1,0)}^{\text{ZZ}}(\hbar)+W_{(3,0)}^{\text{ZZ}}(E)\right)\frac{\rmd E}{\rmd z}(E) = \frac{12 \sqrt{3}}{2 E -1}\hbar^{\frac{3}{2}} +\mathcal{O}\left(\hbar^{\frac{5}{2}}\right).
\end{align}
The equations above demonstrate that equations \eqref{eq:zz4} to \eqref{eq:zz5} hold at least to the $\hbar$ orders we considered. This constitutes a highly non-trivial check, validating the procedure outlined above for computing transseries coefficients of the one-point correlation function via the Gel’fand-Dikii equation.

We can now further expand this check to the coefficients associated with positive multiples of the instanton action $\mathcal{A}_{\text{ZZ}}(x)$. Indeed, using equation \eqref{eq:nonpertopef5}, we may similarly write
\begin{align}
     \hbar\Delta_{z(E)} \frac{Z^{\text{ZZ}}_{(0,1)}(\hbar)}{Z^{\text{pert}}(\hbar)}  &= \hbar\Delta_{z(E)} F^{\text{ZZ}}_{(0,1)}(\hbar) = W_{(0,1)}^{\text{ZZ}}(E)\frac{\rmd E}{\rmd z}(E)  \label{eq:zz8}\\ 
    \hbar \Delta_{z(E)} \frac{Z^{\text{ZZ}}_{(0,2)}(\hbar)}{Z^{\text{pert}}(\hbar)}  &= \left(\hbar\Delta_{z(E)} F^{\text{ZZ}}_{(0,1)}(\hbar)\right) F^{\text{ZZ}}_{(0,1)}(\hbar) +\hbar\Delta_{z(E)} F^{\text{ZZ}}_{(0,2)}(\hbar) \nonumber\\ &= \left(W_{(0,1)}^{\text{ZZ}}(E)F_{(0,1)}^{\text{ZZ}}(\hbar)+W_{(0,2)}^{\text{ZZ}}(E)\right)\frac{\rmd E}{\rmd z}(E) \\ 
      \hbar\Delta_{z(E)} \frac{Z^{\text{ZZ}}_{(0,3)}(\hbar)}{Z^{\text{pert}}(\hbar)} &= \left(\frac{1}{2}W_{(0,1)}^{\text{ZZ}}(E)\left(F_{(0,1)}^{\text{ZZ}}(\hbar)\right)^2 + W_{(0,2)}^{\text{ZZ}}(E)F_{(0,1)}^{\text{ZZ}}(\hbar)+W_{(0,3)}^{\text{ZZ}}(E)\right)\frac{\rmd E}{\rmd z}(E).\label{eq:zz9}
\end{align}
Using the truncated saddle-point expansions \eqref{eq:matrixintegral4}, \eqref{eq:matrixintegral5} and \eqref{eq:matrixintegral6} of Appendix \ref{appendix:Matrixintegrals}, yields
\begin{align}
     & \frac{Z^{\text{pert}}(\hbar)}{Z_{(0,1)}^{\text{ZZ}}(\hbar)} \hbar\Delta_{z(E)} \frac{Z^{\text{ZZ}}_{(0,1)}(\hbar)}{Z^{\text{pert}}(\hbar)}  = -\frac{4 \sqrt{3}}{2E -1}\hbar^{\frac{1}{2}}  -\frac{68 E ^3-51 E +161}{6 (E +1) (2 E -1)^3}\hbar^{\frac{3}{2}} +\mathcal{O}\left(\hbar^{\frac{5}{2}}\right) \\ &
 \frac{Z^{\text{pert}}(\hbar)}{Z_{(0,2)}^{\text{ZZ}}(\hbar)} \hbar\Delta_{z(E)} \frac{Z^{\text{ZZ}}_{(0,2)}(\hbar)}{Z^{\text{pert}}(\hbar)}  =-\frac{8 \sqrt{3}}{2 E -1}\hbar  -\frac{2 \left(37 E  \left(4 E ^2-3\right)+85\right)}{(E +1) (2 E -1)^3}\hbar^{2} +\mathcal{O}\left(\hbar^{3}\right) \\ 
  &
 \frac{Z^{\text{pert}}(\hbar)}{Z_{(0,3)}^{\text{ZZ}}(\hbar)} \hbar\Delta_{z(E)} \frac{Z^{\text{ZZ}}_{(0,3)}(\hbar)}{Z^{\text{pert}}(\hbar)}  =-\frac{12 \sqrt{3}}{2 E -1}\hbar^{\frac{3}{2}} +\mathcal{O}\left(\hbar^{\frac{5}{2}}\right).
\end{align}
On the other hand, using equation \eqref{eq:zz7} along with several other transseries coefficients from Appendix \ref{appendix:Transseriesdata}, we can write
\begin{align}
    & \frac{Z^{\text{pert}}(\hbar)}{Z_{(0,1)}^{\text{ZZ}}(\hbar)}W_{(0,1)}^{\text{ZZ}}(E)\frac{\rmd E}{\rmd z}(E) = -\frac{4 \sqrt{3}}{2E -1}\hbar^{\frac{1}{2}}  -\frac{68 E ^3-51 E +161}{6 (E +1) (2 E -1)^3}\hbar^{\frac{3}{2}} +\mathcal{O}\left(\hbar^{\frac{5}{2}}\right) \\ & \frac{Z^{\text{pert}}(\hbar)}{Z_{(0,2)}^{\text{ZZ}}(\hbar)}\left(W_{(0,1)}^{\text{ZZ}}(E)F_{(0,1)}^{\text{ZZ}}(\hbar)+W_{(0,2)}^{\text{ZZ}}(E)\right)\frac{\rmd E}{\rmd z}(E) = -\frac{8 \sqrt{3}}{2 E -1}\hbar  -\frac{2 \left(37 E  \left(4 E ^2-3\right)+85\right)}{(E +1) (2 E -1)^3}\hbar^{2} +\mathcal{O}\left(\hbar^{3}\right) \\
    &  \frac{Z^{\text{pert}}(\hbar)}{Z_{(0,3)}^{\text{ZZ}}(\hbar)}\left(\frac{1}{2}W_{(0,1)}^{\text{ZZ}}(E)\left(F_{(0,1)}^{\text{ZZ}}(\hbar)\right)^2 + W_{(0,2)}^{\text{ZZ}}(E)F_{(0,1)}^{\text{ZZ}}(\hbar)+W_{(0,3)}^{\text{ZZ}}(E)\right)\frac{\rmd E}{\rmd z}(E) = -\frac{12 \sqrt{3}}{2 E -1}\hbar^{\frac{3}{2}} +\mathcal{O}\left(\hbar^{\frac{5}{2}}\right).
\end{align}
The equations above show that \eqref{eq:zz8} to \eqref{eq:zz9} hold at least to the $\hbar$ orders we have considered. 

As a final, highly non-trivial check, let us consider the leading coefficient of the first bulk transseries sector. Using equation \eqref{eq:nonpertopef4}, we can write
\begin{align}
    \hbar\Delta_{z(E)}\frac{Z^{\text{ZZ}}_{(1,1)}(\hbar)}{Z^{\text{pert}}(\hbar)} &=  \hbar\Delta_{z(E)} F_{(1,0)}^{\text{ZZ}}(\hbar) F_{(0,1)}^{\text{ZZ}}(\hbar) + \hbar\Delta_{z(E)} F_{(0,1)}^{\text{ZZ}}(\hbar) F_{(1,0)}^{\text{ZZ}}(\hbar) + \hbar\Delta_{z(E)}F_{(1,1)}^{\text{ZZ}}(\hbar) \\ &= \left(W_{(1,0)}^{\text{ZZ}}(E) F_{(0,1)}^{\text{ZZ}}(\hbar) + W_{(0,1)}^{\text{ZZ}}(E) F_{(1,0)}^{\text{ZZ}}(\hbar) + W_{(1,1)}^{\text{ZZ}}(E)\right)\frac{\rmd E}{\rmd z}(E).
    \label{eq:zz11}
\end{align}
Using the truncated saddle-point expansion \eqref{eq:matrixintegral7}, we can write
\begin{equation}
    \frac{Z^{\text{pert}}(\hbar)}{Z_{(1,1)}^{\text{ZZ}}(\hbar)} \hbar\Delta_{z(E)} \frac{Z^{\text{ZZ}}_{(1,1)}(\hbar)}{Z^{\text{pert}}(\hbar)}  = \frac{20}{3-6 E }\hbar^{-1}+\mathcal{O}(\hbar^0).
\end{equation}
Alternatively, using equation \eqref{eq:zz10} along with several other transseries coefficients from Appendix \ref{appendix:Transseriesdata} yields
\begin{equation}
    \frac{Z^{\text{pert}}(\hbar)}{Z_{(1,1)}^{\text{ZZ}}(\hbar)}\left(W_{(1,0)}^{\text{ZZ}}(E)F_{(0,1)}^{\text{ZZ}}(\hbar) + W_{(0,1)}^{\text{ZZ}}(E) F_{(1,0)}^{\text{ZZ}}(\hbar) + W_{(1,1)}^{\text{ZZ}}(E)\right)\frac{\rmd E}{\rmd z}(E) = \frac{20}{3-6 E }\hbar^{-1}+\mathcal{O}(\hbar^0).
\end{equation}
Once again, the equation above appears to support \eqref{eq:zz11}, and with it, the validity of the coefficient \eqref{eq:zz10}, derived via the Gel’fand-Dikii equation.



\subsubsection{FZZT-effects}

We now turn to computing the transseries coefficients associated with FZZT-effects. To derive a recursion relation for these coefficients, we substitute the transseries ansatz \eqref{eq:nonperef2} into \eqref{eq:zz12} and collect terms according to powers of $\sigma_{\text{FZZT}_\pm}$. Each power yields an ordinary differential equation that can be solved for the corresponding coefficient, allowing all coefficients to be determined recursively. Doing this quickly reveals the structure
\begin{equation}
   \widehat{ \mathcal{R}}^{\text{FZZT}_\pm}(E,x) = \sum_{g \in \mathbb{N}_0}\widehat{ \mathcal{R}}^{\text{FZZT}_\pm}_{g}(E,x)\hbar^{g}
   \label{eq:fzzt5}
\end{equation}
where the first few coefficients read
\begin{align}
    & \widehat{ \mathcal{R}}^{\text{FZZT}_+}_{0}(E,x) =  \frac{1}{\sqrt{u_0(x)-E }}\\ 
     & \widehat{ \mathcal{R}}^{\text{FZZT}_+}_{1}(E,x) = \frac{2 E -7 u_0(x)}{18 \sqrt{2} u_0(x)^2 (E -u_0(x))^2}\\ 
     & \widehat{ \mathcal{R}}^{\text{FZZT}_+}_{2}(E,x) = \frac{-364 E  u_0(x)+469 u_0(x)^2+100 E ^2}{1296 u_0(x)^4 (u_0(x)-E )^{7/2}}. 
\end{align}

The coefficients associated with the instanton action $-\mathcal{A}_{\text{FZZT}}(E,x)$ follow the rather simple structure
\begin{equation}
   \widehat{\mathcal{R}}^{\text{FZZT}_-}_{g}(E,x) = (-1)^{g}\widehat{\mathcal{R}}^{\text{FZZT}_+}_{g}(E,x).
   \label{eq:fzzt6}
\end{equation}

Using equation \eqref{eq:nonperef3}, we can write
\begin{align}
   & \mathcal{W}^{\text{FZZT}_\pm}(E,x) = \sum_{g \in \mathbb{N}_0} \mathcal{W}_{g}^{\text{FZZT}_\pm}(E,x) \hbar^g
   \label{eq:fzzt7}
\end{align}
where
\begin{align}
  &  \widehat{\mathcal{R}}_{g}^{\text{FZZT}_\pm}(E,x) = \pm\mathcal{A}_{\text{FZZT}}'(E,x)\mathcal{W}_{g}^{\text{FZZT}_\pm}(E,x) + \left(\mathcal{W}_{g-1}^{\text{FZZT}_\pm}\right)'(E,x).
\end{align}
Using the equation above, we can recursively compute the first few coefficients associated with the FZZT sectors of the auxiliary transseries \eqref{eq:zz1}. Doing so yields
\begin{align}
    & \mathcal{W}^{\text{FZZT}_+}_{0}(E,x) = -\frac{1}{2 (E -u_0(x))}\\ 
     &  \mathcal{W}^{\text{FZZT}_+}_{1}(E,x) = \frac{2E -19 u_0(x)}{36 \sqrt{2} u_0(x)^2 (u_0(x)-E )^{5/2}}\\ 
     &  \mathcal{W}^{\text{FZZT}_+}_{2}(E,x) = \frac{7 \left(-148E  u_0(x)+295 u_0(x)^2+28E ^2\right)}{2592 u_0(x)^4 (E -u_0(x))^4}
\end{align}
and
\begin{equation}
    \mathcal{W}^{\text{FZZT}_-}_{g}(E,x) = (-1)^{g+1}\mathcal{W}^{\text{FZZT}_+}_{g}(E,x).
    \label{eq:fzzt8}
\end{equation}

Finally, fixing $x$ to the Fermi surface value \eqref{eq:toprec4} yields
\begin{align}
    & W^{\text{FZZT}_+}_{0}(E) =-\frac{1}{2 (E -1)} \label{eq:fzzt1}\\
     &  W^{\text{FZZT}_+}_{1}(E) =\frac{2E -19}{36 \sqrt{2} (1-E )^{5/2}} \label{eq:fzzt2}\\ 
     &  W^{\text{FZZT}_+}_{2}(E) =\frac{7 \left(28E ^2-148E +295\right)}{2592 (E -1)^4}\label{eq:fzzt3}
\end{align}
and
\begin{equation}
    W^{\text{FZZT}_-}_{g}(E) =(-1)^{g+1}W^{\text{FZZT}_+}_{g}(E). 
    \label{eq:fzzt9}
\end{equation}
More coefficients can be found in Appendix \ref{appendix:Transseriesdata}. 

\paragraph{Match with the WKB expansion:}

Using the uniformization variable $E = 1-2z^2$ (see equation \eqref{eq:chebychev-pure-gravity-x}), we can rewrite the coefficients displayed in equations \eqref{eq:fzzt1}, \eqref{eq:fzzt2}, and \eqref{eq:fzzt3}, as 
\begin{align}
    & W^{\text{FZZT}_+}_{0}(z) =\frac{1}{4 z^2}\\
     &  W^{\text{FZZT}_+}_{1}(z) =-\frac{4 z^2+17}{288 z^5}\\ 
     &  W^{\text{FZZT}_+}_{2}(z) =\frac{7 \left(112 z^4+184 z^2+175\right)}{41472 z^8}.
\end{align}
The equations above exactly match the WKB expansion analytical formulae displayed in equations \eqref{eq:wkb1}, \eqref{eq:wkb2} and \eqref{eq:wkb3}, thereby providing further validation of our method. Moreover, up to an overall minus sign, equation \eqref{eq:fzzt9} agrees with the equation \eqref{eq:wkb4}, derived from the WKB expansion \eqref{eq:wkb5}. The minus sign is immaterial, as it can be absorbed into the multiplicative integration constant associated with the coefficient $\widehat{\mathcal{R}}_0^{\text{FZZT}-}(E,x)$.



\subsubsection{ZZ-FZZT-effects}

\label{subsubsec:zz-fzzt}

Finally, we turn to the computation of transseries coefficients associated with non-perturbative ZZ-FZZT-effects. To derive a recursion relation for these coefficients, we substitute the transseries ansatz \eqref{eq:nonperef2} into \eqref{eq:zz12} and collect terms according to monomials of both $\sigma_{\text{FZZT}_\pm}$ and $\sigma_{\text{ZZ}_\pm}$. Each power yields an algebraic equation that can be solved for the corresponding coefficient, allowing all coefficients to be determined recursively. Doing this quickly reveals the structure
\begin{align}
    & \widehat{ \mathcal{R}}^{\text{ZZ-FZZT}_\pm}_{(n,0)}(E,x) = \hbar^{\frac{n}{2}}\sum_{g \in \mathbb{N}_0} \widehat{ \mathcal{R}}^{\text{ZZ-FZZT}_\pm}_{g,(n,0)}(E,x)\hbar^{g} \label{eq:zz-fzzt1}\\ & \widehat{ \mathcal{R}}^{\text{ZZ-FZZT}_\pm}_{(0,n)}(E,x) = \hbar^{\frac{n}{2}}\sum_{g \in \mathbb{N}_0}\widehat{ \mathcal{R}}^{\text{ZZ-FZZT}_\pm}_{g,(0,n)}(E,x)\hbar^{g}\label{eq:zz-fzzt2}
\end{align}
for all $n \in \mathbb{N}$. As in many previous instances, we compute the transseries coefficients above recursively. We then make use of equation \eqref{eq:nonperef3} to obtain the corresponding auxiliary transseries coefficients, after which we fix $x$ to the Fermi surface value given in \eqref{eq:toprec4}. Doing so yields the structure
\begin{align}
   & \mathcal{W}_{(n,0)}^{\text{ZZ-FZZT}_\pm}(E,x) = \hbar^{\frac{n}{2}}\sum_{g \in \mathbb{N}_0}\mathcal{W}_{g,(n,0)}^{\text{ZZ-FZZT}_\pm}(E,x) \hbar^g 
   \label{eq:zz-fzzt3}\\ 
    & \mathcal{W}_{(0,n)}^{\text{ZZ-FZZT}_\pm}(E,x) = \hbar^{\frac{n}{2}}\sum_{g \in \mathbb{N}_0} \mathcal{W}_{g,(0,n)}^{\text{ZZ-FZZT}_\pm}(E,x) \hbar^g \label{eq:zz-fzzt4}
\end{align}
for all $n \in \mathbb{N}$. Once all steps are carried out, the first few transseries coefficients are given by
\begin{align}
     & W^{\text{ZZ-FZZT}_\pm}_{0,(1,0)}(E)  = \frac{1}{(2E+1)^2\sqrt{1-E}}\left(\frac{5}{\sqrt{6}}-\sqrt{\frac{2}{3}}E\mp2\sqrt{1-E}\right) \\ 
      & W^{\text{ZZ-FZZT}_\pm}_{0,(0,1)}(E)  = -\frac{1}{(2E+1)^2\sqrt{1-E}}\left(\frac{5}{\sqrt{6}}-\sqrt{\frac{2}{3}}E\pm2\sqrt{1-E}\right).
\end{align}
More coefficients can be found in Appendix \ref{appendix:Transseriesdata}.

It is important to remark that, to the best of the authors’ knowledge, there is currently no alternative method available for computing mixed non-perturbative ZZ-FZZT effects. The procedure outlined here not only allows for the explicit computation of the transseries coefficients associated with these effects, but is also remarkably simple, amounting to the recursive solution of linear algebraic equations.

\section{Large order growth of volumes}

\label{sec:largeorder}

A rather compelling application of the framework developed in the last Section consists in proving generic results addressing the large order growth associated with topological expansions of moduli space volumes. This work was already carried out in \cite{eggls23} for the case of JT gravity in which concrete large order growth predictions were obtained and numerically tested for Weil–Petersson volumes. However, similar studies for JT supergravity seem to be mostly absent from the literature. This Section is dedicated to filling this gap. 


\subsection{Transseries and large order growth}

\label{subsec:largeorderformula}
 
As shown in Subsections 2.3 and 6.3 of \cite{abs19}, the large order growth of a transseries sector (such as the perturbative expansion) is directly governed by the remaining transseries sectors. More concretely, let us consider the rather simple toy-transseries
\begin{equation}
    \psi(\hbar;\sigma_1,\sigma_2) = \hbar^{\beta}\sum_{g \in \mathbb{N}_0} a_{g}\hbar^g + \sigma_1\exp\left(-\frac{A_1}{\hbar}\right)\hbar^{\beta_1}\sum_{g \in \mathbb{N}_0} b_{g}\hbar^g+ \sigma_2\exp\left(-\frac{A_2}{\hbar}\right)\hbar^{\beta_2}\sum_{g \in \mathbb{N}_0} c_{g}\hbar^g
\end{equation}
where $\sigma_1,\sigma_2$ denote transseries parameters.

Using a straightforward derivation based on the Stokes phenomenon and Cauchy’s theorem (see Subsection 2.3 of \cite{abs19} for a detailed yet simple account of this derivation), we obtain the following asymptotic relation
\begin{equation}
    a_g \sim \frac{S_1}{\pi \rmi}\frac{\Gamma(g+\beta-\beta_1)}{A_1^{g+\beta-\beta_1}}\sum_{h \in \mathbb{N}_0} A_1^h b_h \prod_{r=1}^h \frac{1}{(g-r)} +\frac{S_2}{\pi \rmi}\frac{\Gamma(g+\beta-\beta_2)}{A_2^{g+\beta-\beta_2}}\sum_{h \in \mathbb{N}_0} A_2^h c_h \prod_{r=1}^h \frac{1}{(g-r)}
    \label{eq:vol1}
\end{equation}
for large $g$ where $S_1,S_2 \in \mathbb{C}$ are problem dependent Stokes constants \cite{abs19}. 

As we can see from the asymptotic relation above, the large order growth of the perturbative coefficients $a_g$ is governed by contributions from both non-perturbative transseries sectors. When $|A_2| > |A_1|$, the second sector is exponentially suppressed relative to the first, and the leading large order behaviour is dominated by the first non-perturbative instanton sector. On the other hand, if $|A_1| = |A_2|$, both sectors contribute at leading order, resulting in a phase interference pattern that manifests as oscillations with a characteristic frequency as the index $g$ varies \cite{eggls23}. 


\subsection{Large order growth of the one-point correlation function}

\label{subsec:largecorrelation}

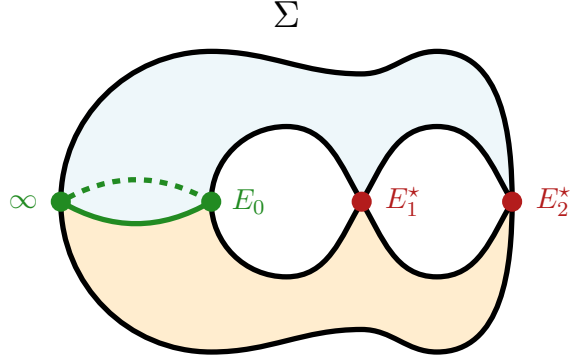
\begin{figure}
    \centering
    \begin{tikzpicture}
        \draw[line width  = 0pt,fill = LightBlue,fill opacity=0.2] (0,0) to[out = 90, in = 180] (2,2)to[out = 0, in = 180] (4,1.7)to[out = 0, in = 180] (5,2)to[out = 0, in = 90] (6,0) to[out = 90+30, in = 0] (5,1)to[out = 180, in = 90-30] (4,0)to[out = 90+30, in = 0] (3,1)to[out = 180, in =90] (2,0)to[out = 180+30, in = -30] cycle;

        \draw[line width  = 0pt,fill = darktangerine,fill opacity=0.2] (0,0) to[out = -90, in = 180] (2,-2)to[out = 0, in = 180] (4,-1.7)to[out = 0, in = 180] (5,-2)to[out = 0, in = -90] (6,0) to[out = -90-30, in = 0] (5,-1)to[out = 180, in = -90+30] (4,0)to[out = -90-30, in = 0] (3,-1)to[out = 180, in =-90] (2,0)to[out = 180+30, in = -30] cycle;

         \draw[line width  = 2pt] (0,0) to[out = 90, in = 180] (2,2)to[out = 0, in = 180] (4,1.7)to[out = 0, in = 180] (5,2)to[out = 0, in = 90] (6,0) to[out = 90+30, in = 0] (5,1)to[out = 180, in = 90-30] (4,0)to[out = 90+30, in = 0] (3,1)to[out = 180, in =90] (2,0);

          \draw[line width  = 2pt] (0,0) to[out = -90, in = 180] (2,-2)to[out = 0, in = 180] (4,-1.7)to[out = 0, in = 180] (5,-2)to[out = 0, in = -90] (6,0) to[out = -90-30, in = 0] (5,-1)to[out = 180, in = -90+30] (4,0)to[out = -90-30, in = 0] (3,-1)to[out = 180, in =-90](2,0);

\draw[line width  = 2pt, color = ForestGreen] (2,0)to[out = 180+30, in = -30] (0,0);
\draw[line width  = 2pt, color = ForestGreen,dashed] (2,0)to[out = 180-30, in = 30] (0,0);

\draw[color = ForestGreen, fill= ForestGreen, line width=1pt] (0,0) circle (0.7ex);
\draw[color = ForestGreen, fill= ForestGreen, line width=1pt] (2,0) circle (0.7ex);
\draw[color = cornellred, fill= cornellred, line width=1pt] (4,0) circle (0.7ex);
\draw[color = cornellred, fill= cornellred, line width=1pt] (6,0) circle (0.7ex);
           
\node[cornellred] at (4+0.55,0){$E_1^\star$};
\node[cornellred] at (6+0.55,0){$E_2^\star$};

        



        \node[ForestGreen] at (-0.5,0){$\infty$};
        \node[ForestGreen] at (2+0.5,0){$E_0$};

        \node at (3,2.5){\scalebox{1.3}{$\Sigma$}};

    \end{tikzpicture}
    \caption{Schematic representation of a spectral curve $\Sigma$ featuring a single cut extending from $\infty$ to $E_0$ (green line) and two saddles (red dots) represented as two distinct pinches.}
    \label{fig:spectralcurveexample}
\end{figure}

Let us now return to the case of a generic double scaled Hermitian one-matrix model whose large $N$ behaviour is governed by a spectral curve $\Sigma$ featuring one cut extending from $\infty$ to $E_0 \in \mathbb{C}$ and $\kappa \in \mathbb{N}$ saddles $\{E_1^\star,\cdots,E_\kappa^\star\}\subset \mathbb{C}$ (see figure \ref{fig:spectralcurveexample} for a schematic depiction of $\Sigma$ for $\kappa = 2$). A particular example is the $(2,2\kappa-1)$ minimal string. 

In this setting, the transseries for the one-point correlation function is expected to feature not just a single symmetric pair of ZZ-type instanton actions, but rather $\kappa$ distinct ones, each representing the non-perturbative effect associated with instanton configurations realised by eigenvalues tunnelling from the cut to distinct saddles (see \cite{mss22} for a detailed account of this interpretation within the contex of the partition function transseries). Indeed, in the particular case of the $(2,3)$ minimal string, $\Sigma$ only has one saddle (see figure \ref{fig:Cycles}) and correspondingly one ZZ-type instanton action \eqref{eq:nonperef10}.

In contrast, the transseries is still expected to involve only a single symmetric pair of FZZT-type instanton actions (corresponding to the FZZT D-brane contribution and its ghost partner), since it depends on a single variable $E$, leaving no natural variable to associate with additional FZZT-type instanton actions. Consequently, the full transseries generally includes $2\kappa$ infinite families of non-perturbative ZZ-sectors and a pair of non-perturbative FZZT-sectors, in addition to all mixed non-perturbative ZZ–FZZT-sectors arising from their combinations (compare with the transseries structure \eqref{eq:nonperef1}). 

Based on the reasoning outlined in Subsection \ref{subsec:largeorderformula}, we expect all of these sectors to govern the large order growth of the perturbative series
\begin{equation}
   W^{\text{pert}}(E) =  \sum_{g \in \mathbb{N}_0}W_{g,1}(E)\hbar^{2g-1}.
\end{equation}

The novelty here lies in the fact that the large order asymptotic relation \eqref{eq:vol1} now depends parametrically on $E$. As a result, the large order growth behaviour of the coefficients can qualitatively change with $E$, since different non-perturbative sectors may dominate depending on how their respective instanton actions evolve with varying $E$. This phenomenon is thoroughly explained and numerically studied in \cite{eggls23}. 

In what follows, we compute the leading coefficient of the asymptotic series governing the large-order growth of the coefficients above, for the FZZT non-perturbative effect and for a generic ZZ non-perturbative effect, linked to one of the saddles $E^\star \in \{E_1^\star,\cdots,E_\kappa^\star\}$. The remaining non-perturbative effects, including mixed ZZ-FZZT effects, can be included in an identical fashion. 

We begin by considering the partial transseries ansatz
\begin{align}
     W_{1}^{\text{partial}}(E;\sigma_{\text{ZZ}},\sigma_{\text{FZZT}}) = &W^{\text{pert}}(E) +   \sigma_{\text{FZZT}} \exp\left(\frac{A_{\text{FZZT}}(E)}{\hbar}\right)\sum_{g \in \mathbb{N}_0} W^{\text{FZZT}}_{g}(E)\hbar^g \nonumber \\ & + \sigma_{\text{ZZ}} \exp\left(-\frac{A_{\text{ZZ}}}{\hbar}\right)\hbar^{\frac{1}{2}}\sum_{g \in \mathbb{N}_0} W^{\text{ZZ}}_{g}(E)\hbar^g 
     \label{eq:vol2}
\end{align}
where (see equations \eqref{eq:nonperef10} and \eqref{eq:vol4})
\begin{align}
    & A_{\text{FZZT}}(E) = \oint_{\gamma(z(E))} \omega_{0,1}(\bullet) = V_{\text{eff}}(E)- V_{\text{eff}}(E_0) \\ 
     &A_{\text{ZZ}} = \oint_B \omega_{0,1}(\bullet) = V_{\text{eff}}(E^*)- V_{\text{eff}}(E_0)
\end{align}
and where $\gamma(z) \subset \Sigma$ is a smooth path connecting $z$ to $\sigma(z)$ and $B\subset \Sigma$ is the $B$-cycle connecting the cut to the saddle $E^\star$ (see figure \ref{fig:Cycles} for examples of these contours for $\kappa = 1$). Plugging the ansatz
\begin{align}
   \widehat{R}_{\text{partial}}(E,x;\sigma_{\text{ZZ}},\sigma_{\text{FZZT}}) = &\widehat{\mathcal{R}}^{\text{pert}}(E,x)  + \sigma_{\text{FZZT}} \exp\left(\frac{\mathcal{A}_{\text{FZZT}}(E,x)}{\hbar}\right)\sum_{g \in \mathbb{N}_0}\widehat{\mathcal{R}}^{\text{FZZT}}_{g}(E,x)\hbar^g \nonumber \\ &+  \sigma_{\text{ZZ}} \exp\left(-\frac{\mathcal{A}_{\text{ZZ}}(x)}{\hbar}\right)\hbar^{\frac{1}{2}}\sum_{g \in \mathbb{N}_0}\widehat{\mathcal{R}}^{\text{ZZ}}_{g}(E,x)\hbar^g  
\end{align}
in the Gel’fand-Dikii resolvent equation \eqref{eq:zz12} and solving for the leading coefficients of the series above yields
\begin{align}
  &\widehat{\mathcal{R}}^{\text{FZZT}}_{0}(E,x) = \frac{1}{\sqrt{u_0(x)-E}} \label{eq:vol27}\\   &\widehat{\mathcal{R}}^{\text{ZZ}}_{0}(E,x)  = -\frac{u^{\text{ZZ}}_{0}(x)}{\sqrt{u_0(x)-E } \left(\mathcal{A}_{\text{ZZ}}'(x)^2+4 E-4 u_0(x) \right)}
\end{align}
where $u_0^{\text{ZZ}}(x)$ denotes the leading coefficient of the specific heat transseries sector associated with the instanton action $-\mathcal{A}_{\text{ZZ}}(x)$. Further resorting to equation \eqref{eq:nonperef3} yields
\begin{align}
     & W^{\text{FZZT}}_{0}(E)  =  \frac{1}{2(E_0-E)} \\ 
     &   W^{\text{ZZ}}_{0}(E) =  \frac{u^{\text{ZZ}}_{0}(\mu)}{8\sqrt{E_0-E^\star}\sqrt{E_0-E } \left( E -E^\star\right)} 
\end{align}
where we used the equations $\partial_x\mathcal{A}_{\text{FZZT}}(E,\mu) = 2\sqrt{E_0-E}$ and $\mathcal{A}_{\text{ZZ}}'(\mu) = 2\sqrt{E_0-E^\star}$. From equations \eqref{eq:vol1} and \eqref{eq:vol2}, it follows that
\begin{align}
     W_{g,1}(E)  \sim & \frac{S_{\text{FZZT}}}{2\pi \rmi }\frac{\Gamma(2g-1)}{A_{\text{FZZT}}^{2g-1}(E)} \left(\frac{1}{E_0-E} +\cdots \right) +\frac{S_{\text{ZZ}}}{8\pi \rmi }\frac{1}{A_{\text{ZZ}}^{2g-\frac{3}{2}}}\Gamma\left(2g-\frac{3}{2}\right) \times \nonumber \\ & \left(\frac{u^{\text{ZZ}}_{0}(\mu)}{\sqrt{E_0-E^\star}\sqrt{E_0-E } \left(E -E^\star\right)}+\cdots \right)
     +\cdots
\label{eq:vol3}
\end{align}
for large $g$ where $S_{\text{FZZT}},S_{\text{ZZ}} \in \mathbb{C}$ are Stokes constants \cite{abs19} and the dots represent both subleading corrections to the asymptotic expansions featuring on the large order relation as well as entire asymptotic series emerging from the transseries sectors not considered in \eqref{eq:vol2}. 

Resorting to a generic expression computed in \cite{gs21}, we can write
\begin{equation}
    S_{\text{ZZ}}u_{0}^{\text{ZZ}}(\mu) = \frac{1}{\sqrt{2\pi}}\frac{1}{\sqrt{V_{\text{eff}}''(E^\star)}}.
\end{equation}
Using the equations above, we can rewrite the asymptotic relation \eqref{eq:vol3} as
\begin{align}
     W_{g,1}(E)  \sim &\frac{S_{\text{FZZT}}}{2\pi \rmi }\frac{\Gamma(2g-1)}{\left(V_{\text{eff}}(E)-V_{\text{eff}}(E_0)\right)^{2g-1}} \left(\frac{1}{E_0-E} +\cdots \right)  +\frac{1}{8\sqrt{2}\pi^{\frac{3}{2}} \rmi} \frac{1}{\left(V_{\text{eff}}(E^\star)-V_{\text{eff}}(E_0)\right)^{2g-\frac{3}{2}}}\times \nonumber\\ & \Gamma\left(2g-\frac{3}{2}\right)\left(\frac{1}{\sqrt{V''_{\text{eff}}(E^\star)(E_0-E^\star)}\sqrt{E_0-E } \left(E-E^\star \right)}+\cdots\right)+\cdots .
    \label{eq:vol10}
\end{align}


\subsection{Results for various moduli space volumes}

\label{subsec:volumes}

Rewriting the asymptotic relation \eqref{eq:vol10} in the uniformization variable $E = E_0-z^2$ yields
\begin{align}
     \widehat{W}_{g,1}(z)  \sim&  -\frac{S_{\text{FZZT}}}{\pi \rmi} \frac{\Gamma(2g-1)}{(V_{\text{eff}}(z)-V_{\text{eff}}(E_0))^{2g-1}} \left(\frac{1}{z}+\cdots\right)-\frac{1}{4\sqrt{2}\pi^{\frac{3}{2}}\rmi} \frac{1}{\left(V_{\text{eff}}(E^\star)-V_{\text{eff}}(E_0)\right)^{2g-\frac{3}{2}}}\times \nonumber\\ & \Gamma\left(2g-\frac{3}{2}\right)\left(\frac{1}{\sqrt{V_{\text{eff}}''(E^\star)(E_0-E^\star)}}\frac{1}{ E_0 -E^\star -z^2}+\cdots\right)+\cdots .
    \label{eq:vol6}
\end{align}
Applying the inverse operator of the Laplace-like integral transform \eqref{eq:int4} to both sides of the asymptotic relation \eqref{eq:vol6} yields
\begin{align}
    V_{g,1}(b) \sim &  -2\frac{S_{\text{FZZT}}}{(2\pi \rmi)^2}\frac{\Gamma\left(2g-1\right)}{b} \left(\oint_{\mathcal{C}_0} \rmd z \frac{\exp\left(bz\right)}{z( V_{\text{eff}}(z)-V_{\text{eff}}(E_0))^{2g-1}}+ \cdots\right) +\frac{1}{4\sqrt{2}\pi^{\frac{3}{2}}\rmi} \Gamma\left(2g-\frac{3}{2}\right)\times \nonumber \\ &\frac{1}{\left(V_{\text{eff}}(E^\star)-V_{\text{eff}}(E_0)\right)^{2g-\frac{3}{2}}}  \left(\frac{1}{\sqrt{V''_{\text{eff}}(E^\star)}}\frac{1}{E_0-E^\star}\frac{\sinh \left(b \sqrt{E_0-E^\star }\right)}{b} + \cdots \right)
    +\cdots 
    \label{eq:vol9}
\end{align}
where $\mathcal{C}_0 \subset \mathbb{C}$ denotes a small closed contour encircling $z = 0$ in the counter-clockwise direction. The asymptotic relation above is expected to hold for any double scaled Hermitian one-matrix model,  including double-scaled  matrix models in the wider Altland-Zirnbauer (AZ) classification\cite{az97}. Before concluding this Section, let us illustrate its application across various JT gravity models.


\paragraph{JT gravity:}

 We start with non-supersymmetric JT gravity in which case, the spectral curve can be written as (see equation \eqref{eq:JT-y})
\begin{equation}
    y(E) = \frac{1}{2\pi \rmi}\sinh(2\pi\sqrt{E}).
\end{equation}
Integrating the expression above yields the effective potential
\begin{equation}
    V_{\text{eff}}(E) = \frac{\rmi}{4 \pi ^3}\left(\sinh (2 \pi \sqrt{E})-2 \pi  \sqrt{E} \cosh (2 \pi  \sqrt{E})\right)
\end{equation}
whose saddles read
\begin{equation}
    E^\star_n =  -\frac{n^2}{4}
\end{equation}
for all $n \in \mathbb{N}$. We can then rewrite the asymptotic relation \eqref{eq:vol9} as
\begin{equation}
    V_{g,1}(b) \sim C_g^{\text{ZZ}}(b)+ C_g^{\text{FZZT}}(b)+\cdots
    \label{eq:vol21}
\end{equation}
where the leading ZZ contribution reads
\begin{equation}
    C_g^{\text{ZZ}}(b) = (-1)^{n+1}\frac{1}{8\sqrt{2} \pi ^{\frac{9}{2}}}\Gamma \left(2 g-\frac{3}{2}\right)\left(\frac{4\pi^2}{n}\right)^{2g} \frac{1}{b}\sinh \left(\frac{b n}{2}\right).
\label{eq:vol30}
\end{equation}

The large-order growth contribution above exactly matches the one displayed in equation (6.6) of \cite{eggls23} upon multiplication by a factor of $2$. This factor arises because the transseries sector associated with $A_{\text{ZZ}}$ contributes to the large-order growth of the perturbative coefficients $W_{g,1}(E)$ through an asymptotic series whose leading coefficient matches the one displayed in the second line of \eqref{eq:vol10}. Thus, accounting for the contribution of this transseries sector as well essentially amounts to doubling the contribution above.

The leading FZZT contribution reads
\begin{equation}
    C_g^{\text{FZZT}}(b) =  -2\frac{S_{\text{FZZT}}}{(2\pi \rmi)^2}\frac{\Gamma(2g-1)}{b}\oint_{\mathcal{C}_0} \rmd z\left(\frac{2 \pi  z \cos (2 \pi  z)-\sin (2 \pi  z)}{4 \pi ^3}\right)^{1-2g} \frac{\exp\left(bz\right)}{z}.
    \label{eq:vol29}
\end{equation}

The contour integration in the expression above yields a polynomial in $b$ for each value of $g$. While the full analytical expression appears to be out of reach, an explicit result can be obtained for the polynomial coefficient of the highest power. Indeed, we can write
\begin{equation}
    \frac{1}{2\pi \rmi}\oint_{\mathcal{C}_0} \rmd z\left(\frac{\sin (2 \pi  z)-2 \pi  z \cos (2 \pi  z)}{4 \pi ^3}\right)^{1-2g} \frac{\exp\left(bz\right)}{z} = -\frac{b^{6g-3}}{\Gamma(6g-2)}\left(\frac{3}{2}\right)^{2g-1}+\cdots 
\end{equation}
where the dots stand for terms proportional to smaller powers of $b$. We can then rewrite the large order growth contribution \eqref{eq:vol29} as
\begin{align}
    C_g^{\text{FZZT}}(b) = \frac{S_{\text{FZZT}}}{\pi \rmi} \frac{\Gamma(2g-1)}{\Gamma(6g-2)}\left(\frac{3}{2}\right)^{2g-1} b^{6g-4}+ \cdots
\end{align}
for large $b$. The large order growth contribution above exactly matches the one displayed in equation (6.11) of \cite{eggls23} upon fixing
\begin{equation}
    S_{\text{FZZT}} = \frac{\rmi}{2}.
    \label{eq:vol28}
\end{equation}

The Stokes constant $S_{\text{FZZT}}$ is not an invariant quantity, as it depends on the choice of integration constant underlying the computation of $\widehat{\mathcal{R}}^{\text{FZZT}}_0(E,x)$. In fixing this coefficient in \eqref{eq:vol27}, we implicitly selected a particular value of $S_{\text{FZZT}}$, which, in the case of JT gravity, turns out to be given by \eqref{eq:vol28}. For different matrix models, the coefficient \eqref{eq:vol27} may be associated with different Stokes constants (as we will soon see).


\paragraph{$\mathcal{N} = 1$ JT supergravity:} We now shift gears and turn to $\mathcal{N}=1$ JT supergravity, whose spectral curve reads:
\begin{equation}
    y(E) = -\frac{\sqrt{2}\cos\left(2\pi\sqrt{-E}\right)}{\sqrt{-E}}.
\end{equation}
Integrating the expression above yields the effective potential
\begin{equation}
    V_{\text{eff}}(E) = \frac{\sqrt{2}\sin (2 \pi  \sqrt{-E})}{\pi}
\end{equation}
whose saddles can be written as
\begin{equation}
    E^\star_n = -\frac{1}{16} (2 n+1)^2
\end{equation}
for all $n \in \mathbb{N}_0$. 

We can rewrite the asymptotic relation \eqref{eq:vol9} as
\begin{equation}
     V_{g,1}(b) \sim C^{\text{ZZ}}_g(b) + C^{\text{FZZT}}_g(b) + \cdots
     \label{eq:vol8}
\end{equation}
where the ZZ contribution reads
\begin{equation}
  C^{\text{ZZ}}_g(b) =  (-1)^{n+1}\frac{1}{(2  n+1)\pi ^{\frac{7}{2}}}\Gamma \left(2 g-\frac{3}{2}\right)\left(\frac{\pi^2}{2}\right)^g\frac{1}{b}\sinh \left(\frac{b}{4}  (2 n+1)\right).
\end{equation}
The ZZ contribution above exactly matches the one obtained in equation (2.13) of \cite{gprs24} upon multiplication by a factor of $2$, accounting for the contribution of the transseries sector associated with the instanton action $A_{\text{ZZ}}$ (see the paragraph below equation \eqref{eq:vol30}).\footnote{The volumes computed in \cite{gprs24,Stanford:2019vob} are defined with an additional factor of $\sqrt{2}$ relative to \eqref{eq:int4}, which must be taken into account in order to obtain precise agreement. This factor arises because \cite{Stanford:2019vob} defines a trumpet partition function that differs from \eqref{eq:trumpet-form} by a factor of $\sqrt{2}$.} 

The FZZT contribution reads
\begin{align}
  C^{\text{FZZT}}_g(b) &=  -2\frac{S_{\text{FZZT}}}{(2\pi \rmi)^2} \frac{\Gamma(2g-1)}{b} \oint_{\mathcal{C}_0}\rmd z \frac{\exp\left(bz\right)}{z}\left(\frac{\sqrt{2}\sin (2 \pi  z)}{\pi}\right)^{1-2g} \nonumber \\ 
  &  = S_{\text{FZZT}}\frac{\Gamma(2g-1)}{2^{g+\frac{1}{2}} \pi^{3-2g}}\oint_{\mathcal{C}_0} \frac{\rmd z}{zb} \frac{\sinh (b z)+\cosh (b z)}{\left(\sin(2\pi z)\right)^{2g-1}}\nonumber \\ & = S_{\text{FZZT}}\frac{\Gamma(2g-1)}{2^{g+\frac{1}{2}} \pi^{3-2g}} \oint_{\mathcal{C}_0} \frac{\rmd z}{zb} \frac{\sinh (b z)}{\left(\sin(2\pi z)\right)^{2g-1}}.
  \label{eq:vol20}
\end{align}
In the last equality, we used the fact that
\begin{equation}
    \oint_{\mathcal{C}_0}\frac{\rmd z}{zb} \frac{\cosh (b z)}{\left(\sin(2\pi z)\right)^{2g-1}} = 0 
\end{equation}
for all $g \in \mathbb{N}$. The FZZT contribution \eqref{eq:vol20} exactly matches the conjecture put forward in equation (E.20) of \cite{Stanford:2019vob} and derived in \cite{gprs24} upon fixing (compare with equation \eqref{eq:vol28})
\begin{equation}
    S_{\text{FZZT}} = \rmi.
\end{equation}

The result of the integral in \eqref{eq:vol20} is a polynomial in $b$ of degree $2g-2$ whose generic analytical expression seems to be out of reach. However, much like in the case of JT gravity, we can get an analytical prediction for the polynomial coefficient of highest power. Indeed, we can write
\begin{equation}
    \frac{1}{2\pi \rmi}\oint_{\mathcal{C}_0} \frac{\rmd z}{z} \frac{\sinh (b z)}{\left(\sin(2\pi z)\right)^{2g-1}} = (2 \pi )^{1-2 g}\frac{b^{2g-1}}{\Gamma (2 g)} +\cdots
\end{equation}
where the dots stand for terms proportional to smaller powers of $b$. Using the equation above, we can rewrite the contribution \eqref{eq:vol20} as  
\begin{equation}
   C^{\text{FZZT}}_g(b) = -\frac{2\sqrt{2}}{\pi}\frac{\Gamma(2g-1)}{\Gamma (2 g)} \left(\frac{1}{2}\right)^{3g}b^{2g-2} + \cdots
\end{equation}
for large $b$.


\paragraph{$\mathcal{N} = 2$ JT supergravity:}

Now, we turn to $\mathcal{N} = 2$ JT supergravity~\cite{Forste:2017apw,Stanford:2017thb,Mertens:2017mtv,Heydeman:2020hhw}. Following the conventions used in \cite{Turiaci:2023jfa,j23b}, we may write the spectral curve as:
\begin{equation}
    y(E) = \frac{\rmi}{4\pi^2 E}\sinh\left(2\pi \sqrt{E-E_0}\right).
\end{equation}
Integrating the spectral curve above yields the effective potential 
\begin{align}
    V_{\text{eff}}(E) &= \frac{\rmi}{8 \pi ^2}e^{-2 \rmi \pi  \sqrt{E_0}}\left(e^{4 \rmi \pi  \sqrt{E_0}} \text{I}_1(E)-e^{4 \rmi \pi  \sqrt{E_0}} \text{I}_2(E)+\text{I}_3(E)-\text{I}_4(E)\right)
    \label{eq:vol31}
\end{align}
where the saddles read
\begin{equation}
    E_n^\star = E_0-\frac{n^2}{4}
    \label{eq:vol22}
\end{equation}
for all $n \in \mathbb{N}$. In equation \eqref{eq:vol31}, we have defined the auxiliary functions
\begin{align}
    &\text{I}_1(E) = \text{Ei}\left(2 \pi  \left(\sqrt{E-E_0}-\rmi \sqrt{E_0}\right)\right) \label{eq:vol34}\\ &\text{I}_2(E) =  \text{Ei}\left(-2 \pi  \left(\sqrt{E-E_0}+\rmi \sqrt{E_0}\right)\right) \\ &\text{I}_3(E) = \text{Ei}\left(2 \pi  \left(\sqrt{E-E_0}+\rmi \sqrt{E_0}\right)\right)\\ &\text{I}_4(E) = \text{Ei}\left(-2 \pi  \sqrt{E-E_0}+2 \rmi \pi  \sqrt{E_0}\right).
    \label{eq:vol35}
\end{align}

We can now rewrite \eqref{eq:vol9} as
\begin{equation}
     V_{g,1}(b) \sim C^{\text{ZZ}}_g(b) + C^{\text{FZZT}}_g(b) + \cdots
     \label{eq:vol24}
\end{equation}
where the ZZ contribution reads
\begin{equation}
    C^{\text{ZZ}}_g(b) = \mathcal{C}^{\text{ZZ}}\Gamma\left(2g-\frac{3}{2}\right)\left(\frac{1}{V_{\text{eff}}(E^\star)}\right)^{2g-\frac{3}{2}}\frac{1}{b}\sinh(\frac{bn}{2}).
\end{equation}
In the equation above, we defined the constant
\begin{equation}
    \mathcal{C}^{\text{ZZ}} =\frac{\rmi^n}{2 \pi  n }\sqrt{\frac{ n^2-4 E_0}{n}}.
    \label{eq:vol32}
\end{equation}
The FZZT contribution reads
\begin{equation}
     C^{\text{FZZT}}_g(b) =  -2\frac{S_{\text{FZZT}}}{(2\pi \rmi)^2}\frac{\Gamma\left(2g-1\right)}{b} \oint_{\mathcal{C}_0} \rmd z \frac{\exp\left(bz\right)}{z( V_{\text{eff}}(z)-V_{\text{eff}}(E_0))^{2g-1}}.
\end{equation}

As in other JT gravity models, this integral cannot be evaluated exactly, yielding a polynomial in $b$ for each value of $g$. Nonetheless, an analytic expression for the coefficient of the highest power of $b$, can still be obtained. Indeed, we can write
\begin{align}
     C^{\text{FZZT}}_g(b) & = -\frac{S_{\text{FZZT}}}{\pi\rmi}  \frac{ \Gamma (2 g-1)}{ \Gamma (6 g-2)}\left(3\pi E_0\right)^{2g-1} b^{6g-4}+ \cdots
\end{align}
for large $b$.


\paragraph{Small $\mathcal{N}  = 4$ JT supergravity:} Now, we address small $\mathcal{N}  = 4$ JT supergravity\cite{Heydeman:2020hhw,Iliesiu:2020qvm}. Following conventions used in refs.\cite{Turiaci:2023jfa,Johnson:2024tgg,Johnson:2025oty}, we can write the relevant spectral curve as
\begin{equation}
    y(E) = \frac{2J\rmi}{\pi E^2}\sinh(2\pi \sqrt{E-E_0})
\end{equation}
where $E_0 = J^2$ and $J\in 2^{-1}\mathbb{Z}_ {>0}$ is the angular momentum defining the representation of the $R$-symmetry group $SU(2)$. Integrating the expression above yields the effective potential
\begin{align}
    V_{\text{eff}}(E) & = \frac{2 \rmi J}{\pi }\left(\frac{\rmi \pi}{2 \sqrt{E_0}}e^{-2 \rmi \pi  \sqrt{E_0}}\left(e^{4 \rmi \pi  \sqrt{E_0}} \text{I}_2(E)-e^{4 \rmi \pi  \sqrt{E_0}} \text{I}_1(E)+\text{I}_3(E)-\text{I}_4(E)\right)-\frac{\sinh \left(2 \pi  \sqrt{E-E_0}\right)}{E}\right)
\end{align}
where the saddles are given by equation \eqref{eq:vol22} and the functions $\text{I}_i(E)$ are defined from equation \eqref{eq:vol34} to \eqref{eq:vol35}. 

We can rewrite the asymptotic relation \eqref{eq:vol9} as
\begin{equation}
     V_{g,1}(b) \sim C^{\text{ZZ}}_g(b) + C^{\text{FZZT}}_g(b) + \cdots
      \label{eq:vol25}
\end{equation}
where the ZZ contribution reads
\begin{equation}
    C^{\text{ZZ}}_g(b) = \mathcal{C}^{\text{ZZ}}\Gamma\left(2g-\frac{3}{2}\right)\left(\frac{1}{V_{\text{eff}}(E^\star)}\right)^{2g-\frac{3}{2}}\frac{1}{b}\sinh(\frac{bn}{2}).
\end{equation}
In the equation above, we defined the constant (compare with equation \eqref{eq:vol32})
\begin{equation}
    \mathcal{C}^{\text{ZZ}} =\frac{\rmi^{1-n} }{8n  \pi  }\frac{n^2-4 E_0}{\sqrt{2n\pi J }}.
    \label{eq:vol33}
\end{equation}
The FZZT contribution reads
\begin{align}
     C^{\text{FZZT}}_g(b) &=  -2\frac{S_{\text{FZZT}}}{(2\pi \rmi)^2}\frac{\Gamma\left(2g-1\right)}{b} \oint_{\mathcal{C}_0} \rmd z \frac{\exp\left(bz\right)}{z( V_{\text{eff}}(z)-V_{\text{eff}}(E_0))^{2g-1}} \nonumber \\ & = -\frac{S_{\text{FZZT}}}{\pi \rmi }\frac{ \Gamma (2 g-1)}{ \Gamma (6 g-2)}\left(\frac{3E_0^2}{8J}\right)^{2 g-1}b^{6g-4}+ \cdots
\end{align}
for large $b$ where the dots denote terms proportional to smaller powers of $b$.


\paragraph{Large $\mathcal{N} = 4$ JT supergravity:} Finally, we turn to large $\mathcal{N} = 4$ JT supergravity \cite{Heydeman:2025vcc}. Following \cite{Johnson:2025oty}, we can write the relevant spectral curve as
\begin{equation}
    y(E) = A_{j_+,j_-}\rmi\frac{\sinh(2\pi \sqrt{E-E_0})}{(E-E^-)(E-E^+)}
    \label{eq:vol23}
\end{equation}
where we defined the constants 
\begin{align}
    & A_{j_+,j_-} = \frac{\alpha^{\frac{3}{2}}j_+j_-\sqrt{2\pi}}{16(1+\alpha)^3} \\ &  E^\pm = \frac{\alpha(j_+ \pm j_-)^2}{(1+\alpha)^2} \\ &
    E_0 = \frac{\alpha j_+^2+j_-^2}{1+\alpha}
\end{align}
for some $j_\pm \in \mathbb{N}_{0}$. Integrating the expression \eqref{eq:vol23} yields the effective potential
\begin{align}
     V_{\text{eff}}(E) =\frac{A_{j_+,j_-}\rmi}{2 (E^--E^+)}&\hspace{3pt}\Bigg(e^{-2 \pi  \sqrt{E^{-}-E_0}} \left(e^{4 \pi  \sqrt{E^{-}-E_0}} \text{I}_2(E)-e^{4 \pi  \sqrt{E^{-}-E_0}} \text{I}_3(E)-\text{I}_1(E)+\text{I}_4(E)\right) \nonumber\\ &+e^{-2 \pi  \sqrt{E^{+}-E_0}} \left(e^{4 \pi  \sqrt{E^{+}-E_0}} \text{I}_7(E)-e^{4 \pi  \sqrt{E^{+}-E_0}} \text{I}_5(E)+\text{I}_6(E)-\text{I}_8(E)\right)\Bigg)
\end{align}
where the saddles are given by equation \eqref{eq:vol22}. In the equation above, we have defined the auxiliary functions
\begin{align}
    &\text{I}_1(E) = \text{Ei}\left(2 \pi  \left(\sqrt{E^{-}-E_0}-\sqrt{E-E_0}\right)\right)\\ 
    &\text{I}_2(E) = \text{Ei}\left(2 \pi  \left(\sqrt{E-E_0}-\sqrt{E^{-}-E_0}\right)\right)\\
    &\text{I}_3(E) = \text{Ei}\left(-2 \pi  \left(\sqrt{E^{-}-E_0}+\sqrt{E-E_0}\right)\right)\\
    &\text{I}_4(E) = \text{Ei}\left(2 \pi  \left(\sqrt{E^{-}-E_0}+\sqrt{E-E_0}\right)\right)\\
    &\text{I}_5(E) = \text{Ei}\left(2 \pi  \left(\sqrt{E-E_0}-\sqrt{E^{+}-E_0}\right)\right)\\
    &\text{I}_6(E) = \text{Ei}\left(2 \pi  \left(\sqrt{E^{+}-E_0}-\sqrt{E-E_0}\right)\right)\\
    &\text{I}_7(E) = \text{Ei}\left(-2 \pi  \left(\sqrt{E-E_0}+\sqrt{E^{+}-E_0}\right)\right)\\
    &\text{I}_8(E) = \text{Ei}\left(2 \pi  \left(\sqrt{E-E_0}+\sqrt{E^{+}-E_0}\right)\right).
\end{align}

We can rewrite the asymptotic relation \eqref{eq:vol9} as
\begin{equation}
     V_{g,1}(b) \sim C^{\text{ZZ}}_g(b) + C^{\text{FZZT}}_g(b) + \cdots
     \label{eq:vol26}
\end{equation}
where the ZZ contribution reads
\begin{equation}
    C^{\text{ZZ}}_g(b) = \mathcal{C}^{\text{ZZ}}\Gamma\left(2g-\frac{3}{2}\right)\left(\frac{1}{V_{\text{eff}}(E^\star)}\right)^{2g-\frac{3}{2}}\frac{1}{b}\sinh(\frac{bn}{2}).
\end{equation}
In the equation above, we defined the constant (compare with equations \eqref{eq:vol32} and \eqref{eq:vol33})
\begin{equation}
    \mathcal{C}^{\text{ZZ}} =\frac{\rmi^{1-n} }{8 \pi ^2 n}\sqrt{\frac{\left(4 E^{-}-4 E_0+n^2\right) \left(4 E^+-4 E_0+n^2\right)}{A_{j_+,j_-}}}.
\end{equation}

The FZZT contribution reads
\begin{align}
     C^{\text{FZZT}}_g(b) &=  -2\frac{S_{\text{FZZT}}}{(2\pi \rmi)^2}\frac{\Gamma\left(2g-1\right)}{b} \oint_{\mathcal{C}_0} \rmd z \frac{\exp\left(bz\right)}{z( V_{\text{eff}}(z)-V_{\text{eff}}(E_0))^{2g-1}} \nonumber \\ & = -\frac{S_{\text{FZZT}}}{\pi\rmi}\frac{ \Gamma (2 g-1)}{ \Gamma (6 g-2)}\left(\frac{3}{4}\frac{(E_0-E^-)(E_0-E^+)}{A_{j_+,j_-}\pi}\right)^{2 g-1}b^{6g-4}+ \cdots
\end{align}
for large $b$ where the dots denote terms proportional to smaller powers of $b$.

\section{Concluding remarks}

\label{sec:conclusion}

In this work, we provided a new prescription for constructing the non-perturbative completion of the one-point correlation function from the Gel’fand-Dikii equation \eqref{eq:gde6}, extending the  method of ref~\cite{Johnson:2024bue} beyond perturbation theory. This was achieved by formulating a transseries ansatz for the Gel’fand-Dikii resolvent \eqref{eq:gde2}, whose coefficients can be solved for recursively by inserting the generic transseries into the Gel’fand-Dikii equation and using the resulting recursion relation. The resulting transseries is then related to that of the one-point correlation function by a simple integration. We have chosen to illustrate this procedure in the relatively simple, yet non-trivial, example of $(2,3)$ minimal string theory, whose associated transseries are  simple, but still rich enough to include all essential features and to clearly demonstrate how this method generalizes to other models of gravity.

The transseries of the one-point correlation function features three distinct types of non-perturbative effects, each associated with different transseries sectors:

\begin{itemize}
    \item Non-perturbative ZZ-effects were already studied in \cite{eggls23} through the lens of non-perturbative topological recursion, a generalisation of standard topological recursion capable of capturing non-perturbative data in the form of transseries coefficients. In this work, we showed that the ZZ non-perturbative transseries sectors obtained via our prescription agree with those computed from non-perturbative topological recursion, providing a set of rather striking and non-trivial consistency checks. It is worth mentioning that our prescription is considerably simpler than non-perturbative topological recursion, as it does not rely on saddle-point expansions of matrix integrals, computations that can be highly intricate and technically demanding \cite{mss22}.

    \item The non-perturbative FZZT-effect was previously studied in \cite{eggls23,Saad:2019lba,os19}. In those works, the prescription for computing the corresponding transseries sector required expanding a correlation function with two determinant insertions by means of the topological recursion. While correct and general, this prescription is not well suited to complicated models, as topological recursion becomes inefficient in such cases. By contrast, our approach allows for a direct and recursive computation of the FZZT transseries sector coefficients directly from the Gel’fand–Dikii equation, and can be implemented efficiently for most models. 

    \item  Non-perturbative ZZ–FZZT effects have not been addressed in the literature to the best of the authors’ knowledge. Nonetheless, our prescription allows for a straightforward computation of these sectors.
\end{itemize}

As a particularly powerful application of our prescription, we derived fully general results for the leading large-order growth of perturbative coefficients associated with the one-point correlation function, as well as with the corresponding one-boundary Weil-Petersson volume for JT gravity and $\mathcal{N}=1,2,4$ JT supergravity. Our results reproduce those obtained in \cite{eggls23}, further provide  proof of a conjecture made in \cite{Stanford:2019vob}, and provide several new formulae. 

A promising direction for future research is to develop analogous prescriptions for extracting non-perturbative data for higher-point correlation functions in the form of transseries coefficients. This  would naturally require  differential equations  whose solutions can be  directly related to the desired correlation functions. Such an equation was recently proposed and studied in ref.~\cite{j25a}.

\section*{Acknowledgements}

We thank Francesco Cominelli, Laura Cuéllar, Jasper Kager, Krishan Saraswat, Ricardo Schiappa, and Mykhaylo Usatyuk for useful discussions. JR is supported by the FCT-Portugal scholarship UI/BD/151499/2021 and by the CAMGSD scholarship BL197/2025-IST-ID. CVJ is supported in part  by US Department of Energy grant  \#DE-SC 0011687, and by the University of California. This
work has been supported by Fundação para a Ciência e Tecnologia through the project 2024.04456.CERN. This paper is partly a result of the ERC-SyG project, Recursive and Exact New Quantum Theory (ReNewQuantum) funded by the European Research Council (ERC) under the European Union’s Horizon 2020 research and innovation programme, grant agreement 810573.  CVJ also thanks Amelia for her support and patience.

\appendix

\section{Transseries data for $(2,3)$ minimal string theory}

\label{appendix:Transseriesdata}

In this appendix, we present the transseries structure along with the first few transseries coefficients for the specific heat \eqref{eq:zz15} and free energy \eqref{eq:nonperef7} associated with $(2,3)$ minimal string theory. Additionally, we display the first few coefficients for the Gel’fand-Dikii resolvent transseries \eqref{eq:nonperef2} and the correlation function transseries \eqref{eq:nonperef1}.

\subsection{Transseries data for the specific heat}

The specific heat $u(x)$ is a solution to the string equation \eqref{eq:stringeq} whose transseries structure reads (see equation \eqref{eq:zz15})
\begin{equation}
    u(x;\sigma_{\text{ZZ}_\pm})  = u^{\text{pert}}(x)+ \sum_{\substack{n_\pm \in \mathbb{N}_0}}' \sigma_{\text{ZZ}_+}^{n_+}\sigma_{\text{ZZ}_-}^{n_-}\exp\left(-\frac{n_+-n_-}{\hbar} \mathcal{A}_{\text{ZZ}}(x)\right)u_{(n_+,n_-)}^{\text{ZZ}}(x)
    \label{eq:transseriesdata7}
\end{equation}
where the perturbative sector can be written as (see equation \eqref{eq:zz17})
\begin{equation}
   u^{\text{pert}}(x) = \sum_{n \in \mathbb{N}_0}u_g(x)\hbar^{2g}.
\end{equation}

To compute the coefficients of the transseries sectors above as well as the instanton action $\mathcal{A}_{\text{ZZ}}(x)$, we simply substitute the transseries ansatz above into the string equation $\eqref{eq:stringeq}$ and expand it in powers of the transseries parameters $\sigma_{\text{ZZ}_\pm}$ and the coupling $\hbar$. This procedure yields an infinite family of differential\footnote{The freedom encoded in the integration constants of these differential equations is analogous to the freedom to transform the transseries parameters $\sigma_{\text{ZZ}_\pm}$ in ways that preserve the transseries structure \eqref{eq:transseriesdata7}. However, the checks we performed in Subsection \ref{subsec:zzeffects} are invariant under such transformations. For further details on this topic, see \cite{asv12}.} and algebraic equations that can be solved for the transseries coefficients. In the following, we report the results of this computation by presenting the first few transseries coefficients. For a detailed overview of the computational procedure, we refer the reader to \cite{gs21,bssv22}, and more recently \cite{krsst25a}.

The classical string equation solution can be written as
\begin{equation}
   u_0(x) =\frac{2^{\frac{5}{4}}}{3^{\frac{1}{2}}}\sqrt{-x}
\end{equation}
while the first few quantum corrections read
\begin{align}
    &u_1(x) = -\frac{4}{27}\frac{1}{u_0(x)^4}\\ 
    &u_2(x) = -\frac{392}{729}\frac{1}{ u_0(x)^9}\\
    &u_3(x) = -\frac{156800}{19683}\frac{1}{u_0(x)^{14}}\\
    &u_4(x) = -\frac{141196832}{531441}\frac{1}{u_0(x)^{19}}\\
    &u_5(x) = -\frac{75325465600}{4782969}\frac{1}{u_0(x)^{24}}.
\end{align}
The instanton action can be written as (see equation \eqref{eq:zz18})
\begin{equation}
    \mathcal{A}_{\text{ZZ}}(x) = \frac{3\sqrt{3}}{5} u_0(x)^{\frac{5}{2}}
\end{equation}
and the ZZ transseries sectors associated with positive multiples of the instanton action $-\mathcal{A}_{\text{ZZ}}(x)$ follow the structure
\begin{equation}
    u_{(n,0)}^{\text{ZZ}}(x) = \hbar^{\frac{n}{2}}\sum_{g \in \mathbb{N}_0} u_{g,(n,0)}^{\text{ZZ}}(x)\hbar^g
\end{equation}
where the first few coefficients read

\begin{align}
u_{0,(1,0)}^{\text{ZZ}}(x) &= -\frac{1}{12}\frac{1}{u_0(x)^{\frac{5}{4}}}   \qquad & u_{0,(2,0)}^{\text{ZZ}}(x)  &= -\frac{1}{288} \frac{1}{u_0(x)^{\frac{5}{2}}}  \nonumber\\
u_{1,(1,0)}^{\text{ZZ}}(x) &= \frac{37}{288 \sqrt{3}} \frac{1}{u_0(x)^{\frac{15}{4}}}       & u_{1,(2,0)}^{\text{ZZ}}(x)  &= \frac{109}{10368 \sqrt{3}}\frac{1}{u_0(x)^5}        \nonumber\\
u_{2,(1,0)}^{\text{ZZ}}(x) &=-\frac{6433}{41472} \frac{1}{u_0(x)^{\frac{25}{4}}}     & u_{2,(2,0)}^{\text{ZZ}}(x)  &= -\frac{11179}{746496}\frac{1}{u_0(x)^{\frac{15}{2}}}       \nonumber\\
u_{3,(1,0)}^{\text{ZZ}}(x)  &=\frac{12741169}{14929920 \sqrt{3}}\frac{1}{u_0(x)^{\frac{35}{4}}}       & u_{3,(2,0)}^{\text{ZZ}}(x)  &= \frac{11258183}{134369280 \sqrt{3}}\frac{1}{u_0(x)^{10}}     \nonumber\\
u_{4,(1,0)}^{\text{ZZ}}(x)  &=-\frac{8854092037}{4299816960}\frac{1}{u_0(x)^{\frac{45}{4}}}       & u_{4,(2,0)}^{\text{ZZ}}(x)  &= -\frac{11222293013}{58047528960 }\frac{1}{u_0(x)^{\frac{25}{2}}}     \nonumber \\
u_{5,(1,0)}^{\text{ZZ}}(x)  &=\frac{1908813972149}{103195607040 \sqrt{3}}\frac{1}{ u_0(x)^{\frac{55}{4}}}  \quad    & u_{5,(2,0)}^{\text{ZZ}}(x)  &= \frac{1122296317499}{696570347520 \sqrt{3}}\frac{1}{ u_0(x)^{15}}   \nonumber\\[2em]  u_{0,(3,0)}^{\text{ZZ}}(x) &= -\frac{1}{5184}\frac{1}{u_0(x)^{\frac{15}{4}}} \nonumber\\ u_{1,(3,0)}^{\text{ZZ}}(x) &=\frac{109}{124416 \sqrt{3}}\frac{1}{u_0(x)^{\frac{25}{4}}} \nonumber\\ u_{2,(3,0)}^{\text{ZZ}}(x) &=-\frac{26317}{17915904}\frac{1}{u_0(x)^{\frac{35}{4}}}\nonumber\\ u_{3,(3,0)}^{\text{ZZ}}(x) &=\frac{169466071}{19349176320 \sqrt{3}}\frac{1}{u_0(x)^{\frac{45}{4}}}\nonumber\\ u_{4,(3,0)}^{\text{ZZ}}(x) &=-\frac{12716405759}{619173642240}\frac{1}{u_0(x)^{\frac{55}{4}}}\nonumber\\ u_{5,(3,0)}^{\text{ZZ}}(x) &=\frac{7559031194533}{44580502241280 \sqrt{3}}\frac{1}{u_0(x)^{\frac{65}{4}}}\nonumber.
\end{align}
Moreover, the ZZ transseries sectors associated with positive multiples of the instanton action $\mathcal{A}_{\text{ZZ}}(x)$ follow the structure
\begin{equation}
     u_{(0,n)}^{\text{ZZ}}(x) = \hbar^{\frac{n}{2}}\sum_{g \in \mathbb{N}_0} u_{g,(0,n)}^{\text{ZZ}}(x)\hbar^g
\end{equation}
where
\begin{equation}
    u_{g,(0,n)}^{\text{ZZ}}(x) = (-1)^gu_{g,(n,0)}^{\text{ZZ}}(x).
    \label{eq:transseriesdata5}
\end{equation}

The first bulk transseries sector can be written as
\begin{equation}
     u_{(1,1)}^{\text{ZZ}}(x) = \hbar\sum_{g \in \mathbb{N}_0}u_{g,(1,1)}^{\text{ZZ}}(x)\hbar^{2g} 
\end{equation}
where the first few transseries coefficients read
\begin{align}
   & u_{0,(1,1)}^{\text{ZZ}}(x) =-\frac{1}{u_0(x)^{\frac{3}{2}}} \\
    & u_{1,(1,1)}^{\text{ZZ}}(x) = -\frac{25}{24}\frac{1}{u_0(x)^{\frac{13}{2}}}\\
     & u_{2,(1,1)}^{\text{ZZ}}(x) = -\frac{300713}{31104}\frac{1}{u_0(x)^{\frac{23}{2}}}\\
      & u_{3,(1,1)}^{\text{ZZ}}(x) =-\frac{4807377125}{20155392}\frac{1}{u_0(x)^{\frac{33}{2}}} \\
       & u_{4,(1,1)}^{\text{ZZ}}(x) = -\frac{7285479010537}{644972544}\frac{1}{u_0(x)^{\frac{43}{2}}}\\
       & u_{5,(1,1)}^{\text{ZZ}}(x) = -\frac{365534447300026375}{417942208512}\frac{1}{u_0(x)^{\frac{53}{2}}}.
\end{align}

\subsection{Transseries data for the free energy}

The free energy is related to the specific heat via the differential equation\footnote{Naturally, this differential equation determines the free energy only up to two integration constants. This ambiguity manifests itself as an overall additive constant in the free energy, which can be safely ignored.}
\begin{equation}
    \frac{\partial^2 }{\partial \mu^2}\bigg\vert_{\mu = x}F(\hbar) = -\frac{2}{\hbar^2}  u(x)
    \label{eq:transseriesdata1}
\end{equation}
which we can use to infer the transseries structure (see equation \eqref{eq:nonperef7}) 
\begin{equation}
    F(\hbar;\sigma_{\text{ZZ}_\pm}) = F^{\text{pert}}(x)+\sum_{n_\pm \in \mathbb{N}_0}'\sigma_{\text{ZZ}_+}^{n_+}\sigma_{\text{ZZ}_-}^{n_-}\exp\left(-\frac{n_+-n_-}{\hbar} A_{\text{ZZ}}\right)F_{(n_+,n_-)}^{\text{ZZ}}(\hbar)
    \label{eq:transseriesdata4}
\end{equation}
where the perturbative sector reads (see equation \eqref{eq:nonperef8}) 
\begin{equation}
    F^{\text{pert}}(\hbar) = \sum_{n \in \mathbb{N}_0} F_g \hbar^{2g-2}. 
\end{equation}

The first few perturbative coefficients are
\begin{align}
    &F_0 = -\frac{3}{80}\\
    &F_1 =0\\
    &F_2 =\frac{7}{810} \label{eq:transseriesdata2}\\
    &F_3 =\frac{245}{6561}\label{eq:transseriesdata3}\\
    &F_4 =\frac{519106}{885735}\\
    &F_5 =\frac{10699640}{531441}.
\end{align}
The reader can readily verify that substituting the equations \eqref{eq:check-recur-2} and \eqref{eq:check-recur-3} into equation \eqref{eq:nonperef5} yields precisely the coefficients \eqref{eq:transseriesdata2} and \eqref{eq:transseriesdata3}. The instanton action reads
\begin{equation}
    A_{\text{ZZ}} = \mathcal{A}_{\text{ZZ}}(\mu) = \frac{3\sqrt{3}}{5}.
\end{equation}
Notice that using equations \eqref{eq:omega_defintion} and \eqref{eq:nonperef10} immediately yields the instanton action above. The ZZ transseries sectors associated with positive multiples of the instanton action $-A_{\text{ZZ}}$ follow the structure
\begin{equation}
   F_{(n,0)}^{\text{ZZ}}(\hbar) = \hbar^{\frac{n}{2}}\sum_{g \in \mathbb{N}_0} F_{g,(n,0)}^{\text{ZZ}}\hbar^g
\end{equation}
where the first few coefficients read

\begin{align}
F_{0,(1,0)}^{\text{ZZ}} &= -\frac{1}{12}   \qquad & F_{0,(2,0)}^{\text{ZZ}} &= -\frac{1}{288} \qquad & F_{0,(3,0)}^{\text{ZZ}} &=-\frac{1}{5184} \nonumber\\
F_{1,(1,0)}^{\text{ZZ}} &= \frac{37}{288 \sqrt{3}}        & F_{1,(2,0)}^{\text{ZZ}} &= \frac{109}{10368 \sqrt{3}}        & F_{1,(3,0)}^{\text{ZZ}} &= \frac{109}{124416 \sqrt{3}} \nonumber\\
F_{2,(1,0)}^{\text{ZZ}} &=-\frac{6433}{41472}       & F_{2,(2,0)}^{\text{ZZ}} &= -\frac{11179}{746496}        & F_{2,(3,0)}^{\text{ZZ}} &= -\frac{26317}{17915904} \nonumber\\
F_{3,(1,0)}^{\text{ZZ}} &=\frac{12741169}{14929920 \sqrt{3}}       & F_{3,(2,0)}^{\text{ZZ}} &= \frac{11258183}{134369280 \sqrt{3}}        & F_{3,(3,0)}^{\text{ZZ}} &= \frac{169466071}{19349176320 \sqrt{3}} \nonumber\\
F_{4,(1,0)}^{\text{ZZ}} &=-\frac{8854092037}{4299816960}       & F_{4,(2,0)}^{\text{ZZ}} &= -\frac{11222293013}{58047528960}        & F_{4,(3,0)}^{\text{ZZ}} &= -\frac{12716405759}{619173642240} \nonumber\\
F_{5,(1,0)}^{\text{ZZ}} &=\frac{1908813972149}{103195607040 \sqrt{3}}   \quad    & F_{5,(2,0)}^{\text{ZZ}} &= \frac{1122296317499}{696570347520 \sqrt{3}}   \quad     & F_{5,(3,0)}^{\text{ZZ}} &= \frac{7559031194533}{44580502241280 \sqrt{3}}\nonumber.
\end{align}

Moreover, the ZZ transseries sectors associated with positive multiples of the instanton action $A_{\text{ZZ}}$ follow the structure
\begin{equation}
     F_{(0,n)}^{\text{ZZ}}(\hbar) = \hbar^{\frac{n}{2}}\sum_{g \in \mathbb{N}_0} F_{g,(0,n)}^{\text{ZZ}}\hbar^g
\end{equation}
where (compare with equation \eqref{eq:transseriesdata5})
\begin{equation}
    F_{g,(0,n)}^{\text{ZZ}} = (-1)^gF_{g,(n,0)}^{\text{ZZ}}.
    \label{eq:transseriesdata6}
\end{equation}

The first bulk transseries sector can be written as
\begin{equation}
     F_{(1,1)}^{\text{ZZ}}(\hbar) = \frac{1}{\hbar}\sum_{g \in \mathbb{N}_0}F_{g,(1,1)}^{\text{ZZ}}\hbar^{2g} 
\end{equation}
where the first few coefficients read
\begin{align}
   & F_{0,(1,1)}^{\text{ZZ}} =\frac{9}{20} \\
    & F_{1,(1,1)}^{\text{ZZ}} = \frac{5}{96}\\
     & u_{2,(1,1)}^{\text{ZZ}} = \frac{15827}{207360}\\
      & F_{3,(1,1)}^{\text{ZZ}} =\frac{6630865}{8957952} \\
       & F_{4,(1,1)}^{\text{ZZ}} =\frac{80060208907}{4299816960}\\
       & F_{5,(1,1)}^{\text{ZZ}} = \frac{1491977335918475}{1671768834048}.
\end{align}

\subsection{Transseries data for the partition function}

The partition function is related to the free energy via the equation
\begin{equation}
    Z(\hbar) = \exp\left(F(\hbar)\right).
\end{equation}
The associated transseries structure follows easily from \eqref{eq:transseriesdata4} by Taylor expanding the exponential is powers of the transseries parameters $\sigma_{\text{ZZ}_\pm}$ and the coupling $\hbar$, yielding (see equation \eqref{eq:nonperef9})

\begin{equation}
Z(\hbar;\sigma_{\text{ZZ}_\pm}) = Z^{\text{pert}}(\hbar)+\sum_{n_\pm \in \mathbb{N}_0}'\sigma_{\text{ZZ}_+}^{n_+}\sigma_{\text{ZZ}_-}^{n_-}\exp\left(-\frac{n_+-n_-}{\hbar} A_{\text{ZZ}}\right)Z_{(n_+,n_-)}^{\text{ZZ}}(\hbar)
\end{equation}
where the perturbative sector reads
\begin{equation}
    Z^{\text{pert}}(\hbar) = \exp\left(F^{\text{pert}}(\hbar)\right).
\end{equation}
The ZZ transseries sectors associated with positive multiples of the instanton action $-A_{\text{ZZ}}$ follow the structure
\begin{equation}
  \frac{ Z_{(n,0)}^{\text{ZZ}}(\hbar)}{Z^{\text{pert}}(\hbar)} = \hbar^{\frac{n^2}{2}}\sum_{g \in \mathbb{N}_0} Z_{g,(n,0)}^{\text{ZZ}}\hbar^g
\end{equation}
where the first few coefficients read
\[
\begin{alignedat}{3}
  Z_{0,(1,0)}^{\text{ZZ}} &= -\frac{1}{12} \qquad &Z_{0,(2,0)}^{\text{ZZ}} &= -\frac{1}{5184 \sqrt{3}} \qquad & Z_{0,(3,0)}^{\text{ZZ}} &= \frac{1}{120932352 \sqrt{3}}\\
Z_{1,(1,0)}^{\text{ZZ}}  &= \frac{37}{288\sqrt{3}}        & Z_{1,(2,0)}^{\text{ZZ}} &= \frac{131}{186624}        & Z_{1,(3,0)}^{\text{ZZ}} &= -\frac{863}{8707129344} \\
Z_{2,(1,0)}^{\text{ZZ}} &= -\frac{6433}{41472}    \quad    & Z_{2,(2,0)}^{\text{ZZ}} &= -\frac{97543}{13436928 \sqrt{3}}   \quad    & Z_{2,(3,0)}^{\text{ZZ}} &= \frac{345763}{139314069504 \sqrt{3}}.
\end{alignedat}
\]
Moreover, the ZZ transseries sectors associated with positive multiples of the instanton action $A_{\text{ZZ}}$ follow the structure
\begin{equation}
     \frac{Z_{(0,n)}^{\text{ZZ}}(\hbar)}{Z^{\text{pert}}(\hbar)} = \hbar^{\frac{n^2}{2}}\sum_{g \in \mathbb{N}_0} Z_{g,(0,n)}^{\text{ZZ}}\hbar^g
\end{equation}
where (compare with equation \eqref{eq:transseriesdata6})
\begin{equation}
    Z_{g,(0,n)}^{\text{ZZ}} = (-1)^{g+1}Z_{g,(n,0)}^{\text{ZZ}}.
\end{equation}

The first bulk transseries sector reads
\begin{equation}
     \frac{Z_{(1,1)}^{\text{ZZ}}(\hbar)}{Z^{\text{pert}}(\hbar)} = \frac{1}{\hbar}\sum_{g \in \mathbb{N}_0}Z_{g,(1,1)}^{\text{ZZ}}\hbar^{2g} 
\end{equation}
where the first few transseries coefficients are
\begin{align}
   & Z_{0,(1,1)}^{\text{ZZ}} =\frac{9}{20} \\
    & Z_{1,(1,1)}^{\text{ZZ}} = \frac{17}{288}\\
     & Z_{2,(1,1)}^{\text{ZZ}} = \frac{20047}{207360}.
\end{align}

\subsection{Transseries data for the Gel’fand-Dikii resolvent}

The Gel’fand-Dikii resolvent transseries structure put forward in the paper reads (see equation \eqref{eq:nonperef2})
\begin{align}
    &\widehat{R}(E,x;\sigma_{\text{ZZ}_\pm},\sigma_{\text{FZZT}_\pm}) = \widehat{\mathcal{R}}^{\text{pert}}(E,x) + \sigma_{\text{FZZT}_+}\exp\left(+\frac{\mathcal{A}_{\text{FZZT}}(E,x)}{\hbar}\right) \widehat{\mathcal{R}}^{\text{FZZT}_+}(E,x) \nonumber\\ & +\sigma_{\text{FZZT}_-}\exp\left(-\frac{\mathcal{A}_{\text{FZZT}}(E,x)}{\hbar}\right)\widehat{\mathcal{R}}^{\text{FZZT}_-}(E,x)  +\sum_{\substack{n_\pm \in \mathbb{N}_0 }}'\sigma_{\text{ZZ}_+}^{n_+}\sigma_{\text{ZZ}_-}^{n_-}\exp\left(-(n_1-n_2) \frac{\mathcal{A}_{\text{ZZ}}(x)}{\hbar}\right) \widehat{\mathcal{R}}_{(n_1,n_2)}^{\text{ZZ}}(E,x)   \nonumber\\ & +\sigma_{\text{FZZT}_+}\exp\left(+\frac{\mathcal{A}_{\text{FZZT}}(E,x)}{\hbar}\right)\sum_{\substack{n_\pm \in \mathbb{N}_0 }}'\sigma_{\text{ZZ}_+}^{n_+}\sigma_{\text{ZZ}_-}^{n_-}\exp\left(-(n_+-n_-) \frac{\mathcal{A}_{\text{ZZ}}(x)}{\hbar}\right) \widehat{\mathcal{R}}_{(n_+,n_-)}^{\text{ZZ-FZZT}_+}(E,x)
   \nonumber \\ & + \sigma_{\text{FZZT}_-}\exp\left(-\frac{\mathcal{A}_{\text{FZZT}}(E,x)}{\hbar}\right)\sum_{\substack{n_\pm \in \mathbb{N}_0}}'\sigma_{\text{ZZ}_+}^{n_+}\sigma_{\text{ZZ}_-}^{n_-}\exp\left(-(n_+-n_-) \frac{\mathcal{A}_{\text{ZZ}}(x)}{\hbar}\right) \widehat{\mathcal{R}}_{(n_+,n_-)}^{\text{ZZ-FZZT}_-}(E,x)
   \label{eq:transseriesdata16}
\end{align}
where the perturbative sector can be written as (see equation \eqref{eq:nonperef12})
\begin{equation}
   \widehat{\mathcal{R}}^{\text{pert}}(E,x) = \sum_{n \in \mathbb{N}_0}\widehat{R}_g(E,x)\hbar^{2g}.
\end{equation}

To compute the coefficients of the transseries sectors above, we simply substitute this transseries ansatz along with the transseries \eqref{eq:transseriesdata7} into the Gel’fand-Dikii equation $\eqref{eq:zz12}$ and expand it in powers of the transseries parameters $\sigma_{\text{ZZ}_\pm},\sigma_{\text{FZZT}_\pm}$ and the coupling $\hbar$. This procedure yields an infinite family of differential and algebraic equations that can be solved for the transseries coefficients. In the following, we report the results of this computation by presenting the first few transseries coefficients, some of which are already displayed in the main body of the paper (particularly in Subsection $\ref{subsec:zzeffects}$).

The perturbative coefficients follow the structure
\begin{equation}
    \widehat{R}_g(E,x)= \frac{p_g(E,u_0(x))}{u_0(x)^{5g-1+\delta_{g,0}} (u_0(x)-E )^{\frac{1}{2}+3g}}
    \label{eq:transseriesdata11}
\end{equation}
for some polynomials $p_g(w,y)$ among which the first couple read
\begin{align}
    p_0(w,y) =& -\frac{1}{2}\\  
    p_1(w,y) =& -\frac{35 y^2}{216}+\frac{7 y w }{54}-\frac{w ^2}{27}\\
    p_2(w,y) =& -\frac{72947 y^5}{46656}+\frac{19943 y^4 w }{5832}-\frac{5509 y^3 w ^2}{1458}+\frac{1778 y^2 w ^3}{729}-\frac{637 y w ^4}{729}+\frac{98 w ^5}{729}\\
    p_3(w,y) =& -\frac{25936015 y^8}{559872}+\frac{211349089 y^7 w }{1259712}-\frac{201553009 y^6 w ^2}{629856}+\frac{10509443 y^5 w ^3}{26244}-\frac{27274415 y^4 w ^4}{78732}\nonumber \\ &+\frac{8122835 y^3 w ^5}{39366}-\frac{531650 y^2 w ^6}{6561}+\frac{372400 y w ^7}{19683}-\frac{39200 w ^8}{19683}\\
    p_4(w,y) =&  -\frac{6087191401795 y^{11}}{2176782336}+\frac{1920751216391 y^{10} w }{136048896}-\frac{322172859295 y^9 w ^2}{8503056}+\frac{73526697335 y^8 w ^3}{1062882}\nonumber \\ &-\frac{1579013834125 y^7 w ^4}{17006112}+\frac{799321975963 y^6 w ^5}{8503056}-\frac{38236515695 y^5 w ^6}{531441}+\frac{21827467333 y^4 w ^7}{531441}\nonumber \\ &-\frac{9017346275 y^3 w ^8}{531441}+\frac{2548810460 y^2 w ^9}{531441}-\frac{441240100 y w ^{10}}{531441}+\frac{35299208 w ^{11}}{531441}.
\end{align}

The ZZ transseries sectors follow the structure (see equations \eqref{eq:zz19} and \eqref{eq:zz20})
\begin{align}
    &\widehat{\mathcal{R}}_{(n,0)}^{\text{ZZ}}(E,x) = \hbar^{\frac{n}{2}}\sum_{g \in \mathbb{N}_0} \widehat{\mathcal{R}}_{g,(n,0)}^{\text{ZZ}}(E,x)\hbar^{g}
    \label{eq:transseriesdata17}\\ 
    &\widehat{\mathcal{R}}_{(0,n)}^{\text{ZZ}}(E,x) = \hbar^{\frac{n}{2}}\sum_{g \in \mathbb{N}_0} \widehat{\mathcal{R}}_{g,(0,n)}^{\text{ZZ}}(E,x)\hbar^{g}\label{eq:transseriesdata18}.
\end{align}
The coefficients in \eqref{eq:transseriesdata17} can be written as
\begin{equation}
    \widehat{\mathcal{R}}_{g,(n,0)}^{\text{ZZ}}(E,x)= \frac{p_{g,(n,0)}^{\text{ZZ}}(E,u_0(x))}{u_0(x)^{\frac{5}{4}n+\frac{5}{2}g-1}(u_0(x)-E)^{ \frac{1}{2}+g+\left\lfloor \frac{g}{2} \right\rfloor}(u_0(x)+2 E )^{1+2g}} 
\end{equation}
for some polynomials $p_{g,(n,0)}^{\text{ZZ}}(w,y)$ among which the first couple read
\begin{align}
     p_{0,(1,0)}^{\text{ZZ}}(w,y) = &-\frac{1}{2} \\ 
     p_{1,(1,0)}^{\text{ZZ}}(w,y) = & \frac{5 w^4}{12 \sqrt{3}}-\frac{5 w^3 y}{12 \sqrt{3}}+\frac{11 w^2 y^2}{16 \sqrt{3}}-\frac{79 w y^3}{24 \sqrt{3}}+\frac{125 y^4}{48 \sqrt{3}}\\ 
     p_{2,(1,0)}^{\text{ZZ}}(w,y) = & \frac{25 w^7}{48}-\frac{25 w^6 y}{48}+\frac{47 w^5 y^2}{288}+\frac{5 w^4 y^3}{48}-\frac{1573 w^3 y^4}{2304}-\frac{38065 w^2 y^5}{2304}+\frac{17539 w y^6}{768}\nonumber \\ &-\frac{26555 y^7}{2304}\\ 
    p_{0,(2,0)}^{\text{ZZ}}(w,y) = & -\frac{1}{12} \\  
    p_{1,(2,0)}^{\text{ZZ}}(w,y) = & \frac{55 w^4}{108 \sqrt{3}}-\frac{55 w^3 y}{108 \sqrt{3}}-\frac{43 w^2 y^2}{144 \sqrt{3}}-\frac{31 w y^3}{108 \sqrt{3}}+\frac{253 y^4}{432 \sqrt{3}} \\
    p_{2,(2,0)}^{\text{ZZ}}(w,y) = & \frac{1325 w^7}{648}-\frac{1325 w^6 y}{648}-\frac{3659 w^5 y^2}{1296}+\frac{389 w^4 y^3}{324}+\frac{8951 w^3 y^4}{3456}-\frac{21371 w^2 y^5}{10368}+\frac{9061 w y^6}{3456}\nonumber \\ &-\frac{25561 y^7}{10368}
    \\ 
    p_{0,(3,0)}^{\text{ZZ}}(w,y) = & -\frac{1}{96} \\  
    p_{1,(3,0)}^{\text{ZZ}}(w,y) = & \frac{215 w^4}{1728 \sqrt{3}}-\frac{215 w^3 y}{1728 \sqrt{3}}-\frac{199 w^2 y^2}{2304 \sqrt{3}}-\frac{13 w y^3}{3456 \sqrt{3}}+\frac{623 y^4}{6912 \sqrt{3}} \\
    p_{2,(3,0)}^{\text{ZZ}}(w,y) = &\frac{40775 w^7}{62208}-\frac{40775 w^6 y}{62208}-\frac{13117 w^5 y^2}{13824}+\frac{26819 w^4 y^3}{62208}+\frac{856663 w^3 y^4}{995328}-\frac{24733 w^2 y^5}{110592}\nonumber \\ &+\frac{135197 w y^6}{995328}-\frac{370583 y^7}{995328}.
\end{align}
The coefficients in \eqref{eq:transseriesdata18} follow the structure (see equation \eqref{eq:negativepositive})
\begin{equation}
    \widehat{\mathcal{R}}_{g,(0,n)}(E,x) = (-1)^g\widehat{\mathcal{R}}_{g,(n,0)}(E,x).
    \label{eq:transseriesdata8}
\end{equation}

The first bulk ZZ transseries sector reads (see equation \eqref{eq:zz14})
\begin{equation}
    \widehat{\mathcal{R}}_{(1,1)}^{\text{ZZ}}(E,x) = \hbar\sum_{g \in \mathbb{N}_0} \widehat{\mathcal{R}}_{g,(1,1)}^{\text{ZZ}}(E,x)\hbar^{2g}
\end{equation}
where the coefficients follow the structure
\begin{equation}
    \widehat{\mathcal{R}}_{g,(1,1)}^{\text{ZZ}}(E,x)= \frac{p_{g,(1,1)}^{\text{ZZ}}(E,u_0(x))}{u_0(x)^{\frac{3}{2}+5g}(u_0(x)-E)^{\frac{3}{2}+3g}(u_0(x)+2 E )^{2+4g}} 
\end{equation}
for some polynomials $p_{g,(1,1)}^{\text{ZZ}}(w,y)$ among which the first couple read
\begin{align}
    p_{0,(1,1)}^{\text{ZZ}}(w,y) =& -w^2+\frac{w y}{2}-y^2\\  
    p_{1,(1,1)}^{\text{ZZ}}(w,y) =& \frac{50 w^9}{3}-25 w^8 y-\frac{250 w^7 y^2}{9}+\frac{821 w^6 y^3}{18}+\frac{433 w^5 y^4}{24}-\frac{8077 w^4 y^5}{144}+\frac{1927 w^3 y^6}{144}-\frac{8405 w^2 y^7}{48}\nonumber \\ & +\frac{15209 w y^8}{144}-\frac{1205 y^9}{36}\\ 
    p_{2,(1,1)}^{\text{ZZ}}(w,y) =&-\frac{601426 w^{16}}{243}+\frac{1503565 w^{15} y}{243}+\frac{316540 w^{14} y^2}{81}-\frac{4402496 w^{13} y^3}{243}-\frac{1509319 w^{12} y^4}{972}\nonumber \\ &+\frac{196049 w^{11} y^5}{8}-\frac{483637 w^{10} y^6}{648}-\frac{24441055 w^9 y^7}{1296}+\frac{4943831 w^8 y^8}{3456}+\frac{351351493 w^7 y^9}{62208}\nonumber \\ &-\frac{303523841 w^6 y^{10}}{31104}+\frac{239809577 w^5 y^{11}}{20736}-\frac{121389055 w^4 y^{12}}{972}+\frac{9287406295 w^3 y^{13}}{62208}\nonumber \\ &-\frac{431276447 w^2 y^{14}}{3456}+\frac{12850623 w y^{15}}{256}-\frac{29596039 y^{16}}{3456}.
\end{align}

The FZZT sectors follow the structure (see equation \eqref{eq:fzzt5})
\begin{equation}
   \widehat{ \mathcal{R}}^{\text{FZZT}_\pm}(E,x) = \sum_{g \in \mathbb{N}_0} \widehat{ \mathcal{R}}^{\text{FZZT}_\pm}_{g}(E,x)\hbar^{g}
\end{equation}
where the coefficients associated with positive instanton action can be written as
\begin{equation}
    \widehat{ \mathcal{R}}^{\text{FZZT}_+}_{g}(E,x) = \frac{p^{\text{FZZT}}_g(E,u_0(x))}{u_0(x)^{2g+\delta_{g,0}+\left\lfloor\frac{g-1}{2}\right\rfloor}(u_0(x)-E)^{\frac{1}{2}+\frac{3}{2}g}}
\end{equation}
for some polynomials $p^{\text{FZZT}}_g(w,y)$ among which the first couple read
\begin{align}
    p^{\text{FZZT}}_0(w,y) = & 1\\ 
     p^{\text{FZZT}}_1(w,y) = & \frac{w}{9 \sqrt{2}}-\frac{7 y}{18 \sqrt{2}}\\ 
      p^{\text{FZZT}}_2(w,y) = &\frac{25 w^2}{324}-\frac{91 w y}{324}+\frac{469 y^2}{1296} \\ 
      p^{\text{FZZT}}_3(w,y) = & -\frac{28}{243\sqrt{2}}  w^4+\frac{5761 w^3 y}{8748 \sqrt{2}}-\frac{46571 w^2 y^2}{29160 \sqrt{2}}+\frac{120659 w y^3}{58320 \sqrt{2}}-\frac{468797 y^4}{349920 \sqrt{2}}\\
      p^{\text{FZZT}}_4(w,y) = &-\frac{602 w^5}{2187}+\frac{1137745 w^4 y}{629856}-\frac{8018423 w^3 y^2}{1574640}+\frac{16738757 w^2 y^3}{2099520}-\frac{45968531 w y^4}{6298560}\nonumber \\ &+\frac{170038421 y^5}{50388480}.
\end{align}

The coefficients associated with the instanton action $-\mathcal{A}_{\text{FZZT}}(E,x)$ follow the rather simple structure (see equation \eqref{eq:fzzt6})
\begin{equation}
   \widehat{\mathcal{R}}^{\text{FZZT}_-}_{g}(E,x) = (-1)^{g}\widehat{\mathcal{R}}^{\text{FZZT}_+}_{g}(E,x).
\end{equation}

Finally, the mixed ZZ-FZZT transseries sectors can be written as (see equations \eqref{eq:zz-fzzt1} and \eqref{eq:zz-fzzt2})
\begin{align}
    & \widehat{ \mathcal{R}}^{\text{ZZ-FZZT}_\pm}_{(n,0)}(E,x) = \hbar^{\frac{n}{2}}\sum_{g \in \mathbb{N}_0}\widehat{ \mathcal{R}}^{\text{ZZ-FZZT}_\pm}_{g,(n,0)}(E,x)\hbar^{g} \label{eq:transseriesdata19}\\ & \widehat{ \mathcal{R}}^{\text{ZZ-FZZT}_\pm}_{(0,n)}(E,x) = \hbar^{\frac{n}{2}}\sum_{g \in \mathbb{N}_0}\widehat{ \mathcal{R}}^{\text{ZZ-FZZT}_\pm}_{g,(0,n)}(E,x)\hbar^{g}
    \label{eq:transseriesdata9}
\end{align}
where the coefficients in \eqref{eq:transseriesdata19} for $n=1$ follow the structure
\begin{align}
    \widehat{ \mathcal{R}}^{\text{ZZ-FZZT}_\pm}_{g,(1,0)}(E,x) = &\frac{p^{\text{ZZ-FZZT}_\pm}_{g}(E,u_0(x))+u_0(x)^{\frac{1}{2}(-1)^{g+1}}(u_0(x)-E)^{\frac{1}{2}}q^{\text{ZZ-FZZT}_\pm}_{g}(E,u_0(x))}{u_0(x)^{\frac{1}{4}+\frac{5}{2}g+\frac{1}{2} g \hspace{1pt}\text{mod} \hspace{1pt}2}(2E+u_0(x))^{1+2g}(u_0(x)-E)^{\frac{3}{2}g+\frac{1}{2}}} 
\end{align}
for some polynomials $p^{\text{ZZ-FZZT}_\pm}_{g}(w,y)$ and $q^{\text{ZZ-FZZT}_\pm}_{g}(w,y)$ among which the first couple read
\begin{align}
    p^{\text{ZZ-FZZT}_+}_{0}(w,y) =& 1 \\ 
     p^{\text{ZZ-FZZT}_+}_{1}(w,y) =& \frac{17 w^4}{9 \sqrt{2}}+\frac{w^3 y}{3 \sqrt{2}}-\frac{91 w^2 y^2}{36 \sqrt{2}}-\frac{35 w y^3}{6 \sqrt{2}}+\frac{131 y^4}{36 \sqrt{2}} \\ 
      p^{\text{ZZ-FZZT}_+}_{2}(w,y) =& -\frac{403 w^7}{648}+\frac{11 w^6 y}{648}-\frac{2839 w^5 y^2}{1296}-\frac{451 w^4 y^3}{648}+\frac{66973 w^3 y^4}{10368}+\frac{362753 w^2 y^5}{10368}\nonumber \\ & -\frac{491953 w y^6}{10368}+\frac{77089 y^7}{3456} \\
       q^{\text{ZZ-FZZT}_+}_{0}(w,y) =& -\sqrt{\frac{2}{3}}\\
        q^{\text{ZZ-FZZT}_+}_{1}(w,y) =& \frac{7 w^3}{18 \sqrt{3}}+\frac{10 w^2 y}{9 \sqrt{3}}+\frac{203 w y^2}{72 \sqrt{3}}-\frac{347 y^3}{72 \sqrt{3}}\\
         q^{\text{ZZ-FZZT}_+}_{2}(w,y) =& \frac{1993 w^7}{108 \sqrt{6}}-\frac{35 w^6 y}{324 \sqrt{6}}-\frac{20521 w^5 y^2}{648 \sqrt{6}}-\frac{4645 w^4 y^3}{324 \sqrt{6}}+\frac{29227 w^3 y^4}{5184 \sqrt{6}}-\frac{94585 w^2 y^5}{5184 \sqrt{6}}\nonumber \\ &+\frac{509149 w y^6}{5184 \sqrt{6}} -\frac{277727 y^7}{5184 \sqrt{6}}. 
\end{align}
Moreover, the polynomials $p^{\text{ZZ-FZZT}_-}_{g}(w,y)$ and $q^{\text{ZZ-FZZT}_-}_{g}(w,y)$ follow the rather simple structure 
\begin{align}
    p^{\text{ZZ-FZZT}_-}_{g}(w,y) &= (-1)^gp^{\text{ZZ-FZZT}_+}_{g}(w,y)  \label{eq:transseriesdata12}\\
    q^{\text{ZZ-FZZT}_-}_{g}(w,y) &=(-1)^{g+1}q^{\text{ZZ-FZZT}_+}_{g}(w,y).\label{eq:transseriesdata13}
\end{align}

The coefficients in \eqref{eq:transseriesdata9} can be written as (compare with equation \eqref{eq:transseriesdata8})
\begin{equation}
    \widehat{ \mathcal{R}}^{\text{ZZ-FZZT}_\pm}_{g,(0,n)}(E,x) = (-1)^g \widehat{ \mathcal{R}}^{\text{ZZ-FZZT}_\mp}_{g,(n,0)}(E,x).
    \label{eq:transseriesdata15}
\end{equation}

\subsection{Transseries data for the correlation function}

The transseries of the correlation function reads (see equation \eqref{eq:nonperef1})
\begin{align}
    &W_1(E;\sigma_{\text{ZZ}_\pm},\sigma_{\text{FZZT}_\pm}) = W^{\text{pert}}(E) + \sigma_{\text{FZZT}_+}\exp\left(+\frac{A_{\text{FZZT}}(E)}{\hbar}\right) W^{\text{FZZT}_+}(E) \nonumber\\ & +\sigma_{\text{FZZT}_-}\exp\left(-\frac{A_{\text{FZZT}}(E)}{\hbar}\right)W^{\text{FZZT}_-}(E)  +\sum_{\substack{n_\pm \in \mathbb{N}_0 }}'\sigma_{\text{ZZ}_+}^{n_+}\sigma_{\text{ZZ}_-}^{n_-}\exp\left(-(n_1-n_2) \frac{A_{\text{ZZ}}}{\hbar}\right) W_{(n_1,n_2)}^{\text{ZZ}}(E)   \nonumber\\ & +\sigma_{\text{FZZT}_+}\exp\left(+\frac{A_{\text{FZZT}}(E)}{\hbar}\right)\sum_{\substack{n_\pm \in \mathbb{N}_0 }}'\sigma_{\text{ZZ}_+}^{n_+}\sigma_{\text{ZZ}_-}^{n_-}\exp\left(-(n_+-n_-) \frac{A_{\text{ZZ}}}{\hbar}\right) W_{(n_+,n_-)}^{\text{ZZ-FZZT}_+}(E)
   \nonumber \\ & + \sigma_{\text{FZZT}_-}\exp\left(-\frac{A_{\text{FZZT}}(E)}{\hbar}\right)\sum_{\substack{n_\pm \in \mathbb{N}_0}}'\sigma_{\text{ZZ}_+}^{n_+}\sigma_{\text{ZZ}_-}^{n_-}\exp\left(-(n_+-n_-) \frac{A_{\text{ZZ}}}{\hbar}\right) W_{(n_+,n_-)}^{\text{ZZ-FZZT}_-}(E)
\end{align}
where the perturbative sector can be written as (see equation \eqref{eq:nonperef11})
\begin{equation}
    W^{\text{pert}}(E) = \sum_{g \in \mathbb{N}_0}W_{g,1}(E)\hbar^{2g-1}.
\end{equation}

To compute the coefficients of the transseries sectors above, we make use of the auxiliary transseries \eqref{eq:zz1} and equate its derivative with the resolvent transseries \eqref{eq:transseriesdata16} (this equation follows from \eqref{eq:nonperef3}). The resulting relation decomposes into an infinite family of recursive algebraic and differential equations, labelled by the monomials constructed from the transseries parameters $\sigma_{\text{ZZ}\pm}$ and $\sigma_{\text{FZZT}_\pm}$. Equations \eqref{eq:zz24}, \eqref{eq:zz25} and \eqref{eq:zz16} provide representative examples of such recursive relations. The final step consists in setting $x$ to the fermi surface value \eqref{eq:toprec4}. In the following, we report the results of this computation by presenting the first few transseries coefficients, some of which are already displayed in the main body of the paper (particularly in Subsection $\ref{subsec:zzeffects}$).

The perturbative coefficients follow the structure (compare with equation \eqref{eq:transseriesdata11})
\begin{equation}
    W_{g,1}(E) = \frac{\mathsf{p}_g(E)}{(1-E)^{3g-\frac{1}{2}}}
\end{equation}
for some polynomials $\mathsf{p}_g(y)$ among which the first couple read
\begin{align}
    \mathsf{p}_0(y) =&\frac{y}{\sqrt{2}}+\frac{1}{2 \sqrt{2}}  \\
     \mathsf{p}_1(y) =& \frac{y}{36 \sqrt{2}}-\frac{5}{72 \sqrt{2}}  \\
      \mathsf{p}_2(y) =&-\frac{7 y^4}{243 \sqrt{2}}+\frac{77 y^3}{486 \sqrt{2}}-\frac{119 y^2}{324 \sqrt{2}}+\frac{3521 y}{7776 \sqrt{2}}-\frac{4291}{15552 \sqrt{2}}  \\
       \mathsf{p}_3(y) =&  \frac{2450 \sqrt{2} y^7}{19683}-\frac{20825 \sqrt{2} y^6}{19683}+\frac{52675 y^5}{6561 \sqrt{2}}-\frac{1408295 y^4}{78732 \sqrt{2}}+\frac{4066615 y^3}{157464 \sqrt{2}}-\frac{2617699 y^2}{104976 \sqrt{2}}\nonumber \\ &+\frac{39255671 y}{2519424 \sqrt{2}}-\frac{25936015}{5038848 \sqrt{2}}.
\end{align}

The ZZ transseries sectors follow the structure (see equations \eqref{eq:zz21} and \eqref{eq:zz22})
\begin{align}
    &W_{(n,0)}^{\text{ZZ}}(E) = \hbar^{\frac{n}{2}}\sum_{g \in \mathbb{N}_0} W_{g,(n,0)}^{\text{ZZ}}(E)\hbar^{g} 
    \label{eq:transseriesdata20}\\ 
   & W_{(0,n)}^{\text{ZZ}}(E) = \hbar^{\frac{n}{2}}\sum_{g \in \mathbb{N}_0} W_{g,(0,n)}^{\text{ZZ}}(E)\hbar^{g}
   \label{eq:transseriesdata21}.
\end{align}
The coefficients in \eqref{eq:transseriesdata20} can be written as
\begin{equation}
    W_{g,(n,0)}^{\text{ZZ}}(E) = \frac{\mathsf{p}_{g,(n,0)}^{\text{ZZ}}(E)}{(2 E +1)^{2 g+1} (1-E )^{\left\lfloor \frac{g}{2}\right\rfloor +g+\frac{1}{2}}}
\end{equation}
for some polynomials $\mathsf{p}_{g,(n,0)}^{\text{ZZ}}(y)$ among which the first couple read
\begin{align}
    \mathsf{p}_{0,(1,0)}^{\text{ZZ}}(y) = & -\frac{1}{2 \sqrt{6}} \\ 
    \mathsf{p}_{1,(1,0)}^{\text{ZZ}}(y) = & -\frac{17 y^3}{36 \sqrt{2}}+\frac{17 y}{48 \sqrt{2}}+\frac{161}{144 \sqrt{2}}\\
    \mathsf{p}_{2,(1,0)}^{\text{ZZ}}(y) = & \frac{1993 y^7}{432 \sqrt{6}}-\frac{1993 y^6}{432 \sqrt{6}}-\frac{1993 y^5}{288 \sqrt{6}}+\frac{1729 y^4}{432 \sqrt{6}}+\frac{38105 y^3}{6912 \sqrt{6}}-\frac{25129 y^2}{2304 \sqrt{6}}+\frac{122707 y}{6912 \sqrt{6}}-\frac{104137}{6912 \sqrt{6}}\\
    \mathsf{p}_{0,(2,0)}^{\text{ZZ}}(y) = & -\frac{1}{24 \sqrt{6}}\\ 
    \mathsf{p}_{1,(2,0)}^{\text{ZZ}}(y) = & -\frac{79 y^3}{648 \sqrt{2}}+\frac{79 y}{864 \sqrt{2}}+\frac{295}{2592 \sqrt{2}}\\ 
    \mathsf{p}_{2,(2,0)}^{\text{ZZ}}(y) = & \frac{2273 y^7}{1296 \sqrt{6}}-\frac{2273 y^6}{1296 \sqrt{6}}-\frac{2273 y^5}{864 \sqrt{6}}+\frac{1589 y^4}{1296 \sqrt{6}}+\frac{49585 y^3}{20736 \sqrt{6}}-\frac{5585 y^2}{6912 \sqrt{6}}+\frac{16859 y}{20736 \sqrt{6}}-\frac{30281}{20736 \sqrt{6}}\\
     \mathsf{p}_{0,(3,0)}^{\text{ZZ}}(y) = & -\frac{1}{288 \sqrt{6}}\\ 
      \mathsf{p}_{1,(3,0)}^{\text{ZZ}}(y) = & -\frac{89 y^3}{5184 \sqrt{2}}+\frac{89 y}{6912 \sqrt{2}}+\frac{233}{20736 \sqrt{2}}\\
       \mathsf{p}_{2,(3,0)}^{\text{ZZ}}(y) = & \frac{57151 y^7}{186624 \sqrt{6}}-\frac{57151 y^6}{186624 \sqrt{6}}-\frac{57151 y^5}{124416 \sqrt{6}}+\frac{41095 y^4}{186624 \sqrt{6}}+\frac{1228463 y^3}{2985984 \sqrt{6}}-\frac{65503 y^2}{995328 \sqrt{6}}\nonumber \\ & -\frac{9851 y}{2985984 \sqrt{6}}-\frac{424639}{2985984 \sqrt{6}}.
\end{align}

The coefficients in \eqref{eq:transseriesdata21} follow the structure (compare with equation \eqref{eq:zz6})
\begin{equation}
     W_{g,(0,n)}^{\text{ZZ}}(E) = (-1)^{g+1} W_{g,(n,0)}^{\text{ZZ}}(E).
\end{equation}

The first bulk ZZ transseries sector can be written as (see equation \eqref{eq:zz23})
\begin{equation}
    W_{(1,1)}^{\text{ZZ}}(E) = \sum_{g \in \mathbb{N}_0} W_{g,(1,1)}^{\text{ZZ}}(E)\hbar^{2g}
\end{equation}
where the coefficients follow the rather simple structure
\begin{equation}
    W_{g,(1,1)}^{\text{ZZ}}(E) = \frac{\mathsf{p}_{g,(1,1)}^{\text{ZZ}}(E)}{(2 E +1)^{4 g+1} (1-E )^{3 g+\frac{1}{2}}}
\end{equation}
for some polynomials $\mathsf{p}_{g,(1,1)}^{\text{ZZ}}(y)$ among which the first couple read
\begin{align}
    \mathsf{p}_{0,(1,1)}^{\text{ZZ}}(y) = & -\frac{3}{2 \sqrt{2}}\\  
     \mathsf{p}_{1,(1,1)}^{\text{ZZ}}(y) = &-\frac{25 y^7}{9 \sqrt{2}}+\frac{25 y^6}{9 \sqrt{2}}+\frac{25 y^5}{6 \sqrt{2}}-\frac{49 y^4}{9 \sqrt{2}}-\frac{41 y^3}{144 \sqrt{2}}-\frac{1391 y^2}{48 \sqrt{2}}+\frac{3173 y}{144 \sqrt{2}}-\frac{1205}{144 \sqrt{2}} \\
      \mathsf{p}_{2,(1,1)}^{\text{ZZ}}(y) = & \frac{15827 y^{14}}{81 \sqrt{2}}-\frac{15827 \sqrt{2} y^{13}}{81}-\frac{15827 \sqrt{2} y^{12}}{81}+\frac{27629 y^{11}}{27 \sqrt{2}}+\frac{80227 y^{10}}{216 \sqrt{2}}-\frac{64309 y^9}{54 \sqrt{2}}\nonumber \\ & -\frac{74263 y^8}{432 \sqrt{2}}+\frac{274823 y^7}{432 \sqrt{2}}-\frac{2941951 y^6}{6912 \sqrt{2}}+\frac{6614567 y^5}{10368 \sqrt{2}}-\frac{213015443 y^4}{20736 \sqrt{2}}+\frac{9308357 y^3}{648 \sqrt{2}}\nonumber \\ & -\frac{95592445 y^2}{6912 \sqrt{2}}+\frac{21517267 y}{3456 \sqrt{2}}-\frac{2690549}{2304 \sqrt{2}}.
\end{align}

The FZZT sectors follow the structure (see equation \eqref{eq:fzzt7})
\begin{equation}
  W^{\text{FZZT}_\pm}(E) = \sum_{g=0}^{+\infty} W^{\text{FZZT}_\pm}_{g}(E)\hbar^{g}.
\end{equation}
The coefficients associated with the instanton action $A_{\text{FZZT}}(E)$ can be written as
\begin{equation}
    W^{\text{FZZT}_+}_{g}(E) = \frac{\mathsf{p}_{g}^{\text{FZZT}}(E)}{(1-E )^{\frac{3 g}{2}+1}}
\end{equation}
for some polynomials $\mathsf{p}_{g}^{\text{FZZT}}(y)$ among which the first couple read
\begin{align}
    \mathsf{p}_{0}^{\text{FZZT}}(y) = & \frac{1}{2} \\  
    \mathsf{p}_{1}^{\text{FZZT}}(y) = & \frac{y}{18 \sqrt{2}}-\frac{19}{36 \sqrt{2}}\\
    \mathsf{p}_{2}^{\text{FZZT}}(y) = & \frac{49 y^2}{648}-\frac{259 y}{648}+\frac{2065}{2592}\\
    \mathsf{p}_{3}^{\text{FZZT}}(y) = & -\frac{1}{243} 7 \sqrt{2} y^4+\frac{9289 y^3}{17496 \sqrt{2}}-\frac{116711 y^2}{58320 \sqrt{2}}+\frac{462119 y}{116640 \sqrt{2}}-\frac{2698997}{699840 \sqrt{2}}\\ 
    \mathsf{p}_{4}^{\text{FZZT}}(y) = & -\frac{595 y^5}{2187}+\frac{2777761 y^4}{1259712}-\frac{24931823 y^3}{3149280}+\frac{68331641 y^2}{4199040}-\frac{50519035 y}{2519424}+\frac{1271229197}{100776960}.
\end{align}
The coefficients associated with the instanton action $-A_{\text{FZZT}}(E)$ follow the rather simple structure (compare with equation \eqref{eq:fzzt8})
\begin{equation}
   W^{\text{FZZT}_-}_{g}(E) = (-1)^{g+1}W^{\text{FZZT}_+}_{g}(E).
\end{equation}

Finally, the mixed ZZ-FZZT transseries sectors can be written as (see equations \eqref{eq:zz-fzzt3} and \eqref{eq:zz-fzzt4}) 
\begin{align}
   & W_{(n,0)}^{\text{ZZ-FZZT}_\pm}(E) = \hbar^{\frac{n}{2}}\sum_{g \in \mathbb{N}_0} W_{g,(n,0)}^{\text{ZZ-FZZT}_\pm}(E) \hbar^g  \label{eq:transseriesdata22}
   \\ 
    & W_{(0,n)}^{\text{ZZ-FZZT}_\pm}(E) = \hbar^{\frac{n}{2}}\sum_{g \in \mathbb{N}_0} W_{g,(0,n)}^{\text{ZZ-FZZT}_\pm}(E) \hbar^g \label{eq:transseriesdata14}
\end{align}
where the coefficients in \eqref{eq:transseriesdata22} for $n = 1$ can be written as
\begin{equation}
    W_{g,(1,0)}^{\text{ZZ-FZZT}_\pm}(E) = \frac{\mathsf{p}_{g}^{\text{ZZ-FZZT}_\pm}(E)+(1-E)^{\frac{1}{2}}\mathsf{q}_{g}^{\text{ZZ-FZZT}_\pm}(E)}{(2 E +1)^{2 g+2} (1-E )^{\frac{3 g}{2}+\frac{1}{2}}}
\end{equation}
for some polynomials $\mathsf{p}_{g}^{\text{ZZ-FZZT}_\pm}(y)$ and $\mathsf{q}_{g}^{\text{ZZ-FZZT}_\pm}(y)$ among which the first couple read
\begin{align}
    \mathsf{p}_{0}^{\text{ZZ-FZZT}_+}(y) =& \frac{5}{\sqrt{6}}-\sqrt{\frac{2}{3}} y\\
    \mathsf{p}_{1}^{\text{ZZ-FZZT}_+}(y) =& \frac{11 y^4}{9 \sqrt{3}}+\frac{29 y^3}{9 \sqrt{3}}+\frac{161 y^2}{12 \sqrt{3}}-\frac{871 y}{18 \sqrt{3}}+\frac{241}{9 \sqrt{3}}\\
    \mathsf{p}_{2}^{\text{ZZ-FZZT}_+}(y) =& \frac{1993 y^8}{108 \sqrt{6}}+\frac{8483 y^7}{648 \sqrt{6}}-\frac{1205 y^6}{36 \sqrt{6}}-\frac{62879 y^5}{1296 \sqrt{6}}-\frac{250477 y^4}{5184 \sqrt{6}}-\frac{2770297 y^3}{10368 \sqrt{6}}\nonumber \\ & +\frac{14415991 y^2}{10368 \sqrt{6}}-\frac{5131201 y}{3456 \sqrt{6}}+\frac{5170399}{10368 \sqrt{6}}\\  
    \mathsf{q}_{0}^{\text{ZZ-FZZT}_+}(y) =& -2\\
    \mathsf{q}_{1}^{\text{ZZ-FZZT}_+}(y) =& -\frac{17 y^4}{9 \sqrt{2}}-\frac{59 y^3}{18 \sqrt{2}}-\frac{61 y^2}{36 \sqrt{2}}+\frac{2003 y}{72 \sqrt{2}}-\frac{527}{24 \sqrt{2}}\\
    \mathsf{q}_{2}^{\text{ZZ-FZZT}_+}(y) =& \frac{539 y^7}{324}+\frac{1813 y^6}{324}+\frac{2065 y^5}{216}+\frac{1255 y^4}{108}+\frac{10601 y^3}{1728}-\frac{63923 y^2}{192}+\frac{2623915 y}{5184}-\frac{1054549}{5184}.
\end{align}
The polynomials $ \mathsf{p}^{\text{ZZ-FZZT}_-}_{g}(y)$ and $\mathsf{q}^{\text{ZZ-FZZT}_-}_{g}(y)$ follow the structure (compare with equations \eqref{eq:transseriesdata12} and \eqref{eq:transseriesdata13})
\begin{align}
   \mathsf{p}^{\text{ZZ-FZZT}_-}_{g}(y) &= (-1)^g\mathsf{p}^{\text{ZZ-FZZT}_+}_{g}(y)  \\
     \mathsf{q}^{\text{ZZ-FZZT}_-}_{g}(y) &= (-1)^{g+1}\mathsf{p}^{\text{ZZ-FZZT}_+}_{g}(y).
\end{align}

The coefficients in \eqref{eq:transseriesdata14} can be written as (compare with equation \eqref{eq:transseriesdata15})
\begin{equation}
    W^{\text{ZZ-FZZT}_\pm}_{g,(0,n)}(E) = (-1)^{g+1} W^{\text{ZZ-FZZT}_\mp}_{g,(n,0)}(E).
\end{equation}

\pagebreak

\section{Matrix integral saddle-point expansions}

\label{appendix:Matrixintegrals}

In this appendix, we compute truncated saddle-point expansions for a number of matrix integrals appearing in the expressions of non-perturbative topological recursion. We consider a generic one-cut Hermitian matrix model whose spectral curve features at least one saddle $E^\star \in \mathbb{C}$. To simplify the presentation of the following computations, for each contour $\gamma(z_1,\cdots,z_n) \subset \Sigma$, we introduce the notation $\eta(E_1,\cdots,E_n) = \gamma(z_1(E_1),\cdots,z_n(E_n))$.


\paragraph{The one-instanton matrix integral:}

Here, we consider the matrix integral
\begin{equation}
    \Delta_{z(E)} \frac{Z_{(1,0)}^{\text{ZZ}}}{Z_{(0,0)}^{\text{ZZ}}} = \frac{Z(t-\hbar)}{Z(t)} \int_{\mathcal{C}^\star} \frac{\rmd E_1}{2\pi} \left(\Delta_{z(E)}\mathbb{S}(\eta(E_1))\right)\psi(\eta(E_1))
\end{equation}
where the contour $\gamma(z)$ is depicted in figure \ref{fig:Cycles}. Performing a saddle-point expansion of the integral above yields
\begin{equation}
     \Delta_{z(E)} \frac{Z_{(1,0)}^{\text{ZZ}}}{Z_{(0,0)}^{\text{ZZ}}} = \frac{Z(t-\hbar)}{Z(t)}\exp\left(-\frac{A_{\text{ZZ}}}{\hbar}\right)\left(C_1(E) \hbar^{\frac{1}{2}} + C_2(E)  \hbar^{\frac{3}{2}} + \mathcal{O}\left(\hbar^{\frac{5}{2}}\right)\right)
     \label{eq:matrixintegral1}
\end{equation}
where
\begin{equation}
    C_1(E) = \frac{\exp\left(\mathbb{S}_1(\eta(E^\star))\right) \Delta_{z(E)} \mathbb{S}_0(\eta(E^\star))}{\sqrt{2 \pi } \sqrt{-\partial_E^2\mathbb{S}_0(\eta(E^\star))}}
\end{equation}
and
\begin{align}
    C_2(E) =& -\frac{1}{24 \sqrt{2 \pi } \left(-\partial_E^2\mathbb{S}_0(\eta(E^\star))\right)^{\frac{7}{2}}}\Big(\exp\left(\mathbb{S}_1(\eta(E^\star))\right) \Big(12 \partial_E^2\mathbb{S}_0(\eta(E^\star)) \Big(-\partial_E^2\mathbb{S}_0(\eta(E^\star)) \partial_E^2\Delta_{z(E)}\mathbb{S}_0(\eta(E^\star)) \nonumber \\& +\left(\partial_E^3\mathbb{S}_0(\eta(E^\star))-2 \partial_E\mathbb{S}_1(\eta(E^\star)) \partial_E^2\mathbb{S}_0(\eta(E^\star))\right) \partial_E\Delta_{z(E)}\mathbb{S}_0(\eta(E^\star))+2 \partial_E^2\mathbb{S}_0(\eta(E^\star))^2 \Delta_{z(E)}\mathbb{S}_1(\eta(E^\star))\Big) \nonumber\\ &+\Big(12 \partial_E\mathbb{S}_1(\eta(E^\star)) \partial_E^3\mathbb{S}_0(\eta(E^\star)) \partial_E^2\mathbb{S}_0(\eta(E^\star))-12 \partial_E\mathbb{S}_1(\eta(E^\star))^2 \partial_E^2\mathbb{S}_0(\eta(E^\star))^2+3 \Big(8 \mathbb{S}_2(\eta(E^\star))\times \nonumber \\ & \partial_E^2\mathbb{S}_0(\eta(E^\star))^2 -4 \mathbb{S}''_1(\eta(E^\star)) \partial_E^2\mathbb{S}_0(\eta(E^\star))+\partial_E^4\mathbb{S}_0(\eta(E^\star))\Big) \partial_E^2\mathbb{S}_0(\eta(E^\star))-5 \partial_E^3\mathbb{S}_0(\eta(E^\star))^2\Big)\times \nonumber\\& \Delta_{z(E)}\mathbb{S}_0(\eta(E^\star))\Big)\Big).
\end{align}
%


\paragraph{The two-instanton matrix integral:}

Here, we consider the matrix integral
\begin{equation}
    \Delta_{z(E)} \frac{Z_{(2,0)}^{\text{ZZ}}}{Z_{(0,0)}^{\text{ZZ}}} = \frac{Z(t-2\hbar)}{Z(t)} \int_{\mathcal{C}^\star} \frac{\rmd E_1}{2\pi}\int_{\mathcal{C}^\star} \frac{\rmd E_2}{2\pi}  \left(\Delta_{z(E)}\mathbb{S}(\eta(E_1,E_2))\right)(E_1-E_2)^2\psi(\eta(E_1,E_2))
\end{equation}
where the contour $\gamma(z_1,z_2)$ is depicted in figure \ref{fig:nCycle}. Performing a saddle-point expansion of the integral above yields
\begin{equation}
     \Delta_{z(E)} \frac{Z_{(2,0)}^{\text{ZZ}}}{Z_{(0,0)}^{\text{ZZ}}} = \frac{Z(t-2\hbar)}{Z(t)}\exp\left(-\frac{2A_{\text{ZZ}}}{\hbar}\right)\left(C_1(E) \hbar + C_2(E)  \hbar^{2} + \mathcal{O}\left(\hbar^{\frac{5}{2}}\right)\right)
      \label{eq:matrixintegral2}
\end{equation}
where
\begin{equation}
    C_1(E) = \frac{\exp\left(\mathbb{S}_1(\eta(E^\star,E^\star))\right) \Delta_{z(E)} \mathbb{S}_0(\eta(E^\star,E^\star))}{2\pi \partial_{E_1}^2\mathbb{S}_0(\eta(E^\star,E^\star))}
\end{equation}
and
\begin{align}
    C_2(E) = &\frac{\exp\left(\mathbb{S}_1(\eta(E^\star,E^\star))\right)}{24 \pi  \partial_{E_1}^2\mathbb{S}_0(\eta(E^\star,E^\star))^5} \Big(12 \partial_{E_1}^2\mathbb{S}_0(\eta(E^\star,E^\star)) \Big(-2 \partial_{E_1}\mathbb{S}_1(\eta(E^\star,E^\star)) \partial_{E_1}^2\mathbb{S}_0(\eta(E^\star,E^\star)) \times \nonumber \\ & \partial_{E_1}\Delta_{z(E)} \mathbb{S}_0(\eta(E^\star,E^\star))+\partial_{E_1}^2\mathbb{S}_0(\eta(E^\star,E^\star)) \Big(\partial_{E_2} \partial_{E_1}\Delta_{z(E)} \mathbb{S}_0(\eta(E^\star,E^\star))-2 \partial_{E_1}^2\Delta_{z(E)}\mathbb{S}_0(\eta(E^\star,E^\star))\Big) \nonumber \\& +2 \partial_{E_1}^3\mathbb{S}_0(\eta(E^\star,E^\star)) \Delta_{z(E)}\partial_{E_1} \mathbb{S}_0(\eta(E^\star,E^\star))+\partial_{E_1}^2\mathbb{S}_0(\eta(E^\star,E^\star))^2 \Delta_{z(E)} \mathbb{S}_1(\eta(E^\star,E^\star))\Big)\nonumber \\ & +\Big(-12 \partial_{E_1}\mathbb{S}_1(\eta(E^\star,E^\star))^2 \partial_{E_1}^2\mathbb{S}_0(\eta(E^\star,E^\star))^2+24 \partial_{E_1}\mathbb{S}_1(\eta(E^\star,E^\star)) \partial_{E_1}^3\mathbb{S}_0(\eta(E^\star,E^\star)) \times \nonumber \\ & \partial_{E_1}^2\mathbb{S}_0(\eta(E^\star,E^\star))+3 \Big(4 \partial_{E_1}^2\mathbb{S}_0(\eta(E^\star,E^\star)) \Big(\partial_{E_2}\partial_{E_1}\mathbb{S}_1(\eta(E^\star,E^\star))+\mathbb{S}_2(\eta(E^\star,E^\star)) \partial_{E_1}^2\mathbb{S}_0(\eta(E^\star,E^\star)) \nonumber \\ & -2 \partial_{E_1}^2\mathbb{S}_1(\eta(E^\star,E^\star))\Big)+3 \partial_{E_1}^4\mathbb{S}_0(\eta(E^\star,E^\star))\Big) \partial_{E_1}^2\mathbb{S}_0(\eta(E^\star,E^\star))-17 \partial_{E_1}^3\mathbb{S}_0(\eta(E^\star,E^\star))^2\Big)\times \nonumber \\ &\Delta_{z(E)} \mathbb{S}_0(\eta(E^\star,E^\star))\Big).
\end{align}
%


\paragraph{The three-instanton matrix integral:}
Here, we consider the matrix integral
\begin{align}
     \Delta_{z(E)} \frac{Z_{(3,0)}^{\text{ZZ}}}{Z_{(0,0)}^{\text{ZZ}}} &= \frac{1}{6}\frac{Z(t-3\hbar)}{Z(t)} \int_{\mathcal{C}^\star} \frac{\rmd E_1}{2\pi}\int_{\mathcal{C}^\star} \frac{\rmd E_2}{2\pi}  \int_{\mathcal{C}^\star} \frac{\rmd E_3}{2\pi} \left(\Delta_{z(E)}\mathbb{S}(\eta(E_1,E_2,E_3))\right) \times \nonumber\\ &  (E_1-E_2)^2(E_1-E_3)^2(E_2-E_3)^2\psi(\eta(E_1,E_2,E_3))
\end{align}
where the contour $\gamma(z_1,z_2,z_3)$ is given by equation \eqref{eq:nonpertopef2}. Performing a saddle-point expansion of the integral above to leading order in $\hbar$ yields
\begin{equation}
     \Delta_{z(E)} \frac{Z_{(3,0)}^{\text{ZZ}}}{Z_{(0,0)}^{\text{ZZ}}}  = \frac{Z(t-3\hbar)}{Z(t)}\exp\left(-\frac{3A_{\text{ZZ}}}{\hbar}\right)\left(C_1(E) \hbar^{\frac{7}{2}}+\mathcal{O}\left(\hbar^{\frac{9}{2}}\right)\right)
      \label{eq:matrixintegral3}
\end{equation}
where
\begin{equation}
    C_1(E) = \frac{\exp\left(\mathbb{S}_1(\eta(E^\star,E^\star,E^\star))\right) \Delta_{z(E)} \mathbb{S}_0(\eta(E^\star,E^\star,E^\star))}{\sqrt{2}\pi^{\frac{3}{2}}\left(-\partial_{E_1}^2\mathbb{S}_0(\eta(E^\star,E^\star,E^\star))\right)^{\frac{9}{2}}}.
\end{equation}


\paragraph{The one-anti-instanton matrix integral:}

Here, we consider the matrix integral
\begin{equation}
    \Delta_{z(E)} \frac{Z_{(0,1)}^{\text{ZZ}}}{Z_{(0,0)}^{\text{ZZ}}} = \frac{Z(t+\hbar)}{Z(t)} \int_{\bar{\mathcal{C}}^\star} \frac{\rmd E_1}{2\pi} \left(\Delta_{z(E)}\mathbb{S}(\eta(E_1))\right)\psi(\eta(E_1))
\end{equation}
where the contour $\gamma(z)$ is depicted in figure \ref{fig:AntiCycle}. Performing a saddle-point expansion of the integral above yields
\begin{equation}
     \Delta_{z(E)} \frac{Z_{(0,1)}^{\text{ZZ}}}{Z_{(0,0)}^{\text{ZZ}}} = \frac{Z(t+\hbar)}{Z(t)}\exp\left(\frac{A_{\text{ZZ}}}{\hbar}\right)\left(C_1(E) \hbar^{\frac{1}{2}} + C_2(E)  \hbar^{\frac{3}{2}} + \mathcal{O}\left(\hbar^{\frac{5}{2}}\right)\right)
      \label{eq:matrixintegral4}
\end{equation}
where
\begin{equation}
    C_1(E) = -\frac{\exp\left(\mathbb{S}_1(\eta(E^\star))\right) \Delta_{z(E)} \mathbb{S}_0(\eta(E^\star))}{\sqrt{2 \pi } \sqrt{-\partial_E^2\mathbb{S}_0(\eta(E^\star))}}
\end{equation}
and
\begin{align}
    C_2(E) =& -\frac{1}{24 \sqrt{2 \pi } \left(-\partial_E^2\mathbb{S}_0(\eta(E^\star))\right)^{\frac{7}{2}}}\Big(\exp\left(\mathbb{S}_1(\eta(E^\star))\right) \Big(12 \partial_E^2\mathbb{S}_0(\eta(E^\star)) \Big(-\partial_E^2\mathbb{S}_0(\eta(E^\star)) \partial_E^2\Delta_{z(E)}\mathbb{S}_0(\eta(E^\star)) \nonumber \\& +\left(\partial_E^3\mathbb{S}_0(\eta(E^\star))-2 \partial_E\mathbb{S}_1(\eta(E^\star)) \partial_E^2\mathbb{S}_0(\eta(E^\star))\right) \partial_E\Delta_{z(E)}\mathbb{S}_0(\eta(E^\star))+2 \partial_E^2\mathbb{S}_0(\eta(E^\star))^2 \Delta_{z(E)}\mathbb{S}_1(\eta(E^\star))\Big) \nonumber\\& +\Big(12 \partial_E\mathbb{S}_1(\eta(E^\star)) \partial_E^3\mathbb{S}_0(\eta(E^\star)) \partial_E^2\mathbb{S}_0(\eta(E^\star))-12 \partial_E\mathbb{S}_1(\eta(E^\star))^2 \partial_E^2\mathbb{S}_0(\eta(E^\star))^2+3 \Big(8 \mathbb{S}_2(\eta(E^\star))\times \nonumber \\ &  \partial_E^2\mathbb{S}_0(\eta(E^\star))^2 -4 \mathbb{S}''_1(\eta(E^\star)) \partial_E^2\mathbb{S}_0(\eta(E^\star))+\partial_E^4\mathbb{S}_0(\eta(E^\star))\Big) \partial_E^2\mathbb{S}_0(\eta(E^\star))-5 \partial_E^3\mathbb{S}_0(\eta(E^\star))^2\Big) \times \nonumber\\& \Delta_{z(E)}\mathbb{S}_0(\eta(E^\star))\Big)\Big).
\end{align}


\paragraph{The two-anti-instanton matrix integral:}

Here, we consider the matrix integral
\begin{equation}
    \Delta_{z(E)} \frac{Z_{(0,2)}^{\text{ZZ}}}{Z_{(0,0)}^{\text{ZZ}}} = \frac{Z(t+2\hbar)}{Z(t)} \int_{\bar{\mathcal{C}}^\star} \frac{\rmd E_1}{2\pi}\int_{\bar{\mathcal{C}}^\star} \frac{\rmd E_2}{2\pi}  \left(\Delta_{z(E)}\mathbb{S}(\eta(E_1,E_2))\right)(E_1-E_2)^2\psi(\eta(E_1,E_2))
\end{equation}
where the contour $\gamma(z_1,z_2)$ is depicted in figure \ref{fig:AntiCycle}. Performing a saddle-point expansion of the integral above yields
\begin{equation}
     \Delta_{z(E)} \frac{Z_{(0,2)}^{\text{ZZ}}}{Z_{(0,0)}^{\text{ZZ}}} = \frac{Z(t+2\hbar)}{Z(t)}\exp\left(\frac{2A_{\text{ZZ}}}{\hbar}\right)\left(C_1(E) \hbar + C_2(E)  \hbar^{2} + \mathcal{O}\left(\hbar^{\frac{5}{2}}\right)\right)
      \label{eq:matrixintegral5}
\end{equation}
where
\begin{equation}
    C_1(E) = -\frac{\exp\left(\mathbb{S}_1(\eta(E^\star,E^\star))\right) \Delta_{z(E)} \mathbb{S}_0(\eta(E^\star,E^\star))}{2\pi \partial_{E_1}^2\mathbb{S}_0(\eta(E^\star,E^\star))}
\end{equation}
and
\begin{align}
    C_2(E) = &\frac{\exp\left(\mathbb{S}_1(\eta(E^\star,E^\star))\right)}{24 \pi  \partial_{E_1}^2\mathbb{S}_0(\eta(E^\star,E^\star))^5} \Big(12 \partial_{E_1}^2\mathbb{S}_0(\eta(E^\star,E^\star)) \Big(-2 \partial_{E_1}\mathbb{S}_1(\eta(E^\star,E^\star)) \partial_{E_1}^2\mathbb{S}_0(\eta(E^\star,E^\star)) \times \nonumber \\ & \partial_{E_1}\Delta_{z(E)} \mathbb{S}_0(\eta(E^\star,E^\star))+\partial_{E_1}^2\mathbb{S}_0(\eta(E^\star,E^\star)) \Big(\partial_{E_2} \partial_{E_1}\Delta_{z(E)} \mathbb{S}_0(\eta(E^\star,E^\star))-2 \partial_{E_1}^2\Delta_{z(E)}\mathbb{S}_0(\eta(E^\star,E^\star))\Big) \nonumber \\& +2 \partial_{E_1}^3\mathbb{S}_0(\eta(E^\star,E^\star)) \Delta_{z(E)}\partial_{E_1} \mathbb{S}_0(\eta(E^\star,E^\star))+\partial_{E_1}^2\mathbb{S}_0(\eta(E^\star,E^\star))^2 \Delta_{z(E)} \mathbb{S}_1(\eta(E^\star,E^\star))\Big)\nonumber \\ & +\Big(-12 \partial_{E_1}\mathbb{S}_1(\eta(E^\star,E^\star))^2 \partial_{E_1}^2\mathbb{S}_0(\eta(E^\star,E^\star))^2+24 \partial_{E_1}\mathbb{S}_1(\eta(E^\star,E^\star)) \partial_{E_1}^3\mathbb{S}_0(\eta(E^\star,E^\star)) \times \nonumber \\ & \partial_{E_1}^2\mathbb{S}_0(\eta(E^\star,E^\star))+3 \Big(4 \partial_{E_1}^2\mathbb{S}_0(\eta(E^\star,E^\star)) \Big(\partial_{E_2}\partial_{E_1}\mathbb{S}_1(\eta(E^\star,E^\star))+\mathbb{S}_2(\eta(E^\star,E^\star)) \partial_{E_1}^2\mathbb{S}_0(\eta(E^\star,E^\star)) \nonumber \\ & -2 \partial_{E_1}^2\mathbb{S}_1(\eta(E^\star,E^\star))\Big)+3 \partial_{E_1}^4\mathbb{S}_0(\eta(E^\star,E^\star))\Big) \partial_{E_1}^2\mathbb{S}_0(\eta(E^\star,E^\star))-17 \partial_{E_1}^3\mathbb{S}_0(\eta(E^\star,E^\star))^2\Big)\times \nonumber \\ &\Delta_{z(E)} \mathbb{S}_0(\eta(E^\star,E^\star))\Big).
\end{align}


\paragraph{The three-anti-instanton matrix integral:}
Here, we consider the matrix integral
\begin{align}
     \Delta_{z(E)} \frac{Z_{(0,3)}^{\text{ZZ}}}{Z_{(0,0)}^{\text{ZZ}}} =& \frac{1}{6}\frac{Z(t+3\hbar)}{Z(t)} \int_{\bar{\mathcal{C}}^\star} \frac{\rmd E_1}{2\pi}\int_{\bar{\mathcal{C}}^\star} \frac{\rmd E_2}{2\pi}  \int_{\bar{\mathcal{C}}^\star} \frac{\rmd E_3}{2\pi} \left(\Delta_{z(E)}\mathbb{S}(\eta(E_1,E_2,E_3))\right) \times \nonumber\\ &  (E_1-E_2)^2(E_1-E_3)^2(E_2-E_3)^2\psi(\eta(E_1,E_2,E_3))
\end{align}
where the contour $\gamma(z_1,z_2,z_3)$ is given by equation \eqref{eq:nonpertopef3}. Performing a saddle-point expansion of the integral above to leading order in $\hbar$ yields
\begin{equation}
     \Delta_{z(E)} \frac{Z_{(0,3)}^{\text{ZZ}}}{Z_{(0,0)}^{\text{ZZ}}}  = \frac{Z(t+3\hbar)}{Z(t)}\exp\left(\frac{3A_{\text{ZZ}}}{\hbar}\right)\left(C_1(E) \hbar^{\frac{7}{2}}+\mathcal{O}\left(\hbar^{\frac{9}{2}}\right)\right)
      \label{eq:matrixintegral6}
\end{equation}
where
\begin{equation}
    C_1(E) = -\frac{\exp\left(\mathbb{S}_1(\eta(E^\star,E^\star,E^\star))\right) \Delta_{z(E)} \mathbb{S}_0(\eta(E^\star,E^\star,E^\star))}{\sqrt{2}\pi^{\frac{3}{2}}\left(-\partial_{E_1}^2\mathbb{S}_0(\eta(E^\star,E^\star,E^\star))\right)^{\frac{9}{2}}}.
\end{equation}


\paragraph{The one-instanton-anti-instanton matrix integral:} Finally, we consider the matrix integral
\begin{align}    \Delta_{z(E)}\frac{Z^{\text{ZZ}}_{(1,1)}}{Z^{\text{ZZ}}_{(0,0)}}  = &\int_{\mathcal{C}^\star} \frac{\rmd E_1}{2\pi}  \int_{\bar{\mathcal{C}}^\star} \frac{\rmd \bar{E}_1}{2\pi}\frac{1}{(E_1-\bar{E}_1)^2}\Delta_{z(E)}\mathbb{S}(\eta(E_1,\bar{E}_1))\psi(\eta(E_1,\bar{E}_1)) \nonumber \\  &+\int_{\mathcal{C}_{\text{res}}} \frac{\rmd E_1}{2\pi}  \int_{\bar{\mathcal{C}}^\star} \frac{\rmd \bar{E}_1}{2\pi} \frac{1}{(E_1-\bar{E}_1)^2}\Delta_{z(E)}\mathbb{S}(\eta(E_1,\bar{E}_1))\psi(\eta(E_1,\bar{E}_1))
\end{align}
where the contour $\gamma(z_1,z_2)$ is depicted in figure \ref{fig:BulkCycle}. Performing a saddle-point expansion of the integral above to leading order in $\hbar$ yields
\begin{equation}
\Delta_{z(E)}\frac{Z^{\text{ZZ}}_{(1,1)}}{Z^{\text{ZZ}}_{(0,0)}} = C_1(E)\hbar^{-1} + \mathcal{O}\left(\hbar\right)
 \label{eq:matrixintegral7}
\end{equation}
where 
\begin{equation}
    C_1(E) = \frac{\rmi}{2\pi}\int_{E_0}^{E^\star}\rmd \bar{E}_1\exp\left(\mathbb{S}_1(\eta\left(\bar{E}_1,\bar{E}_1\right))\right)\partial_{E_1}\Delta_{z(E)}\mathbb{S}_0(\eta(\bar{E}_1,\bar{E}_1)).
\end{equation}

\section{The Borel summation: A review}

\label{appendix:BorelSummation}

Perturbative methods are widely employed throughout physics, from quantum mechanics to quantum field theory and string theory. Despite their power and broad applicability in analytically treating interacting theories, perturbative expansions suffer from a fundamental drawback: the resulting power series are often merely asymptotic.

By definition, asymptotic series have zero radius of convergence in the complex $\hbar$-plane and therefore cannot be summed by conventional means. To extract meaningful information from such series, two main approaches are commonly considered:

\begin{itemize}
    \item In weak-coupling regimes (small $|\hbar|$), one can often obtain a reliable numerical approximation to the exact result by summing only the first few terms of the expansion and truncating the series just before the coefficients begin to grow, thereby achieving optimal accuracy. This procedure is known as ``optimal truncation''.

    \item One can attempt to determine the analytic function (in $\hbar$) whose asymptotic expansion reproduces the series under consideration. Although more laborious, this approach yields the exact solution, valid across all regimes, from weak (small $|\hbar|$) to strong (large $|\hbar|$) coupling. The systematic procedure used to construct such functions is known as the Borel summation procedure.
\end{itemize}

In this appendix, we present a concise introduction to the Borel summation procedure (the second approach). For a more pedagogical and comprehensive review, we refer the reader to \cite{abs19}.
\begin{figure}[h]
    \centering
    \begin{tikzpicture}

    \node[draw,line width = 2pt,align=center,rounded corners = 5pt] at (-3.5,-2) {  
    $\phi(\hbar) = \displaystyle\sum_{g \in \mathbb{N}_0}\phi_g \hbar^g$} ;

    \draw[->,line width = 2pt] (-3.5,-1) -- (-3.5,1);

     \node[align=center] at (-3.1,0){$\mathcal{B}$};

    \node[draw,line width = 2pt,align=center,rounded corners = 5pt] at (-3.5,2) {  
    $\mathcal{B}\left[\phi\right](s) = \displaystyle\sum_{g \in \mathbb{N}_0}\widehat{\phi}_g s^g$} ;

    \draw[->,line width = 2pt] (-1.3,2) -- (1.55,2);

    \node[align=center] at (0.11,2.7){Analytical \\  extension};

    \node[draw,line width = 2pt,align=center,rounded corners = 5pt] at (4,2) {Analytic function $\Phi(s)$};

    \draw[->,line width = 2pt] (3.5,1.3) -- (3.5,-1.07);

    \node[draw,line width = 2pt,align=center,rounded corners = 5pt] at (3.6,-2){ $\mathcal{S}\left[\phi\right](\hbar) = \displaystyle\int_{0}^{+\infty}\rmd se^{-\frac{s}{\hbar}}\Phi(s)$};

\draw[->,line width = 2pt] (-1.57,-2) -- (0.75,-2);

\node[align=center] at (-0.5,-2.40) {$\mathcal{S}$} ;

\node[align=center] at (3.85,0.25) {$\mathcal{L}$} ;

\node[violet] at (0,0){\scalebox{1.4}{Borel summation}};

\def\hspacee{0};

\draw[line width = 2pt,->] (-8+\hspacee,4) -- (-5+\hspacee,4);
\draw[line width = 2pt,->] (-6.5+\hspacee,2.5) -- (-6.5+\hspacee,5.5);
\draw[line width = 2pt,color = violet, fill = violet,fill opacity=0.2] (-6.5+\hspacee,4) circle(0.5);
\node[draw,line width = 1.8pt] at (-5.5+0.4+\hspacee,5+0.4) {$s$};
\node[violet] at (-5.8+\hspacee,4.6) {$\text{D}_{\text{R}}$};

\def\hspacee{13.5};

\draw[line width = 2pt,->] (-8+\hspacee,4) -- (-5+\hspacee,4);
\draw[line width = 2pt,->] (-6.5+\hspacee,2.5) -- (-6.5+\hspacee,5.5);
\node[draw,line width = 1.8pt] at (-5.5+0.4+\hspacee,5+0.4) {$s$};

\draw[fill = red,color = red] (7+0.35,4+0.35) circle (0.1);
\filldraw[rounded corners=5pt, fill=violet, fill opacity = 0.2,color = violet, line width = 2pt]
    (5.5+0.5,2.5+0.5) rectangle (5.5+3-0.5,5.5-0.5);

\def\hspacee{14};
\def\vspace{-8};

\draw[line width = 2pt] (-8+\hspacee,4+\vspace) -- (-6.5+\hspacee,4+\vspace);
\draw[line width = 2pt,color = Orange,->] (-6.5+\hspacee,4+\vspace) -- (-5+\hspacee,4+\vspace);
\draw[line width = 2pt,->] (-6.5+\hspacee,2.5+\vspace) -- (-6.5+\hspacee,5.5+\vspace);
\node[draw,line width = 1.8pt] at (-5.5+0.4+\hspacee,5+0.4+\vspace) {$s$};

\draw[fill = red,color = red] (7+0.35+0.5,4+0.35+\vspace) circle (0.1);
\filldraw[rounded corners=5pt, fill=violet, fill opacity = 0.2,color = violet, line width = 2pt]
    (5.5+0.5+0.5,2.5+0.5+\vspace) rectangle (5.5+3-0.5+0.5,5.5-0.5+\vspace);

    \node at (-6,-4.2) {$|\phi_g| \le C {\color{violet}\text{R}}^g g! \hspace{3pt} g \in \mathbb{N}_0 $};

\end{tikzpicture}
    \caption{Schematic summary of the Borel summation procedure. The red disk indicates a possible location of a singularity of $\Phi(s)$. The Laplace transform integration contour is shown as an orange line and the analyticity domain of the Borel transform is shaded in purple.}
    \label{fig:Borelsummation}
\end{figure}
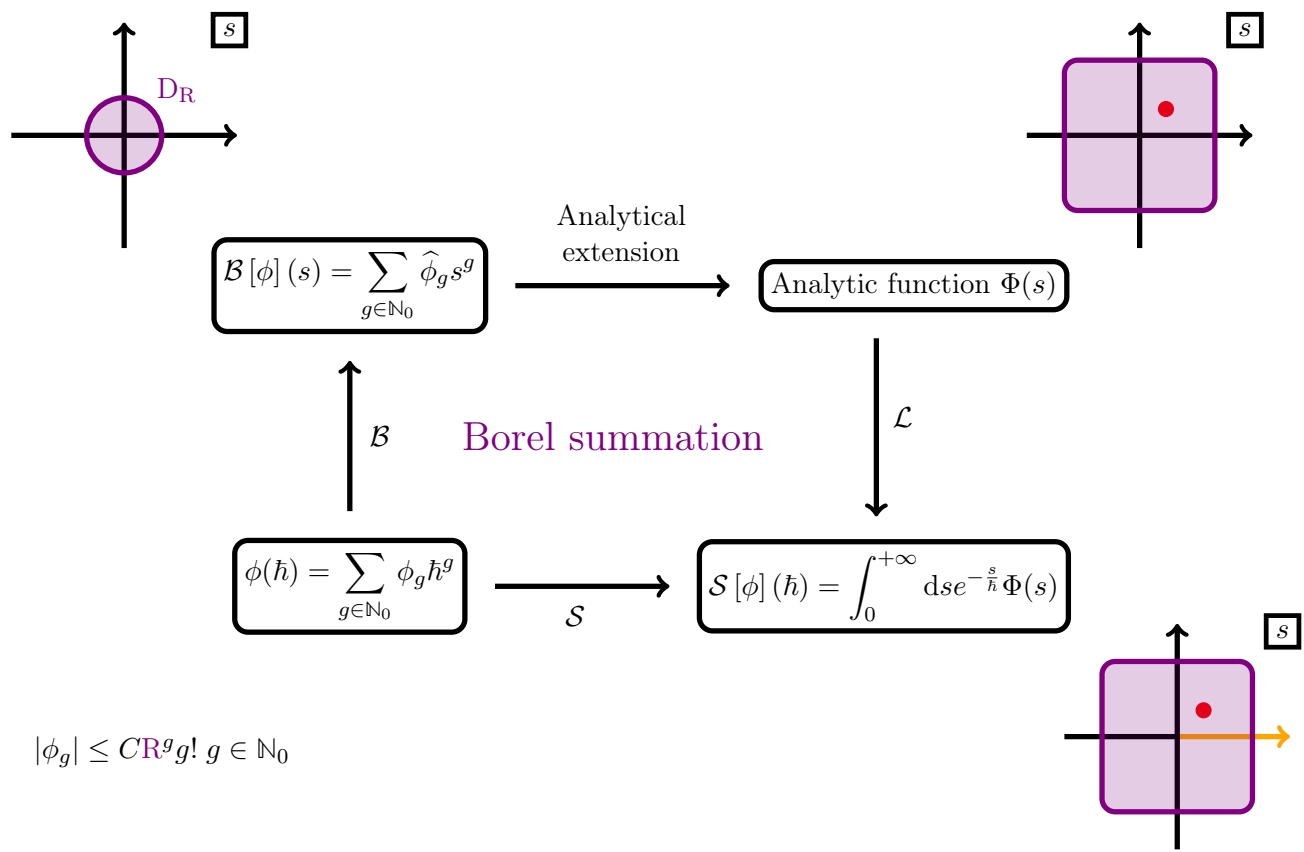

The starting point is a 1-Gevrey formal series \cite{s14}, denote here as
\begin{equation}
    \phi(\hbar) = \sum_{g\in \mathbb{N}_0}\phi_g \hbar^g.
    \label{eq:borelsummation2}
\end{equation}
Being a 1-Gevrey formal series means that the coefficients $\phi_g$ obey the 1-Gevrey estimate
\begin{equation}
    |\phi_g| \le C \text{R}^g g! \hspace{5pt} , \hspace{3pt} g \in \mathbb{N}_0
    \label{eq:borelsummation1}
\end{equation}
for some $C,\text{R} >0$. The first step of the Borel summation procedure is to apply the Borel transform to $\phi(\hbar)$. The Borel transform, denoted here by $\mathcal{B}$, maps the original series to a new one in which the potential factorial growth of the coefficients has been removed. Concretely, the resulting series takes the form
\begin{equation}
    \mathcal{B}[\phi](s) \coloneqq\sum_{g\in \mathbb{N}_0}\widehat{\phi}_gs^g
\end{equation}
where we defined the coefficients
\begin{equation}
  \widehat{\phi}_g =   \frac{\phi_g}{g!}.
\end{equation}

Using the estimate \eqref{eq:borelsummation1}, we can readily write the inequality
\begin{equation}
    |\widehat{\phi}_g| \le C \text{R}^g  \hspace{5pt} , \hspace{3pt} g \in \mathbb{N}_0
\end{equation}
which tell us that $\mathcal{B}[\phi](s)$ has a convergence radius equal to $\text{R}$. 

The Borel transform has turned our asymptotic series into a converging power series, well defined in the disk $\text{D}_{\text{R}} = \left\{s \in \mathbb{C} \hspace{1pt} \vert \hspace{1pt} |s| < \text{R}\right\}\subset \mathbb{C}$. The second step of the Borel summation procedure consists in analytically continuing this function to the complex $s$-plane (the Borel plane), thereby yielding a function $\Phi(s)$ that is analytic except for isolated singularities, such as poles and/or branch points. At this stage, we are closer to the goal of resuming \eqref{eq:borelsummation2}, having obtained a well-defined analytic function constructed solely from the information contained in the asymptotic series. However, this function does not yet satisfy the requirement of admitting an asymptotic expansion identical to \eqref{eq:borelsummation2}. This problem is addressed in the last step of the Borel summation procedure.

The final step consists in applying the Laplace transform to $\Phi(s)$, which can only be carried out provided there are no singularities on the positive real axis, in which case \eqref{eq:borelsummation2} is said to be Borel-summable. The resulting function reads
\begin{equation}
    \mathcal{S}[\phi](\hbar) \coloneqq \mathcal{L}[\Phi](\hbar) = \int_{0}^{+\infty}\rmd se^{-\frac{s}{\hbar}}\Phi(s).
\end{equation}
Term by term in the asymptotic series \eqref{eq:borelsummation2}, the Laplace transform defined above acts as the formal inverse of the Borel transform $\mathcal{B}$, yielding an analytic function whose asymptotic expansion exactly reproduces \eqref{eq:borelsummation2}. We refer the reader to \cite{s14} for a detailed proof of this fact.

The entire Borel summation procedure is schematically summarized in Figure~\ref{fig:Borelsummation}. If the asymptotic series \eqref{eq:borelsummation2} is not Borel-summable, namely, if $\Phi(s)$ exhibits a singularity on the positive real line, we are forced to deform the Laplace transform integration contour slightly above or below the positive real axis in order to avoid the singularity. This deformation introduces an ambiguity in the non-perturbative completion. Resolving this ambiguity requires understanding how the Borel summation procedure changes when the Laplace integration contour is obstructed by singularities in the Borel plane. The set of points at which these obstructions are met form lines in the complex $\hbar$-plane, commonly referred to as Stokes lines. Extending the resummation beyond these lines requires shifting the transseries parameters in accordance with the appropriate connection formulae \cite{bssv22}, an effect known as the Stokes phenomenon.

\pagebreak





\providecommand{\href}[2]{#2}\begingroup\raggedright\endgroup

\end{document}